\documentclass{aa}  

\usepackage{natbib}
\bibpunct{(}{)}{;}{a}{}{,} 
\usepackage[dvipsnames]{xcolor}
\usepackage{graphicx}
\usepackage{txfonts}
\usepackage{lscape}
\DeclareUnicodeCharacter{2212}{\textendash}

\newcommand{\nus}{\emph{NuSTAR} } 
\newcommand{\xmm}{\emph{XMM-Newton} } 
\newcommand{\sft}{\emph{Swift}} 
\newcommand {\pexr}{\texttt{PEXRAV} }

\newcommand {\pexmn}{\texttt{\texttt{PEXMON} }}
\newcommand {\bors}{\texttt{\texttt{borus02} }}
\newcommand {\xill}{\texttt{\texttt{XILLVER} }}

\usepackage{tikz}
\tikzset{
  treenode/.style = {shape=rectangle, rounded corners,
                     draw, align=center,
                     top color=white, bottom color=blue!20},
  root/.style     = {treenode, font=\Large, bottom color=red!30},
  env/.style      = {treenode, font=\ttfamily\normalsize},
  dummy/.style    = {circle,draw}
}

\begin{document}

   \title{Constraining the X-ray reflection in low accretion rate AGN using \emph{XMM-Newton}, \nus and \sft}


   \author{Diaz, Y
          \inst{1}, Hernández-García L.\inst{2,1}, Arévalo P.\inst{1}, López-Navas E.\inst{1}, Ricci C.\inst{3,4,5}, Koss M.\inst{6}, Gonzalez-Martin O.\inst{7}, Balokovi\'{c} M.\inst{8,9}, Osorio-Clavijo N.\inst{7}, Garc\'ia J.A.\inst{10} and Malizia A.\inst{11}}

   \institute{$^{1}$Instituto de F\'isica y Astronom\'ia, Facultad de Ciencias, Universidad de Valparaíso, Gran Bretaña No. 1111, Playa Ancha, Valparaíso, Chile\\
   $^{2}$Millennium Institute of Astrophysics (MAS), Nuncio Monse\~nor S\'otero Sanz 100, Providencia, Santiago, Chile \\
   $^{3}$ N\'ucleo de Astronom\'ia de la Facultad de Ingenier\'ia, Universidad Diego Portales, Av. Ej\'ercito Libertador 441, Santiago, Chile \\
		$^{4}$ Kavli Institute for Astronomy and Astrophysics, Peking University, Beijing 100871, China \\
		$^{5}$ George Mason University, Department of Physics \& Astronomy, MS 3F3, 4400 University Drive, Fairfax, VA 22030, USA\\
		$^{6}$ Eureka Scientific, 2452 Delmer Street Suite 100, Oakland, CA 94602-3017, USA \\
		$^{7}$ Instituto de Radioastronom\'ia and Astrof\'isica (IRyA-UNAM), 3-72 (Xangari), 8701, Morelia, Mexico \\
		$^{8}$ Yale Center for Astronomy \& Astrophysics, 52 Hillhouse Avenue, New Haven, CT 06511, USA \\
		$^{9}$ Department of Physics, Yale University, P.O. Box 208120, New Haven, CT 06520, USA \\
		$^{10}$ Cahill Center for Astronomy and Astrophysics, California Institute of Technology, Pasadena, CA 91125, USA \\
		$^{11}$ OAS-INAF, Via P. Gobetti 101, 40129 Bologna, Italy \\
              \email{yaherlyn.diaz@postgrado.uv.cl}
             }

   \date{Received 4 August 2022; accepted 25 October 2022}

 
  \abstract
   {An interesting feature in active galactic nuclei (AGN) accreting at low rate is the weakness of the reflection features in their X-ray spectra, which can result from the gradual disappearance of the torus with decreasing accretion rates. It has been suggested that low luminosity AGN (LLAGN) would have a different reflector configuration compared with high luminosity AGN, either covering a smaller fraction of the sky or simply having less material. Additionally, we note that the determination of the spectral index ($\Gamma$) and the cut-off energy of the primary power-law emission is affected by the inclusion of reflection models, showing the importance of using them to study the accretion mechanism, especially in the case of the LLAGN that have previously shown a high dispersion on the relation between $\Gamma$ and the accretion rate. }
   {Our purpose is to constrain the geometry and column density of 
   the reflector in a sample of LLAGN covering a broad X-ray range of energy combining data from \emph{XMM-Newton}+ \emph{NuSTAR} + \emph{Swift}. The spectral analysis also allows us to investigate the accretion mechanism in LLAGN.}
   {We use \emph{XMM-Newton}+ \emph{NuSTAR} + \emph{Swift} observations of a hard X-ray-flux limited sample of 17 LLAGN from BASS/DR2 with accretion rates $\lambda_{Edd}$=L$_{\rm Bol}$/L$_{\rm Edd}$<10$^{-3}$. We fit all spectra using the reflection model for torus (\texttt{BORUS}) and accretion disk (\texttt{XILLVER}) reflectors.   }
   { We found a tentative correlation between the torus column density and the accretion rate, LLAGN shows a lower column density compared with the high-luminosity objects. We also confirm the relation 
    between $\Gamma$ and $\lambda_{Edd}$, with a smaller scatter than previously reported, thanks to the inclusion of high-energy data and the reflection models. Our results are consistent with a break at $\lambda_{Edd}\sim10^{-3}$, suggestive of a different accretion mechanism compared with higher accretion AGN.}
   {}

\titlerunning{X-ray reflection in LLAGN}
\authorrunning{Díaz et al. }
\maketitle
%

\section{Introduction}

Active galactic nuclei (AGN) emit over the entire electromagnetic spectrum and are powered by accretion onto a supermassive black hole \citep{1984REES}. The central engine emits UV photons which interact with energetic electrons in the so-called corona \citep{1994Nandra} producing the X-ray emission (e.g., \citealt{1993haardt}). This emission interacts with the circumnuclear dust and gas, producing the obscuration observed in the spectra. The dust emission can be observed in the infrared energy range, as a result of the thermalization of the UV photons by the dust. In the optical, the nuclear emission will be obscured, removing  the continuum and the broad components in the emission lines \citep{1981osterbrock}.  On the other hand, the gas will absorb and scatter the X-ray continuum producing the X-ray absorption, most noticeable at energies below 10 keV in the X-ray spectrum \citep{2011brightmanNandra}. Note that in general, the absorption in the optical and in the X-rays occurs together \citep{2007maineri, 2012malizia, 2014merloni,2015davies, 2017mike}.

Obscuration gives evidence of material in the line of sight, which could be associated with the torus. Gas that is not in the line of sight of the observer can also imprint some features on the X-ray spectrum. Between 10 keV and up to hundreds of keV there is a reflection hump created by X-rays being reflected at the accretion disk \citep{2006Fabian} or more distant material, like the torus \citep{2011brightmanNandra}. Furthermore, the most robust emission line seen in the X-rays, the Fe K$\alpha$ emission line (e.g., \citealt{2006Fabian}), can be related to circumnuclear material, being broad and exhibiting relativistic effects due to its creation close to the supermassive black hole (SMBH), and narrow, presumably originating from more distant material.  These reflection features are therefore a useful tool to study the configuration of the accretion disk and the torus. To better understand the properties of the reflector, many models have been developed, like \texttt{BORUS} \citep{2018balokovic} where the reprocessing medium is assumed to be a sphere with a conical cut-off at both poles, approximating a torus with variable covering factor, \texttt{cTORUS} \citep{2014liu}, similar to \texttt{BORUS} but clumpy and with the half-opening angle of the torus fixed at 60 degrees, \texttt{MYTORUS} \citep{2009Murphy} that proposes a toroidal geometry where the covering fraction is fixed to 0.5, \texttt{RXTorus} \citep{2017Apaltani} a model that assumes absorption and reflection from a torus with a varying ratio of the minor to major axis or \texttt{XILLVER} \citep{2013garcia} that calculates the reflected spectrum from the surface of an X-ray illuminated, ionized accretion disk by solving the equations of radiative transfer, energy balance, and ionization equilibrium in a Compton-thick and plane parallel medium.

It is not clear how the reflecting structure is formed, but clues can be gathered from the relation between reflection strength and the nuclear accretion rate. From the observational point of view, the torus in the infrared (IR) becomes weaker in the low luminosity regime (i.e., for low accretion rates - below 10$^{-3}$, \citealt{2017Gonzalez-martin}). Furthermore, in the X-rays, it has been seen that the Compton thin absorption (N$_{\rm H}$<1.5x10$^{24}$ cm$^{-2}$) is less frequent in objects with low accretion rates: the fraction of Compton-thin obscured sources (10$^{22}$<N$_{\rm H}$<10$^{24}$ cm$^{-2}$) decreases in the low luminosity regime \citep{2017Riccii, 2022natalia}, while the fraction of Compton thick sources apparently remains constant. Both Compton thick and thin absorbers can produce reflection features, with different shapes and strengths. In addition, \cite{2022natalia} found that in a sample of 81 AGN, $\sim$13$\%$ of the objects are lacking reflection signatures and the remaining galaxies (with a detected reflection component) should be highly obscured. In the following, in this work, we attempt to measure the global distribution of gas around the nucleus, whether in the line of sight or not, through their contribution to the reflection. In particular, we aim to establish whether the changes in the gas configuration become flattered or overall optically thinner as the accretion rate goes down.  

Additionally, by modeling the X-ray reflection we are able to study the continuum emission, estimating the coronal parameters: the power-law ($\Gamma$) and the high energy cut-off (E$_{\rm cut}$). It has been shown that the slope of the power law depends on the accretion rate with changes at intermediate accretion rates ($L_{Bol}/L_{Edd}$=$\lambda_{Edd}$ $\sim$10$^{-3}$), pointing to a change in the accretion mechanism, for example between a corona on a thin disk to an advection dominated accretion flow  \citep[ADAF,][]{1994narayan}. The relationship toward low accretion rates is usually seen with a lot of scattering, which can be intrinsic or due to observational uncertainties \citep{2006shem, 2009Gu, 2011younes, 2015yang, 2018She}. Our second objective is to re-evaluate this relationship in the low accretion rate range, through detailed modeling of the reflection and broadband X-ray data, using observations from \emph{XMM-Newton}+\emph{NuSTAR}+\emph{Swift}).

This paper is organized as follows: in Sect. \ref{sect:sample} we present details of the observations and sample. The data reduction is reported in Sect. \ref{sect:data_reduction}. The methodology followed during this work is shown in Sect. \ref{sect:metodo}. All the results are reported in Sect. \ref{sect:results}. The implications of our X-ray spectral analysis are discussed in Sect. \ref{Sec:discussion}. Finally, a summary of our findings is presented in Sect. \ref{sect:conclusion}.

\section{Sample and data}
\label{sect:sample}

Hard X-rays (E$\geq$10 keV) are not significantly affected by obscuration, at least up to $N_{\rm H}\sim 10^{24}\rm\,cm^{-2}$ \citep{2015ricci}, which allows us to obtain a highly complete AGN sample. Our work focuses on LLAGN selected through their hard-band X-ray emission as identified in the \emph{Swift}/BAT 70-month catalogue \citep{2013baumgartner} on board the \textit{Neil Gehrels Swift Observatory} \citep{gehrels2004}. BAT operates in the 14–195 keV energy band.  The BAT AGN Spectroscopic Survey (BASS) is a survey that provides high-quality multi-wavelength data for the BAT AGN, including black hole mass measurements \citep{2017mike} and X-ray spectroscopy modeling \citep{2017ricciapJS}. The first data release (DR1) of the BASS project \citep{2017mike} includes 642 of \emph{Swift}/BAT AGN and the second release of optical spectroscopy (BASS/DR2) will also soon be publicly available \citep{Koss_DR2_overview, Oh_DR2_NLR}.

Our sample of galaxies was selected from the BASS/DR2 with accretion rates $\log(\lambda_{Edd})\leq$-3.0 obtaining in total a sample of 24 AGN. We used the HEASARC\footnote{http://heasarc.gsfc.nasa.gov/} archive to search simultaneous and not simultaneous \emph{NuSTAR} and \emph{XMM–Newton} data public until August 2020. This analysis provided data with both telescopes for 16 sources. We include the proprietary data of the galaxy NGC\,5033 (PI: Diaz Y.; $\log(\lambda_{\rm Edd})=$-4.0), an AGN also contained in the BASS/DR2.
 
 Our final sample of LLAGN contains 17 objects, 11 of which are classified as Seyfert 2 (i.e. only narrow lines are visible in the optical spectrum), and six are classified as Seyfert 1.9 (a broad component is visible in H$\alpha$ but not in H$\beta$) in the BASS/DR2. Table  \ref{table:sample} shows the general properties of our sample. Notes for the individual galaxy are in the Appendix \ref{app:notes_object} and Table \ref{table:observations} shows the log of the observations.

\begin{table*}
\caption{General properties of the sample galaxies}             
\label{table:sample}      
\centering          
\begin{tabular}{c c c c c c c c c c} 
\hline\hline       
Name & RA & DEC & Type & Redshift & N$_{\rm gal}$ & $M_{\rm BH}$ & $L_{\rm Bol}$ & $\lambda_{\rm Edd}$ & \\
 & (J2000) & (J2000) & & & (10$^{20}$ cm$^{-2}$) & M$_{\odot}$ & ($\log$) & ($\log$) \\
 (1) & (2) & (3) & (4) & (5) & (6) & (7) & (8) & (9)
\\ 
\hline   

NGC\,3998 & 179.484 & 55.454 & Sy1.9 & 0.003 & 20.09 & 8.93$^{\rm L}$ & 42.29 & -4.74 \\

NGC\,3718 & 173.145 & 53.068 & Sy1.9 & 0.003 & 20.03 &  8.14 & 41.74 & -4.49 \\

NGC\,4258* & 184.740 & 47.304 & Sy1.9 & 0.001 & 20.08 &  7.56$^{\rm L}$ & 41.39 & -4.28 \\

NGC\,5033 & 198.364 &  36.593 & Sy1.9 & 0.002 & 20.00 & 7.68 & 41.78 & -4.00 \\

ESO\,253-G003* & 81.325 & -46.00 & Sy2 &  0.042 & 20.62 & 9.84+ & 43.89 & -3.95 \\

NGC 1052 & 40.270 & -8.256 & Sy2 & 0.005 & 20.49 & 8.67 & 42.83 & -3.94 \\

NGC\,2655 & 133.907 & 78.223 & Sy2 & 0.004 & 20.32 & 8.20 & 42.43 & -3.87 \\

NGC\,3147* & 154.223 & 73.400 & Sy2 & 0.009 & 20.54 & 8.81 & 43.10 & -3.81 \\

NGC\,2110* & 88.047 & -7.456 & Sy2 & 0.007 & 21.27 & 9.38 & 43.81 & -3.67 \\

LEDA\,96373* & 111.610 & -35.906 & Sy2 & 0.029 & 21.47 & 9.21 & 43.80 & -3.51 \\

NGC\,2992 & 146.425 & -14.326 & Sy1.9 & 0.007 & 20.72 & 8.33 & 43.13 & -3.30 \\

M\,51 & 202.484 & 47.230 & Sy2 & 0.001 & 20.19 & 6.59 & 41.40 &  -3.29 \\

NGC\,2273* & 102.536 & 60.845 & Sy2 & 0.006 & 20.84 & 7.99 & 42.84 & -3.25 \\

HE\,1136-2304 & 174.713 & -23.360 & Sy1.9 & 0.027 & 20.63 & 9.39 & 44.28 & -3.21 \\

IGRJ\,11366-6002 & 174.175 & -60.052 & Sy1 & 0.014 & 21.81 & 8.56 & 43.51 & -3.15 \\

IC\,4518A & 224.421 & -43.132 & Sy2 & 0.016 & 20.96 & 8.79 & 43.83 & -3.06 \\

NGC\,7674* & 351.986 & 8.779 & Sy2 & 0.028 & 20.70 & 9.18 & 44.28 &  -3.0 \\


\hline 
\end{tabular}
\tablefoot{ 
(Col. 1) Name, (Col. 2 and 3) right ascension  and declination in Equatorial (J2000.0) from \emph{Swift} BAT 105-month hard X-ray survey \citep{2018oh}, (Col. 4, 5, 7, 8 and 9) optical classification from BASS/DR2, redshift, Black hole mass using the velocity dispersion method, bolometric luminosity and accretion rate $\lambda_{Edd}$=L$_{\rm Bol}$/L$_{\rm Edd}$ from BASS/DR2 survey ($^{\rm L}$ identify masses taken from literature by the BASS/DR2 survey and the symbol + means from MgII). (Col. 6) represents the galactic absorption \citep{1990Dickey}. Objects marked with * are the galaxies with non-simultaneous observations with \xmm\ and \nus .  }

\end{table*}

\section{Data Reduction}
\label{sect:data_reduction}

Data reduction was performed following the methodology explained in this section. Details on the observations can be found in Table \ref{table:observations}.  

\subsection{\emph{XMM-Newton data}}

This satellite has two X-ray instruments, a grating spectrometer, and the European Photon Imaging Camera (EPIC). The EPIC instrument has three detectors, two MOS \citep{2001turner} and one PN CCDs \citep{2001struder}, we only used the observations from the EPIC-PN because of its higher throughput \citep{2001struder} and because of inclusion of the EPIC-MOS spectra resulted in too much statistical weight to the low energy range data points compared to the \nus and \sft/BAT data.
We processed the Observation Data Files (ODFs) from the European Photon Imaging Camera (EPIC) PN detector using the Science Analysis System (SAS version 17.0.0). We followed standard procedures to obtain calibrated and concatenated event lists, filter them for periods of high background flaring activity, and extract light curves and spectra. Source events were extracted using a circular region of 49 arcsec centered on the target, and background events were extracted from a circular region of 98 arcsec on the same chip far from the source. 
We verified the photon pile-up is negligible in the filtered event list with the XMMSAS task \textsc{epatplot}. We generated response matrix files (RMFs) and ancillary response files (ARFs) and rebinned the spectra in order to include a minimum of 25 counts in each background-subtracted spectral channel and to not oversample the intrinsic energy resolution by a factor larger than 3.

\subsection{\emph{NuSTAR data}}

\emph{Nuclear Spectroscopic Telescope Array} (\emph{NuSTAR}) was successfully launched in 2012 June, \cite{2013harrison}. \emph{NuSTAR} has two identical co-aligned telescopes, each consisting of an independent set of X-ray mirrors and a focal-plane detector, referred to as focal plane modules A and B (FPMA and FPMB) that operate in the energy range 3--79 keV. The data reduction was performed with \textsc{nustardas v1.6.0}, available in the \emph{NuSTAR} Data Analysis Software. The event data files were calibrated with the \textsc{nupipeline} task using the response files from the Calibration Database \textsc{caldb} v.20180409 and \textsc{HEASOFT} version 6.25. With the \textsc{nuproducts} script, we generated both the source and background spectra, plus the ARF and RMF files. For both focal plane modules (FPMA, FPMB), we used a circular extraction region of radius 49 arcsec centered on the position of the source. The background selection was made, taking a region free of sources of twice the radius of the target and located in the same detector quadrant. Spectral channels were grouped with the \textsc{ftools} task \textsc{grppha} to have a minimum of 20 counts per spectral bin in the 3.0 -- 79.0 keV energy range. 

\subsection{\emph{Swift} data}

The Neil Gehrels \emph{Swift} Observatory was launched on November 20, 2004. It carries three instruments to enable the most detailed observations: the \emph{Swift Burst Alert Telescope} (BAT; \citealt{2005barthelmy}; bandpass: 15-350 keV), the X-ray Telescope (XRT; \citealt{burrows2005}; bandpass: 0.3-10 keV), and the UV/Optical Telescope (UVOT 170–650 nm). In this work, we are focusing on \emph{Swift}/BAT and \emph{Swift}/XRT instruments.

\begin{itemize}
    \item \underline{\emph{Swift/BAT:}} \\

We retrieved the binned and calibrated spectra, together with the response matrices for our targets, from the \emph{Swift}/BAT 105-month All-sky Hard X-Ray Catalog reported in \cite{2018kyu}. The observations were taken with the Burst Alert Telescope (BAT) on board the  \emph{Swift} observatory. This survey has a sensitivity of 8.4$\times$10$^{-12}$ erg s$^{-1}$ cm$^{-2}$ in the 14 – 195 keV bands over 90$\%$ of the sky, with eight-channel spectra averaged over the 105-month duration of the survey. The complete analysis pipeline is described in the \emph{Swift}/BAT 22 All-sky Hard X-Ray Survey \citep{tueller2010}.  \\

\item \underline{\emph{Swift/XRT:}} \\

For three sources (NGC\,7674, ESO\,253-G003, and IGRJ\,11366-6002) there are no simultaneous \emph{XMM-Newton} and \emph{NuSTAR} observations. We explored the \emph{Swift}/XRT archived and found simultaneous observations for the three of them.  The data reduction of
the \emph{Swift}/XRT in the Photon Counting mode was performed by following standard routines described by the UK \emph{Swift} Science Data Centre (UKSSDC) and using the software in HEASoft version 6.30.1. Calibrated event files were produced using the routine {\sc xrtpipeline}, accounting for bad
pixels and effects of vignetting, and exposure maps were also created. Source and background spectra were extracted from circular regions with 25 arcsec and 50 arcsec radius. The {\sc xrtmkarf} task was used to create the corresponding ancillary response files. The response matrix files were obtained from the HEASARC CALibration DataBase. The spectra were grouped to have a minimum of 20 counts per bin using the {\sc grppha} task.

After the data reduction, we find that both NGC\,7674 and ESO\,253-G003 present very low counts, preventing us from doing a proper spectral fit, so these will not be used in the analysis.      

\end{itemize}

\section{Methodology}
\label{sect:metodo}


The analysis of the data comprises two steps: (1) a Combination of \xmm and \nus observations; and (2) homogeneous spectral fitting of the sample. All the spectra have been fitted using \rm{\texttt{xspec}} version 12.10.0 \citep{1996arnaud} and all the errors reported throughout the paper correspond to 90$\%$ of confidence level.

\subsection{Combination of the \xmm and \nus observations}

In this work, we have \nus observations, with an energy range from 3 to 79 keV, vital to study the Compton hump which is a key signature of the reflection. Additionally, we have \xmm data, which provides the best combination of sensitivity, bandpass, and spectral resolution at energies ranging from 0.5 - 10.0 keV. 
Objects with simultaneous observations with \xmm and \nus were fitted with all model parameters tied between the different spectra, except for a free cross-normalization factor. Objects with non-simultaneous observations (denoted with the symbol * in our work) were tested for  spectral variability between the observation epochs.  In order to detect spectral variability, we simultaneously fitted the \xmm + \nus spectra in the overlapping 3.0 -- 10.0 keV range for each object with a power-law model under neutral absorption. In cases where the spectrum was not well-fitted with this model, we added a Gaussian component centered at 6.4 keV and studied the improvement of the fit. 

At first, all parameters were tied between the spectra of the different epochs/instruments. If this model produced a satisfactory fit ($\chi^{2}_{\nu}\leq$1.2) the source is considered non-variable and treated in the same way as the objects with simultaneous observations. Two objects are well-fitted with a tied normalization of the power law (LEDA\,96373* and NGC\,2273*).  For the remaining objects (NGC\,4258*, NGC\,3147*, NGC\,2110* and IC\,4518A*) we found that the normalization of the power-law varying between epochs resulted in a satisfactory fit. For these objects, in the subsequent fitting, the normalization of the power law was left free between the epochs, but the remaining parameters were tied. For one object, (ESO\,253-G003*) allowing the slope of the power law to vary freely improved the fit significantly according to the F-test. Given the spectral variability in this source, we had to leave most parameters untied between the epochs, and therefore the inclusion of the lower energy spectrum would not constrain the reflection model further. For this reason, in this source, we used the high-energy spectra only. 
Finally, one object (NGC\,7674) could not be fitted well with a free slope and normalization, as this is a known changing look AGN and its spectrum changed significantly in shape between observations \citep{2005bianchi}, so we retained only the \nus spectrum for the following analysis. The best model and the final configuration for each object are summarized in Table \ref{Table:not_simul}. 

\subsection{Spectral analysis}

The aim of our spectral analysis is to quantify how much reflecting material \emph{can} exist around the central engine. For this purpose, we allow different types of reflectors and maximize the freedom of the fitted parameters of all required models, even if the fit results in unconstrained values for many of them. We tested the significance of the reflection component in the 3--195 keV band in all objects, returning significant detections in most of them. These tests are summarized in Appendix \ref{apnx:reflex_comp}.

 For all spectral fits, we included a multiplicative constant normalization between FPMA, FPMB, EPIC-PN, and  \sft/BAT to account for calibration uncertainties between the instruments. We started with a baseline model and added different components until a satisfactory fit was obtained. We have selected three broad components in order to parametrize three scenarios:

\begin{enumerate}
    \item \textbf{Cut off Power-law model (cPL) obscured by neutral material:} a single power law model, which corresponds to the primary emission of a non-thermal source. The column density, N$_{\rm H, los}$, is added as a free parameter to take the absorption by matter along our line of sight to the target into account. The free parameters in this model are the column density, N$_{\rm H, los}$, the slope of the power law, $\Gamma$, the high energy cut-off, E$_{\rm cut}$ and the normalization. \\
    
    \item \textbf{Reflection models (Refl):} When the X-ray continuum is scattered by the surrounding gas, it can produce fluorescent emission lines (most notably Fe K$\alpha$ 6.4 keV) and a broad hump-like continuum peaking around 10--30 keV.  The reflection was modeled with three possible scenarios: \\
    
    \begin{itemize}
    
    \item A neutral reflector with a semi-infinite column density modelled with \texttt{PEXMON} \citep{2007nandra}. This model assumes the existence of optically thick and cold material, distributed in a slab and covering a given fraction of the X-ray source. The \texttt{PEXMON} model includes  fluorescence, adding some spectral features, such the emission lines FeK$\alpha$ and Fek$\beta$, following the Monte Carlo calculations by \cite{1991george}. This model represents both the reflected and intrinsic emission defined with $\Gamma$ and the high energy cut-off (E$_{\rm cut}$) and the reflection fraction, R$_{\rm f}$. The free parameters in this model are the reflection fraction, R$_{\rm f}$ (to account for the reflection component and the contribution from the intrinsic power-law continuum), the spectral index, $\Gamma$, the high energy cut-off, E$_{\rm cut}$, the inclination and the normalization. \\

    \item A smooth spherical distribution of neutral gas, with conical cavities along the polar direction, is modeled with \texttt{BORUS} \citep{2018balokovic}. This model calculates the reprocessed continuum of photons that are propagated through a cold and static medium. \texttt{BORUS} is similar to the torus model \texttt{BNtorus} of \cite{2011brightmanNandra} but it has additional free parameters (E$_{\rm cut}$, $A_{\rm Fe}$), additional chemical elements included, calculation extending to higher energies and line-of-sight component separated out. Furthermore, this model has a variable covering factor which is an advantage compared with other models, as \texttt{MYTORUS} \citep{2009Murphy} that proposes a toroidal geometry where the covering fraction is fixed to 0.5. In this work, we used the geometry of a smooth spherical distribution of gas, with conical cavities along the polar directions (\texttt{borus02}). The column density and the inclination of the torus are free parameters in this model. \texttt{borus02} includes fluorescent emission lines, according to fluorescent yields for K$\alpha$ and K$\beta$ lines from \cite{1979krause}, for all elements up to zinc (Z < 31).
    The reflected spectrum of this torus is calculated for a cut-off power-law illuminating continuum, where E$_{\rm cut}$, $\Gamma$ and normalization are free parameters. We modeled the direct coronal emission separately with a cut-off power law under a neutral absorber as described above. We have set as free parameters the column densities along the line-of-sight, N$_{\rm H, los}$, the inclination, Cos($\theta_{incl}$), the covering factor, CF, the column density of the reflector, $\log(\rm{N}_{\rm{H,refl}})$, the spectral index of the primary emission, $\Gamma$, the high energy cut off, E$_{\rm cut}$ and the normalization of the reflector tied to the primary emission.  \\

     \item The accretion disk reflection modeled with \texttt{XILLVER} \citep{2013garcia} where the coronal spectrum is a power-law with an exponential cut-off described by the photon index, $\Gamma$ and the high energy cut-off, $E_{\rm{cut}}$. Another important parameter is the ionization parameter, $\xi$, defined as the incident flux divided by the density of the disk. This parameter is described by $\log(\xi)$ ranging from 0 for a neutral disk to 4.7\,erg\,cm$^{-2}$\,s$^{-1}$ for a heavily ionized disk (see \citealt{2013garcia}, for a more detailed description). Other parameters in this model are the iron abundance, $A_{\rm Fe}$ relative to the solar value (assumed to be solar in this work), redshift, reflection fraction, R$_{\rm f}$, and the inclination. Also, this model takes into account both the reflected continuum and the FeK$\alpha$. The free parameters in this model are the spectral index, $\Gamma$, the high energy cut-off, E$_{\rm{cut}}$, the ionization degree, $\log(\xi)$, the inclination, incl, the reflection fraction, R$_{\rm f}$ (to normalize the reflection component relative to the  intrinsic power-law continuum) and the normalization.  \\

    \end{itemize}
    \item \textbf{Soft X-ray emission (SE):} When the combination of the above models does not produce a good fit, we explore if the addition of spectral component(s) improves the fit. The following spectral components are considered: \\
     \begin{itemize}
    
     \item An absorbed scattered power-law: an absorbed power-law \texttt{PL} to model the scattered emission that is deflected by ionized gas. The photon index, $\Gamma$, of the scattered component is tied to the primary power law. We set as free parameters  the column density, N$_{\rm H,ext}$ and the normalization of the scattered component but restricted to be less than 5$\%$ of the main one. \\
     
     \item Thermal emission: An optically-thin thermal component, modeled by \texttt{MEKAL} in \texttt{xspec}, to model the soft excess observed below 1 keV, and potentially due to either star formation processes and/or thermal emission from a hot interstellar medium. We kept the hydrogen column density, abundance, and switch at their default values (1, 1, and 1 respectively) and we let the temperature, ionization, and normalization free to vary. \\
    
        \item An ionized absorber (ab): a warm absorber was modelled with \texttt{zxipcf} within \texttt{xspec}. This model uses a grid of \texttt{xstar} photoionized absorption models (calculated assuming a microturbulent velocity of 200 km s$^{\rm -1}$) for the absorption, and it assumes an absorbent covering some fraction of the source, cf$_{\rm W}$ \citep{2008reeves}. \texttt{zxipcf} has as free parameters the column density, N$_{\rm H,W}$, the ionization state, $\log(\xi_{\rm W})$, the covering fraction, cf$_{\rm W}$, and redshift. We set the covering fraction to cf$_{\rm W}$=1 to mimic an absorber covering all the sky. We let as a free parameter N$_{\rm {H,W}}$ and $\log(\xi_{\rm W})$.\\

     \end{itemize}

\end{enumerate}

\begin{figure}
    \centering
    \includegraphics[width=8.0cm]{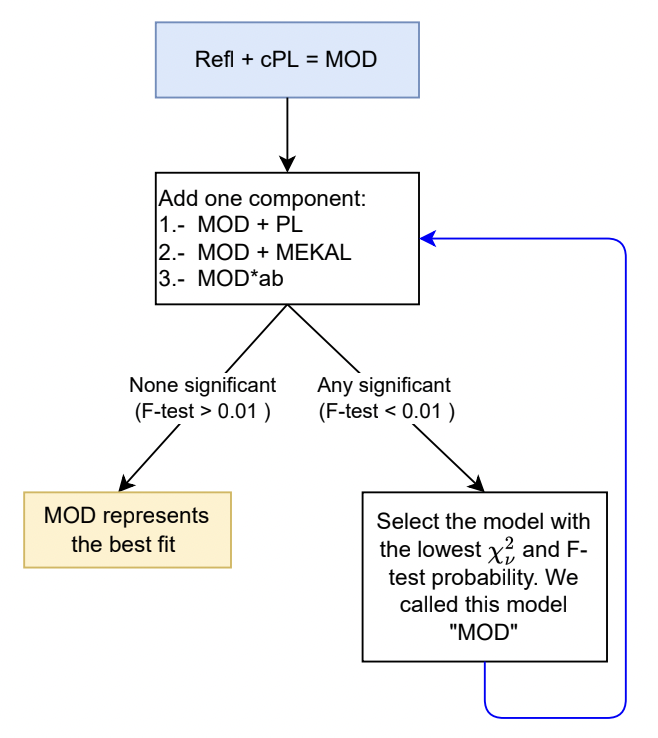}
    \caption{Schematic view of the methodology followed to fit the data. Note that the loop in blue iterate a maximum of four times. For a detailed explanation of the method, we refer the reader to the text.}
    \label{fig:method}
\end{figure}

 We started our analysis by fitting a baseline model that is defined as \texttt{MOD = Refl + cPL} to the data. Then we added one SE emission or absorption component (we tested one by one: \texttt{MOD + PL}, \texttt{MOD + MEKAL} and \texttt{MOD*ab}) and explore if the inclusion of these components improves the fit (using an F-test in the case of the scattered power-law and MEKAL, and we evaluate the improvement of the fit using the values of the $\chi^2$ and a visual inspection of the residuals in the case of the ionized absorber). If any of the improvements was significant, we selected the model that returned the lowest value of $\chi^2/$d.o.f.\footnote{d.o.f. means the degree of freedom} and select it as the new baseline model and the process of including and testing an additional SE component was repeated. When none of the additional SE components provided a significant improvement, the iteration stopped. Up to 4 iterations were necessary for each object and reflection model.
 The method is represented in Fig. \ref{fig:method}. The process was repeated separately for each reflection model. Thus, we report up to three best-fitting models for each object.

The models that were selected to fit the data are represented in \texttt{xspec} as:
\vskip 0.3cm

\boxed{\rm{C} \times  N_{\rm H,Gal} \times ab \times (N_{{\rm H},ext} \times SE +  N_{{\rm H},los}\times  $ cPL $ +  N_{{\rm H, los}} \times \rm{Refl})} 

\vskip 0.5cm

Where $\rm C$ represents the cross-calibration constant between different instruments,  N$_{\rm H, Gal}$ is the Galactic absorption (\texttt{phabs} in \texttt{xspec}) predicted using N$_{\rm H}$ tool within \rm{FTOOLS} \citep{1990Dickey, 2005karberla}. ``ab'' is the ionized absorption component modelled with \texttt{zxipcf}, in cases where this component is used, otherwise is equal to unity. Two absorbing column densities are used, which will be called here N$_{\rm H,ext }$ and N$_{\rm H,los}$ (\texttt{zphabs} in \texttt{xspec}). N$_{\rm H,los}$ is assumed to cover the nuclear components (power-law and disk reflection)\footnote{In case of a torus-like reflection, the absorber is not acting in the torus-like reflector} and N$_{\rm H,ext}$ covers the SE component\footnote{In case of MEKAL, the absorber is not acting in this component}. Moreover, cPL is a cut-off power-law (\texttt{cutoffpl} in \texttt{xspec}) representing the primary X-ray emission and ``Refl'' represents the  different reflection models used.

Note that we imposed the following conditions to the resulting best-fit $\Gamma$>0.5,  N$_{\rm H, Gal}\leq$ N$_{\rm H,ext }$ and N$_{\rm H,los}$>N$_{\rm H,ext }$. In the case of NGC 1052, additional Gaussian lines were required at soft energies from a visual inspection, we included S XIV at 2.4 keV and Si XIII at 1.85 keV, also in agreement with \citep{2020natalia1052}. They were added as a narrow Gaussian line with fixed centroid energy and a width fixed at 0.01 keV.

\section{Results}
\label{sect:results}

We refer the reader to the following sections and tables for details on the analysis. Comparison with previous works and our results on individual objects can be found in Appendix \ref{app:notes_object}. The coronal parameters (i.e., $\Gamma$, E$_{\rm cut}$, and $\chi^2$) are listed in Table \ref{table:corona}. The reflection parameters, i.e., R$_f$ and inclination for \texttt{PEXMON}; $\log(\rm{N}_{\rm{H,refl}})$, CF and inclination for \texttt{borus02}; $\log(\xi)$, R$_{\rm f}$ and inclination for \texttt{XILLVER} are listed in Table \ref{table:reflection_parameters}. In Table \ref{table:soft} we show all the additional components required for the fit with each of the reflection models, i.e., the column density of the neutral absorbers in the line of sight to the extended and nuclear components and the temperature of the optically thin thermal emission components. Additional parameters, i.e., the column density and ionization parameter of the ionized absorbers in the line of sight and the normalization of the scattered power law can be seen in Table \ref{table:other}. All cross-calibration constants are listed in Table \ref{table:cab}. The plots of the spectra with the best-fit models and their residuals can be found in Appendix \ref{app:plots}.

\subsection{Models} 

\label{sect:models}

 In this work, we used three reflection models (\texttt{PEXMON}, \texttt{borus02}, and \texttt{XILLVER}) that were used to fit the spectrum of each of the sources in the sample, i.e., each of the sources is fitted by three different models.  

The simplest model used in our work (\texttt{PEXMON}) is a good representation of the data, however, we will focus on models that can explore different reflector geometries 
as \texttt{borus02} and \texttt{XILLVER}. To decide which model provides the best description of the observations, we estimate the ``evidence ratio'' using the Akaike information criterion (AIC) for both models. This evidence ratio allows us to compare if one model is better than another one, it is defined using as $\epsilon$=W(AIC$_{\rm{torus}}$)/W(AIC$_{\rm{disk}}$) where W(AIC$_{\rm{torous}}$)
and W(AIC$_{\rm{disk}}$) are the ``Akaike weight'' (see \citealt{2016emmanopolus} for more details). The evidence ratio is a measure of the relative likelihood of the torus versus the disk model. The torus model is 200 times more likely than the disk model when $\epsilon\leq$0.0067. The disk model is 200 times more likely than the
torus model when $\epsilon\geq$150. The evidence ratio is listed in Table \ref{table:AIC}.

\begin{table}
\caption{Best model results according to Akaike criterion}             
\label{table:AIC}      
\centering          
\begin{tabular}{c c c } 
\hline\hline       
Name & $\epsilon$ & Model \\
\\ 
\hline   

NGC\,3998 & 8.06E-01 & T/D \\

NGC\,3718 & 1.68E+00 & T/D \\

NGC\,4258* & 2.57E-16 & Torus \\

NGC\,5033 & 5.67E-04 & Torus \\

ESO\,253-G003* & 1.20E+00 & T/D \\

NGC 1052 & 3.86E-59 & Torus \\

NGC\,2655 & 3.86E-02 & T/D\\

NGC\,3147* & 6.44E-02 & T/D \\

NGC\,2110* & 1.65E-53 & Torus \\

LEDA\,96373* & 8.97E-26 & Torus \\

NGC\,2992 & 1.77E-74 & Torus \\

M\,51 & 1.68E-04 & Torus \\

NGC\,2273* &  4.41E+07 & D \\

HE\,1136-2304 & 5.31E-15 & Torus \\

IGRJ\,11366-6002 & 2.49E+00 & T/D \\

IC\,4518A &  1.18E-06 & Torus\\

NGC\,7674* & 2.23E+02 & D \\

\hline 
\end{tabular}
\tablefoot{ Evidence ratio for the Akaike method and resulting model best to each source. T/D represents the cases when either torus or disk models provide equally good fits. T represents the torus model (\texttt{borus02}) and D the disk model (\texttt{XILLVER}). Objects marked with * are the galaxies with non-simultaneous observations with \xmm\ and \nus . }

\end{table}

For nine (53$\%$) objects (NGC\,4258,NGC\,1052, NGC\,2110, LEDA\,96373, NGC\,2992, M\,51, HE\,1136-2304, IC\,451A, and NGC\,5033) \texttt{borus02} is preferred. Then in the following sections, we choose this model as the best representation of the data in these objects. On the other hand, two (12$\%$) objects (NGC\,2273 and NGC\,7674) are well fitted with a disk (\texttt{XILLVER}) model and in six (35$\%$) objects (NGC\,3998, NGC\,3718, ESO\,253-G003, NGC\,2655, NGC\,3147, and IGRJ\,11366-6002) both models fit similarly well the data.

Since it is not possible to distinguish between a reflection dominated by a torus or a disk, we will treat them separately in the following sections. When referring to the torus case, we refer to all \texttt{borus02} models for the whole sample (15 galaxies: indistinguishable and distinguishable cases). In the disk case, we refer to the \texttt{XILLVER} model, considering the indistinguishable cases and the cases where the disk is a good representation of the data (8 galaxies in total), since in the other cases a torus model is not a good representation of the data.

\subsubsection{X-Ray Continuum Properties}

\begin{figure}
\centering
    \includegraphics[width=0.40\textwidth]{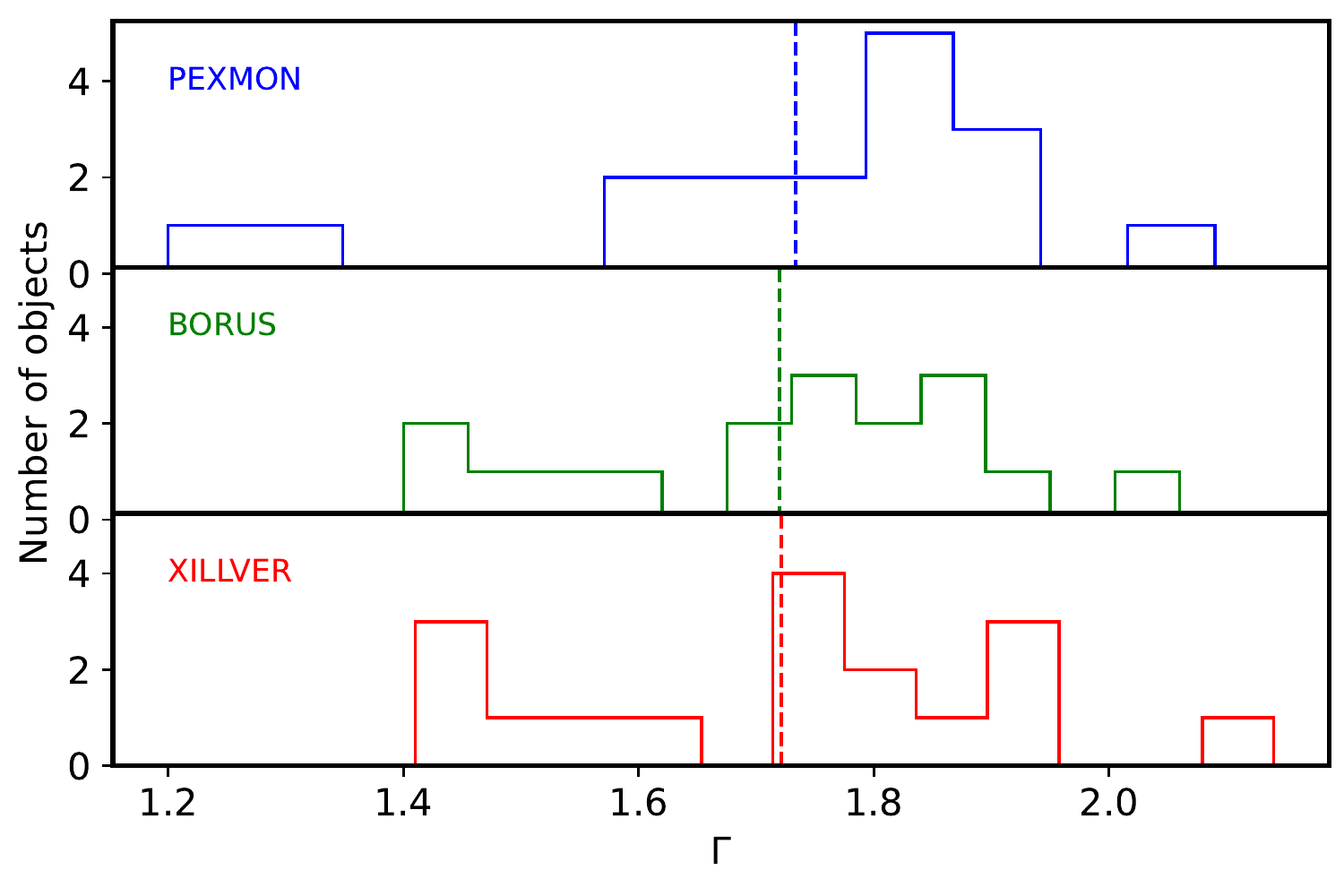}
    \caption{ Comparison between the spectral index estimation, $\Gamma$, between the models. The dotted lines represent the mean values. }
     \label{fig:photon_index}
\end{figure}

The X-ray continuum of AGN is described by a power law with a high-energy cutoff. The free parameters of this component are the spectral index ($\Gamma$) and the high energy of the cut-off (E$_{\rm{cut}}$). In Fig. \ref{fig:photon_index} we show the histogram derived from the broadband spectral analysis with each reflection model. Note that the spectral index between both the reflected and intrinsic emission are tied.  We find that the mean values (dashed vertical lines) of $\Gamma$ for the sample using \texttt{PEXMON}, \texttt{borus02} and \texttt{XILLVER} are consistent  (1.73$\pm$0.21, 1.72$\pm$0.17 and 1.72$\pm$0.20 respectively). Note that the simplest model used in our analysis is \texttt{PEXMON}, which shows values of the photon index consistent with more geometrical models. 

Considering the torus model (15 galaxies), we found median values of $\Gamma$= 1.76 and $\sigma$=0.16, with values ranging between [1.40, 2.06]. Another important parameter that could be estimated with the \nus data is the high energy cut-off (E$_{\rm cut}$). This parameter can be considered as an indicator of the temperature of the X-ray corona. Consequently, its knowledge provides information about the dynamics of the corona and the physical processes occurring within it. Nevertheless, this parameter is poorly constrained. A lower (upper) limit of E$_{\rm cut}$ could be determined for eight (two) sources. The five AGN for which E$_{\rm cut}$ could be determined (NGC\,3998, ESO\,253-G003, NGC\,2110, NGC\,2992, and NGC\,5033) have a mean value of E$_{\rm cut}$=193.28 keV with a standard deviation of $\sigma$=99.19 keV.

Furthermore, taking into account the disk model (8 galaxies), we found median values of $\Gamma$=1.71  and $\sigma$=0.23, with values ranging between [1.40, 2.06]. Regarding the high energy cut-off, we could find five (one) lower (upper) limits and for six objects we obtained a mean value of E$_{\rm cut}$=371.47.58 keV with a standard deviation of $\sigma$=619.99 keV.

\subsubsection{Soft band spectral fit}

In the soft  (0.3 -- 10.0 keV) energy band, we added a thermal (MEKAL in \texttt{xspec}), scattered power-law, absorption by ionized gas (also referred to as “warm absorption”) (modelled with \texttt{zxipcf} in \texttt{xspec}) or a combination of these components to improve the spectral fit.

When considering the torus model (15 objects), six objects (NGC\,3998, ESO\,253-G003, NGC\,3147, M\,51, IGRJ\,11366-6002, and NGC\,5033) do not require an additional component to improve the fit. Two objects (NGC\,3718 and HE\,1136-2304) required a \texttt{MEKAL} component (with kT=0.88$^{**}_{0.67}$ keV and kT=0.59$^{0.67}_{0.51}$ keV, respectively). Two objects are well-fitted with a combination of \texttt{MEKAL}, power-law, and warm absorber (NGC 1052 with \texttt{MEKAL+PL} and NGC IC\,451A with \texttt{MEKAL*ab}). Composite models are needed for five galaxies (NGC\,4258, NGC\,2655, NGC\,2110, LEDA\,96373, and NGC\,2992). In the cases where two \texttt{MEKAL} were required, the values of the temperatures are in the range kT$_{\rm 1}$ = [0.58 - 0.62] keV with a mean value of kT$_{1}$=0.60 keV and $\sigma$=0.02 keV, and kT$_{\rm 2}$ = [0.15 - 0.22] keV with a mean value of KT$_{2}$=0.19 keV and $\sigma$=0.03 keV. The mean value of the ionized absorber is N$_{\rm{H,W}}$=1.66 $\times$10$^{\rm{22}}$ cm$^{\rm{-2}}$ and $\sigma$=1.41 $\times$10$^{\rm{22}}$ cm$^{\rm{-2}}$ 
The degree of ionization is in the range [-1.14, 4.30] with the mean $\log(\xi_{\rm W})$=1.31 and $\sigma$=1.99.

In relation to the disk model (8 objects), five galaxies do not require any component to improve the spectral fit (NGC\,3998, ESO\,253-G003, NGC\,3147, IGRJ\,11366-6002 and NGC\,7674). One galaxy (NGC\,3718) require a \texttt{MEKAL} component to improve the fit. Two galaxies (NGC\,2655 and NGC\,2273) are well-fitted with a composite model, \texttt{MEKAL*ab}.

\subsubsection{Line-of-sight column density}

Absorption of X-rays by neutral material is the result of the combined effect of Compton scattering and photoelectric absorption. The Compton scattering and the photoelectric absorption were modelled using \texttt{CABS} and \texttt{ZPHABS}  in \texttt{xspec} respectively. In \texttt{ZPHABS}, we fixed the redshift at the value of each source. The only free parameter is the column density, which is tied in all fits (i.e., $\rm N_{\rm{H-ZPHABS}}=\rm N_{\rm{H-CABS }} = \rm N_{\rm{H-los} }$).

According to the torus model, we can classify six galaxies as unobscured ($\log(\rm{N_{\rm{H, los}}})$<22) (NGC\,3998, NGC\,3147, NGC\,2992, HE\,1136-2304, IGRJ\,11366-6002 and NGC\,5033) with values between $\log(\rm{N_{\rm{H, los}}})$=[20.0, 21.89] and eight galaxies (NGC\,3718, NGC\,4258, ESO\,253-G003, NGC\,1052, NGC\,2655, NGC\,2110, IC\,451A, and LEDA\,96373) as obscured (22<$\log$($\rm{N}_{\rm H}$)<24.18) with values ranging $\log(\rm{N_{\rm{H, los}}})$=[22.01, 24.09]. According to our spectral analysis, one galaxy (M\,51) in our sample can be classified as Compton thick (CT) (using as a  threshold N$_{\rm{H}}=1.5\times 10^{24}$cm$^{-2}$, or $\log(\rm{N_{\rm{H, los}}})$=24.18). The mean values of spectral index, column density in the line of sight, and column density of the torus are reported in Table \ref{table:result_group}. All the parameters are consistent between the groups. Note that the values of the cross-calibration constant between the groups are consistent. 

Regarding the disk model, two galaxies (NGC\,3998 and IGRJ\,11366-6002) can be classified as unobscured. Six galaxies (NGC\,3718, ESO\,253-G003, NGC\,2655, NGC\,3147, NGC\,2273, and NGC\,7674) as obscured with values $\log(\rm{N}_{\rm{H, los}})$=[22.03, 23.36]. The mean values of the spectral index, the column density in the line of sight, the ionization degree of the accretion disk, and the reflection fraction are reported in Table \ref{table:result_group} and showed values consistent between the categories. Note that according to a reflection dominated by an accretion disk, none of the galaxies in our sample can be classified as CT.

\begin{table*}
\caption{Mean values and standard deviation of the spectral parameters for the subgroups with the torus and the disk models.}             
\centering          
\begin{tabular}{c|c c c c c c} 
\hline      
 
Group & $\Gamma\pm\sigma$ & $\log(\rm{N}_{\rm{H, los}})\pm\sigma$ & $\log(\rm{N}_{\rm{H,refl}})\pm\sigma$ \\
\hline

Unobscured (6) & 1.80$\pm$0.12 & 20.85$\pm$0.56 & 23.45$\pm$0.68  \\

Obscured (8) & 1.74$\pm$0.17 & 22.99$\pm$0.65 & 23.65$\pm$0.58 \\

\hline 
\end{tabular}

\vskip 2.5mm
\begin{tabular}{c|c c c c c c} 
 
\hline      

Group & $\Gamma\pm\sigma$ & $\log(\rm{N}_{\rm{H, los}})\pm\sigma$ & $\log(\xi)\pm\sigma$ & R$_{\rm{f}}\pm\sigma$  \\

\hline

Unobscured (2) & 1.85$\pm$0.05 & 21.90$\pm$1.32 & 3.61$\pm$0.47 &3.97$\pm$0.12  \\

Obscured (6) & 1.74$\pm$0.26 & 22.90$\pm$0.44 & 2.43$\pm$0.77 & 6.59$\pm$3.29 \\

\hline 
\end{tabular}
\tablefoot{Group, The standard deviation(s) and the mean value of the following parameters: For the Torus: $\Gamma$, column density in the line of sight in log units and column density of the torus like reflector in units of log. For a disk: $\Gamma$, column density in the line of sight in log units, ionization degree, and the reflection fraction. The parentheses show the number of AGN in each category.
}
\label{table:result_group}      

\end{table*}

\subsubsection{The reflection component}

The reflection features observed in the hard X-ray spectra of AGN may be caused by neutral and distant material such as the torus or by the ionized material of the accretion disk. 

For the case where the reflection is dominated by the torus, the mean value of our sample for the column density for this structure is $\log(N_{H, refl})$=23.69 and $\sigma$=0.76 with values between [22.50, 25.40]. Four objects (NGC 1052, M\,51, IGRJ\,11366-6002, and IC451A) show a column density of the torus consistent with a Compton thick structure. Another important parameter derived from the torus reflector model is the covering factor. We were only able to determine a lower (upper) limit for three (six) objects (a lower limit for NGC\,3998, NGC\,2655, and IC451A, and an upper limit for NGC\,3718, NGC\,4258, ESO\,253-G003, NGC\,3147, NGC\,2992, and M\,51). This parameter was determined for six ($40\%$) objects (NGC 1052, NGC\,2110, LEDA\,96373, HE\,1136-2304, IGRJ\,11366-6002, and NGC\,5033) with a mean value of CF =0.59 and $\sigma$=0.25. The half-opening angle of the polar cutouts, cos($\theta_{incl}$), is also measured with the torus model. However, we obtain only lower (upper) limits for seven (three) objects and properly constrained it for five sources. 

To the disk-like reflection, we constrain the value of the ionization degree of the accretion disk in six galaxies (NGC\,3718, ESO\,253-G003, NGC\,3147, NGC\,2273, IGRJ\,11366-6002 and NGC\,7674), we found median values of ionization degree of the disk of $\log(\xi)$=2.44 and $\sigma$=0.81. In one (one) objects, we only obtain upper (lower) limit (NGC\,3998 a lower limit and NGC\,2655 an upper limit). Regarding the reflection fraction, R$_{\rm f}$, we performed a test by fixing the value of $\log(\xi)$ to 0 and comparing the reflection fraction obtained with this \xill configuration and \texttt{PEXMON}. For the sample, we found consistent values between both models. However, since our goal is to restrict the accretion disk features, we left the ionization degree as a free parameter. We obtain  six lower limits, one upper limit (NGC\,2655) and it is constrained in one case (NGC\,7674). This model also allows us to estimate the inclination, and this parameter is constrained in five sources with mean value Incl=70.10 deg and $\sigma$=27.23 deg and four upper limits.

\subsubsection{Flux and Luminosity}

\begin{figure*}
\centering
    \includegraphics[width=0.850\textwidth]{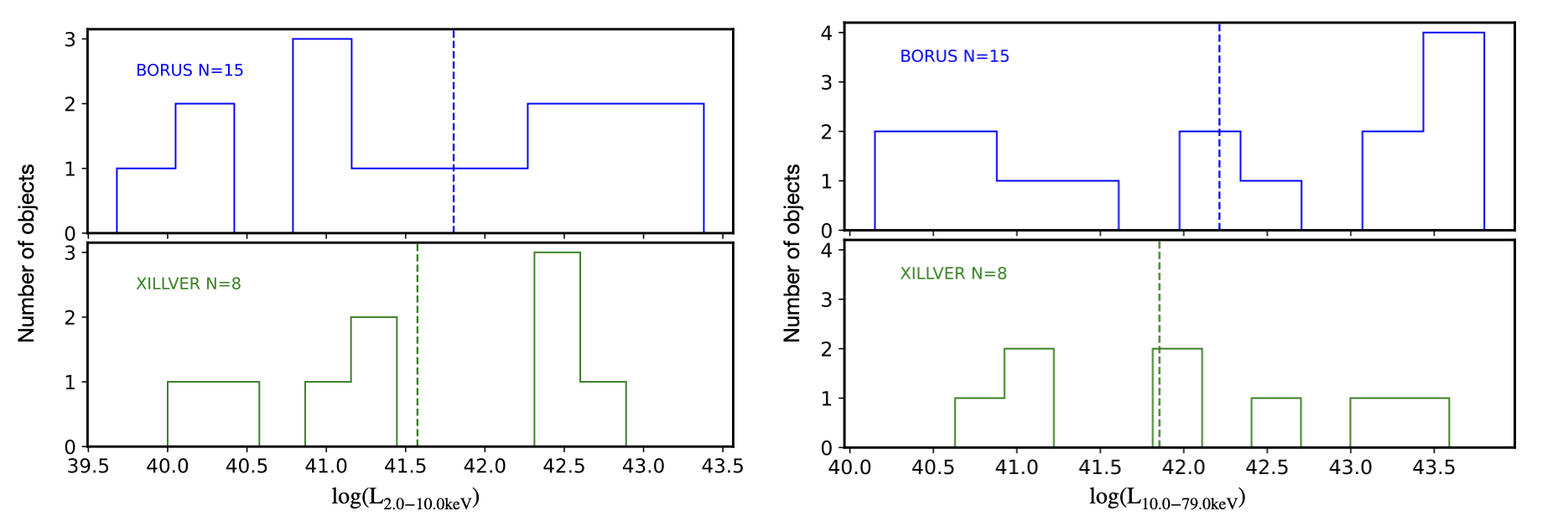}
    \caption{Histogram of the intrinsic luminosity in the band 2.0 --  10.0 keV (left) and 10.0 -- 79.0 (right) in the \bors and \xill with number of objects in each group.  }
     \label{fig:lumin}
\end{figure*}

We computed the X-ray flux and luminosity in two energy bands: 2.0 -- 10.0 keV and 10.0-- 79.0 keV using \texttt{xspec}. Note that the redshift of the sources was taken from NASA/IPAC Extragalactic Database (NED). The values can be seen in Table \ref{table:luminos}. Taking into account a torus model, the mean value of the intrinsic luminosities in the \bors case are  $\log(L_{2.0-10.0})$=41.73 and $\sigma$=1.16 $\log(L_{10.0-79.0})$=42.14 with $\sigma$=1.29. In the \xill case, we found  $\log(L_{2.0-10.0})$=41.57 with $\sigma$=1.06 and $\log(L_{10.0-79.0})$=41.85 with $\sigma$=1.24, and they are equivalent. The distribution of the intrinsic luminosity obtained in both cases can be seen in Fig. \ref{fig:lumin}.

\section{Discussion}
\label{Sec:discussion}

We have performed the X-ray spectral analysis of an AGN sample with accretion rates, $\log(L_{\rm Bol}/L_{\rm Edd})\leq$ -3 selected from the BASS/DR2 that have available \nus + \xmm + \emph{Swift} data. Models from a neutral reflector (\texttt{PEXMON}), reflection from an ionized accretion disk (\texttt{XILLVER}) and from the torus (\texttt{borus02}) have been used to fit the data. This sample is composed of 17 objects, and our main results are summarized as follows:

\begin{enumerate}
    \item In our sample, six (35$\%$) objects are equally well-fitted with a disk or with a torus-like reflector. For nine (53$\%$) galaxies, the torus reflection model is the best representation of the data. In two cases (12$\%$) the disk model well fits the data.

    \item When modeling the reflection with \texttt{borus02}, seven objects are well-fitted by a single neutrally-absorbed cut-off power-law plus reflection (i.e., no components are required in the soft band). When modeling the reflection with \texttt{XILLVER} instead, five objects can be well modeled in the same way. The remaining objects require the addition of a MEKAL and/or scattered power law, an ionized absorber, or a combination of two or more of these components. 

    \item According to the torus model, six sources can be classified as unobscured ($\log$(N$_{\rm H}$)<22), eight galaxies as obscured (22<$\log$(N$_{\rm H}$)<24.18) and one object have a column density in the line of sight consistent with a Compton thick source ($\log$(N$_{\rm H}$)>24.18). According to the disk reflection, two (six) objects can be classified as unobscured (obscured). These classifications are consistent among the models, except in the case of NGC\,3147 (unobscured according to the torus and obscured with the disk).

\end{enumerate}

The high quality and broad spectral coverage available combining \xmm+\nus+\emph{Swift}  allowed us to put constraints on spectral parameters related to the accretion mechanism and reflection of LLAGN. Our analysis covers energies above 10.0 keV, where the reflection has an important role in the spectral fit, and considering this feature in the X-ray spectral analysis can affect the estimation of the coronal parameters (see \citealt{2020diaz}). In the following, we discuss the physical interpretations of the results presented in this paper.

\subsection{ Determination of the  $L_{\rm Bol}/L_{\rm Edd}$}
\label{sec:bolo_corre}

The selection of the sample presented in this work was based on sources with $\log(L_{\rm Bol}/L_{\rm Edd})\leq$ -3 according to those values reported in BASS/DR2. However, because variability is one of the properties that characterize AGN, we will estimate these accretion rates using the data that have been analyzed here. 

To estimate $L_{\rm Bol}/L_{\rm Edd}$, we follow the relation given in \citet{2010eracleous}, which uses the black hole mass and bolometric luminosities.  According to \cite{2017mike}, the black hole masses available for the BASS sources were determined using different methods. For 14 of our sources, these were estimated using the velocity dispersion method, from the M$_{\rm BH}$-$\sigma_{*}$ relation by \cite{2013kormendy}. Two galaxies have M$_{\rm BH}$ taken from the literature (NGC\,3998 via the M-$\sigma$ relation and NGC\,4258 by a rotating H$_2$O maser disk) and for one source it was estimated from the MgII emission line (ESO\,253-G003). The uncertainties on these M$_{\rm BH}$ determinations are 
$\sim$0.3-0.4 dex (as explained in the BASS/DR1 paper - \citealt{2017mike}). We conservatively assume that the typical uncertainty on M$_{\rm BH}$ is 0.4 dex.

The other key parameter is the bolometric luminosity; the best method to estimate it is by the integrated area under the spectral energy distribution (SED). However,  observations from a variety of telescopes are necessary in order to build up a complete and detailed SED. The bolometric correction is another way to estimate it, which depends on the X-rays luminosity, and it is the one we will be using in this work. For instance, BASS/DR1 \citep{2017mike} focused on the bolometric correction derived by \cite{2007vasu, 2009vasu}, where L$_{\rm bol}$/L$_{\rm 2-10 keV}$ = 20 for $L_{\rm Bol}/L_{\rm Edd}\leq$ 0.4, and L$_{\rm Bol}$/L$_{\rm 2-10 keV}$ = 70 for $L_{\rm Bol}/L_{\rm Edd}\geq$ 0.4.

As the bolometric luminosity is fundamental in the estimation of the accretion rate, we examine an alternative determination of $L_{\rm Bol}/L_{\rm Edd}$ based on the available X-ray luminosity estimated using the data of our sample of AGN.

\begin{figure}
\label{Fig:gamma_tor_1}
    \centering    

    \includegraphics[width=9.2cm]{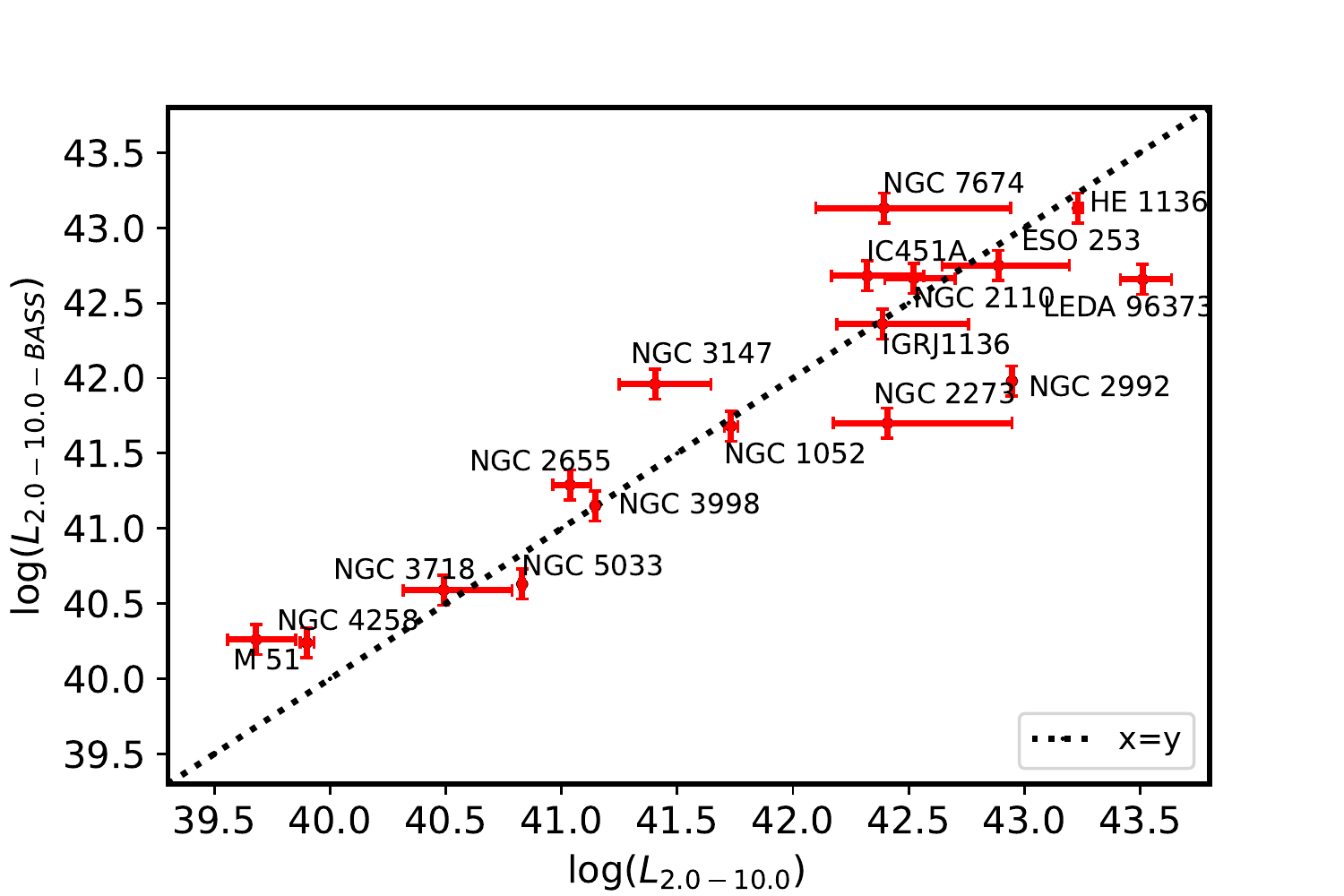}

\caption{Intrinsic luminosity in the 2.0-10.0 keV range from BASS/DR2 and our work. Dotted black line represents x=y. }
\label{fig:l2_10}
\end{figure}

We use the intrinsic luminosity in the 2--10 keV rest-frame energy range, L$_{\rm 2.0-10.0 keV}$, derived from the best-fitting spectral models of the X-ray data. Note that in the case of indistinguishable cases, we use the values from the \bors model. Using the \xill model, the results are the same. A comparison between our $L_{\rm 2.0-10.0 keV}$ calculation and the BASS/DR2 is presented in Fig. \ref{fig:l2_10}, showing differences in the luminosity, possibly related with variability. The difference between the fluxes measured by BASS/DR2 integrated 70 months and the flux measured in the short exposures with \nus that we use here can be quite large for highly variable sources such as NGC\,2992 \citep{2000Gilli, 2010shu, 2017lore, 2018marinucci, 2020marinucci} and LEDA\,96373 \citep{2009landi}.

To be consistent between the state of each AGN when measuring $\Gamma$ and other parameters, we recalculate $L_{\rm Bol}/L_{\rm Edd}$ using the fluxes measured here and refine it by changing the bolometric correction as described below.

We use our L$_{\rm 2.0-10.0 keV}$ calculation in combination with the bolometric correction K(2.0-10.0 keV) from \cite{2020duras}, who used a sample of $\sim$1000 type 1 and type 2 AGN from five different AGN surveys for which they performed a SED -fitting. They reported a bolometric correction as a function of 2.0-10.0 keV X-ray luminosity. The resulting K(2.0-10.0 keV) are slightly smaller than those used previously (L$_{\rm bol}$/L$_{\rm 2-10 keV}$ = 20), with a median value of K(2.0-10.0 keV)= 15.60 with a scatter of $\sim$0.37 dex \citep{2020duras}. The values of bolometric luminosity and Eddington ratio are given in table \ref{table:lumi_kcorrect}. The errors in the bolometric luminosity correspond to the error propagation of M$_{\rm BH}$ (0.4 dex), K(2.0-10.0 keV) (0.37 dex), and L$_{\rm 2-10 keV}$ (estimated with \texttt{xspec}). In the following analysis, we will use these $L_{\rm Bol}/L_{\rm Edd}$ values to minimize the effects of source variability.

\subsection{Accretion mechanism: The $\Gamma$ vs $L_{\rm Bol}/L_{\rm Edd}$ relation}

It has been suggested that the accretion mechanism in LLAGN ($L_{\rm Bol}/L_{\rm Edd}$<10$^{-3}$) is different from that in more powerful AGN (e.g., Seyferts) and similar to that of X-ray binaries (XRB) in their low/hard state \citep{2005yamakoa, 2009Gu, 2011younes, 2011xu, 2014yuan, 2016Lore}. 

Some authors, following the relations obtained for XRB, have studied the accretion mechanism using the relation between the spectral index $\Gamma$ and the accretion rate $\lambda_{\rm Edd}$, finding a positive correlation between these quantities at high accretion rates, suggesting a geometrically thin and optically thick disk, known as the standard model for accretion disks \citep{1973shakura, 1999Pkora}. A negative correlation has also been found at low accretion rates, indicating radiatively inefficient accretion (e.g., \citealt{yuan2007}). In this configuration, the accretion disk becomes truncated near the SMBH, with a geometrically thick and optically thin disk at lower radii and a thin disk at higher radii. However, these correlations show a large scatter \citep{2006shem, 2009Gu, 2011younes, 2015yang, 2017benny, 2018She}, with $\Gamma$ values between [1,3] \citep{2009Gu, 2011younes} and [0.5, 3.5] \citep{2018She}. The high scatter in the spectral index estimate is still not understood - it could be due to the sensitivity of the measurements or to the intrinsic properties of the galaxies.

\begin{figure*}
    \centering    

    \includegraphics[width=13cm]{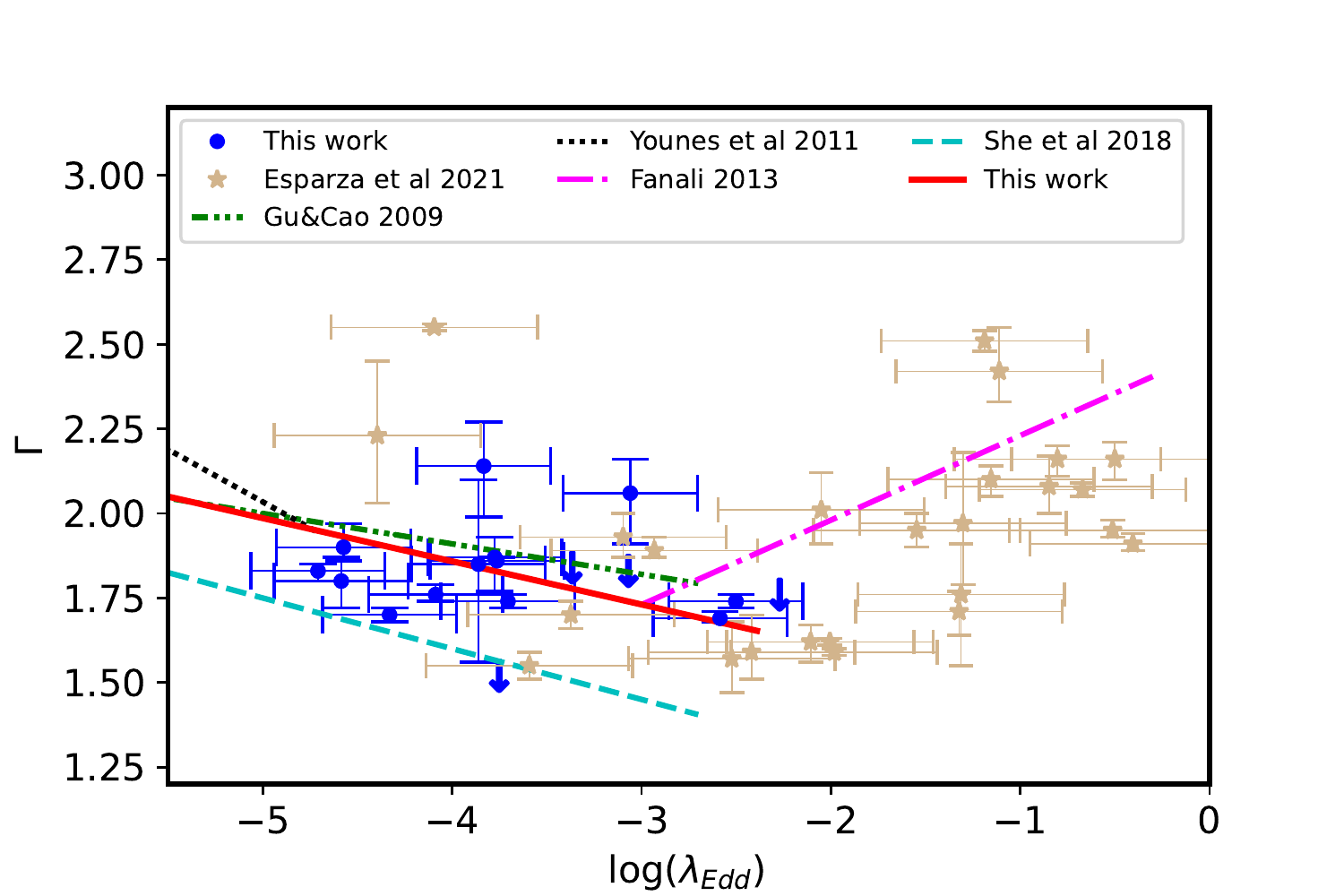}
\caption{Correlation between the spectral index, $\Gamma$, from individual fits, vs. the Eddington ratio, $\log(\lambda_{\rm Edd}) =\log(L_{\rm Bol}/L_{\rm Edd})$, for our sample of galaxies of the best fit models. 
The dot and dashed green line represent the relation given by \cite{2009Gu}, the orange dotted represents \cite{2011younes}, the magenta dashed line is the relation obtained by \cite{2018She}, while the solid black line is the correlation obtained in this work.  The purple dashed line corresponds to the relation found by \cite{2013fanali}. The blue points represented the binned data. The pink points and light yellow stars are the data point of the best fit model in this work and the ones obtained by \cite{2021esparza}.
}
\label{fig:gamma_acrat}
\end{figure*}

Thanks to the excellent statistics of \nus in combination with \xmm, we were able to better constrain the spectral index $\Gamma$ in our low accretion rate sample. In Fig. \ref{fig:gamma_acrat} we show the relation between $\Gamma$ and $\lambda_{\rm Edd}$ using the best fitting reflection model (\texttt{borus02}). We have added data from \cite{2021esparza}, who studied the torus configuration of 36 AGN using \nus and \textit{Spitzer} data and estimated the spectral parameters using the same reflection model used in this work (\texttt{borus02}). We applied the same bolometric correction to these data (see Sect. \ref{sec:bolo_corre}). In Fig. \ref{fig:gamma_acrat}, the blue points correspond to this work and the light yellow stars represent the data points from \cite{2021esparza}.

 In order to check whether there exists a break in this relation, as in XRB, we will test different scenarios. We started fitting a 1st-degree polynomial to the data, using the \texttt{polyfit} tool in python to perform this analysis.  However, previous works reported a break in the correlation (see, for example, \citealt{2018She}), where a 1st-degree polynomial with a negative slope is found on one side and a positive slope on the other. Then, we used \texttt{piecewise-regression}\footnote{https://github.com/chasmani/piecewise-regression} tool in python \citep{Pilgrim2021}, to test the existence of the breakpoint. This package fits breakpoint positions and linear models for the different fit segments, and it gives confidence intervals for all the model estimates (see \citealt{muggeo2003estimating} for more details). We found a breakpoint at $\log(\lambda_{\rm{Edd, break}})$=-2.39 with $\sigma=$0.45, in agreement with what was previously obtained by \cite{2018She}, who proposed a break at -2.5. To determine which model (with a breakpoint or not) best represents the data, we use the Bayesian information criterion (BIC), where the model with the lowest BIC value is considered the best. With no break, BIC is -113.2, and with a break, it is -115.5. This suggests that the model with a breakpoint at $\log(\lambda_{\rm{Edd, break}})$ represents well the data, i.e., AGN seems to follow the same relation as XRB. We also explored,  the possibility of more breakpoints, and it does not change the previous result.  
 
 In Figure \ref{fig:gamma_acrat} we also plot the relations given by other authors for comparison. For high luminosity AGN ($\log(\lambda_{\rm Edd})$ >-2.39), we compare with \cite{2013fanali}, who studied a sample of 71 type 1 AGN using \xmm data (purple dashed line). In the low luminosity branch ($\log(\lambda_{\rm Edd})$ < -2.39), we compare with \cite{2009Gu}, which used a sample of 55 LLAGN using \emph{Chandra} and \emph{XMM-Newton} data (green dashed line); \cite{2018She} used a sample of 314 AGN with \emph{Chandra} (cyan dashed line); and \cite{2011younes} used \emph{Chandra} and \emph{XMM-Newton} data from a sample of 13 LINER with accretion rates below -4.5 (black dashed line). 

In this work, we have shown that the inclusion of \emph{XMM-Newton} + \nus data and reflection models in the spectral fit improves the estimation of the spectral index - as also reported in \cite{2021hinkle} - which could improve the scatter compared to what was previously found by \cite{2009Gu, 2011younes, 2018She}. For details on the improvement of the uncertainties in the spectral index estimation, see Appendix \ref{apnx:scatter_error}. Indeed, in Fig. \ref{fig:gamma_acrat} can be seen that our results, when compared with previous studies, seem to agree with the correlations found by \cite{2009Gu}, \cite{2018She}, and \citet{2011younes}, but the effect of the large scatter in previous studies can be appreciated. The same is true for the high-accretion branch, where the relation of \cite{2013fanali} (at $\log(\lambda_{\rm Edd})$ > -2.39) fits well the data of \cite{2021esparza}.

To determine whether there is a relation between $\Gamma$ and $\lambda_{Edd}$, we use the tool \texttt{pymccorrelation} in Python \citep{1986isobe, 2014curran, 2020privon} to test the relationship between two variables.  This tool is able to calculate Pearson's r, Spearman's $\rho$, and Kendall's $\tau$ correlation coefficients. In this work, we use the Kendall $\tau$ correlation test, a non-parametric method for measuring the degree of association of two variables in censored data (upper/lower limits) and taking into account the uncertainties in the parameters (see \citealt{1996akritas, 1986isobe} for a detailed explanation of the calculus). A Kendall's $\tau$ close to zero indicates that there is no trend, and if they are perfectly related, Kendall's $\tau$ becomes 1.0 (or -1.0 for an anti-correlation). For the LLAGN, $\log(\lambda_{Edd}$) < -2.39, Kendall's correlation coefficient is $\tau$=-0.27. However, possibly because of the small number of sources, the associated p-value is 0.06, so the correlation is not formally significant and confirmation would require a larger sample. Note that our result is also consistent with a flat correlation for these objects. In the high luminosity branch ($\log(\lambda_{Edd})$ > -2.39), we obtain $\tau$=0.39 and a corresponding p-value of check 0.02, consistent with a positive correlation for high accreting sources. Thus, it appears that our sample provides evidence of a $\Gamma$-$\lambda_{Edd}$ relation that is consistent with previous studies, although at lower statistical significance. In any case, the change in correlation between these parameters at $\log(\lambda_{Edd}) \sim$-2.39 highlights the change in accretion physics between high- and low-luminosity AGN, consistent with previous studies (\citealt{2006shem, 2011younes} and reference therein).

Despite the small number of sources in our sample, we study the anti-correlation of the sample presented here using the tool \texttt{linregress} in Python. Then, for the low-luminosity branch where $\log(\lambda_{\rm Edd})$ < -2.39:

\begin{equation*}
    \Gamma=(-0.128\pm0.272)\times\log(\lambda_{\rm Edd})+(1.348\pm0.075)
\end{equation*}

Our work allowed us to identify the change in correlation between the spectral index and the accretion rate at $\log(\lambda_{\rm Edd})\sim$-2.39, which is highly suggestive of a change in accretion physics in AGN. We recall that a larger sample of sources combining \emph{XMM--Newton} and \emph{NuSTAR} data and fitting physical reflection models would be very useful to confirm this relation.

\subsection{Reflection}

An important feature in the spectra of AGN is the reflection that imprints its mark at X-ray energies. The shape of this reflection component is characterized by the FeK$\alpha$ emission line and the Compton hump, peaking at $\sim$ 30 keV \citep{1990Pounds}.
The gas producing the X-ray reflection in AGN could be related to the accretion disk, a neutral reflector such as the torus, or a combination of both emissions. Because we cannot separate these scenarios, in the following we will analyze the scenarios in which each of the structures dominates the X-ray spectra.

We started our analysis, by studying the torus-like reflector. Previous studies suggest a relation between the torus properties and the accretion rate. For example, \cite{2013muller}, used observations of 
\emph{VLT}/SINFONI AO-assisted integral-field spectroscopy of H2 1–0 S(1) emission of four LLAGN (NGC\,1052, NGC\,2911, NGC\,3169 and NGC\,1097), and found that on scales of 50–150 pc, the spatial distribution and kinematics of the molecular gas are consistent with a rotating thin disk, where the ratio of rotation (V) to dispersion ($\sigma$) exceeds the unity. However, in the central 50 pc in their sample, the observations reveal a geometrically and optically thick structure of molecular gas (V/$\sigma$<1 and $\rm N_{\rm H}$>10$^{23}$ cm$^{-2}$). This can be associated with the outer extent of any smaller-scale obscuring structure. In contrast to Seyfert galaxies, the molecular gas in LLAGN has V/$\sigma$<1 over an area that is $\sim$nine times smaller and column densities that are on average $\sim$three times smaller.  They interpret these results as evidence for a gradual disappearance of the nuclear-obscuring structure, also in agreement with what was previously found by \cite{2017Gonzalez-martin} using a sample of 109 AGN using \emph{IRS/Spitzer} observations.

\begin{figure}
    \centering

     \includegraphics[width=9.5cm]{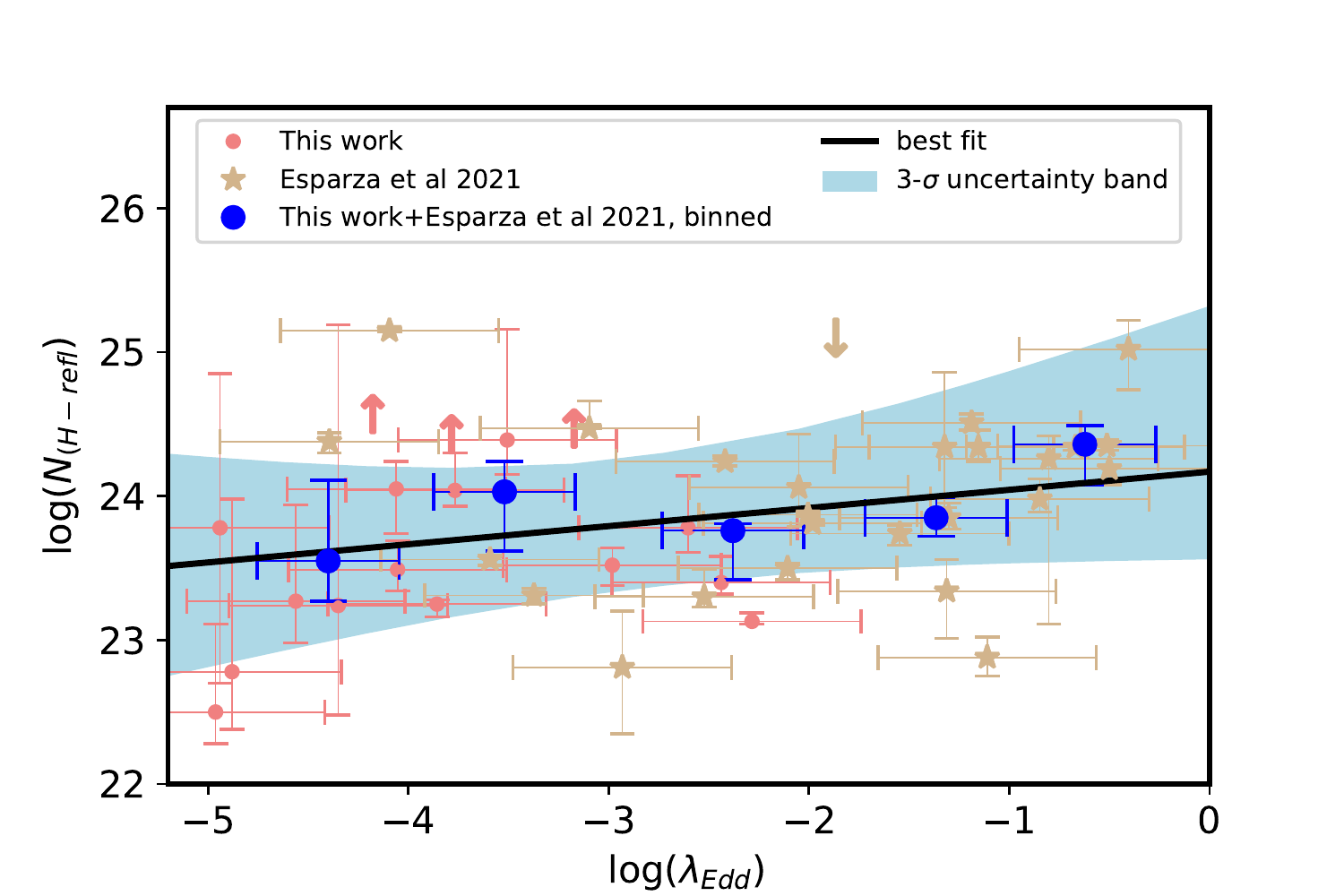}
    
    \caption{Relation between the column density of the torus-like reflector (in log) vs. the Eddington ratio, $\lambda_{\rm Edd} =L_{\rm bol}/L_{\rm Edd}$, for the sample of this work. The pink points and light yellow stars are the data point of the best fit model of this work (\texttt{borus02}) and the one obtained by \cite{2021esparza}. The blue points represented the binned data point for a bin size equal to 0.5. The black solid line represents the best fit and the light blue zone the 3$\sigma$ confidence level.}
    \label{fig:reflex}
\end{figure}

Later, \cite{2017Riccii} found that the probability of a source being obscured in the X-rays (covering factor of gas) depends primarily on the Eddington ratio instead of on absolute luminosity. They propose that the radiation pressure on dusty gas is responsible for regulating the distribution of obscuring material around the central black hole. At high accretion rates, radiation pressure expels the obscuring material in the form of outflows \citep{2006Fabian}. However, this work was made for the line of sight (LOS) column density, which is different from the torus column density (N$_{\rm H-LOS}$ $\neq$ $\rm N_{\rm{H-refl}}$). Here we will analyze, the relation between the column density of the torus-like reflector and the Eddington ratio. We plot this relation in Fig. \ref{fig:reflex} and the values obtained in this work are presented in Table \ref{table:reflection_parameters}. The pink circles and light yellow stars are the data points of the best fit model (\texttt{borus02} in the indistinguishable cases) in this work and the ones obtained by \cite{2021esparza}, respectively. The blue points represent the binned data points for a bin size equal to 0.5 dex in $\lambda_{\rm Edd}$.

Using Kendall’s tau correlation coefficient we found a correlation coefficient of $\tau$=0.22 and p-value of 0.04 for a torus-like reflector $\log(N_{H,refl})$ and $\lambda_{\rm Edd}$, suggestive of a correlation – but confirmation is required using a larger sample. As the parameters seem to be positively correlated, we perform a linear regression of the data using \texttt{polyfit} in python, and we found the following relation:

\begin{equation*}
    \log(N_{H,refl})=(0.126\pm0.303)\times \lambda_{\rm Edd}+(24.166\pm0.102)
\end{equation*}

Therefore, we find that our data is consistent with the scenario where lower accretion rate objects have, on average, lower column density material in their surroundings. However, due to the size of the sample, our correlation shows a high dispersion, then it is also consistent with being flat at 3$\sigma$.  We note that our torus fits allow for a free covering factor, so the lower column densities are not a consequence of a fixed covering factor in the model and a geometrically thinner reflector in lower accretion rate objects. For the LLAGN ($\log(\lambda_{\rm Edd})$<-2.39) we obtain a mean value of the torus column density $\log(N_{H,refl}$ cm$^{-2})$=23.76 with $\sigma$=0.74 and in the high luminosity regime, $\log(N_{H,refl}$ cm$^{-2})$=24.09 with a standard deviation $\sigma$=0.56. Consequently, our result is in line with the results in the infrared, which suggested a gradual disappearance of the torus \citep{2013muller, 2017Gonzalez-martin} and in agreement with the scenario proposed by \cite{2022ricci_BASSDR2}, where it is expected that LLAGN has lower column density. They proposed a model in which AGN move in the obscuration−accretion rate plane during their life cycle. The growth of AGN begins with an unobscured AGN accreting at $\log(\lambda_{\rm Edd})\leq$-4. Then, an accretion event then takes place, in which the SMBH is fueled, and as a result the accretion rate, column density, and covering fraction all increase. As a consequence, obscured AGN are preferentially observed. When the Eddington limit for dusty gas is reached, the covering factor and the column density will decrease, leading to an unobscured AGN being typically observed. As the remaining fuel is depleted, the SMBH goes back into a quiescent phase (see \citealt{2022ricci_BASSDR2} for more details). Even with small statistics, our results can be interpreted within the framework of this evolutionary model, in which radiation pressure regulates their evolution.

\begin{figure}
    \centering
    \includegraphics[width=9.5cm]{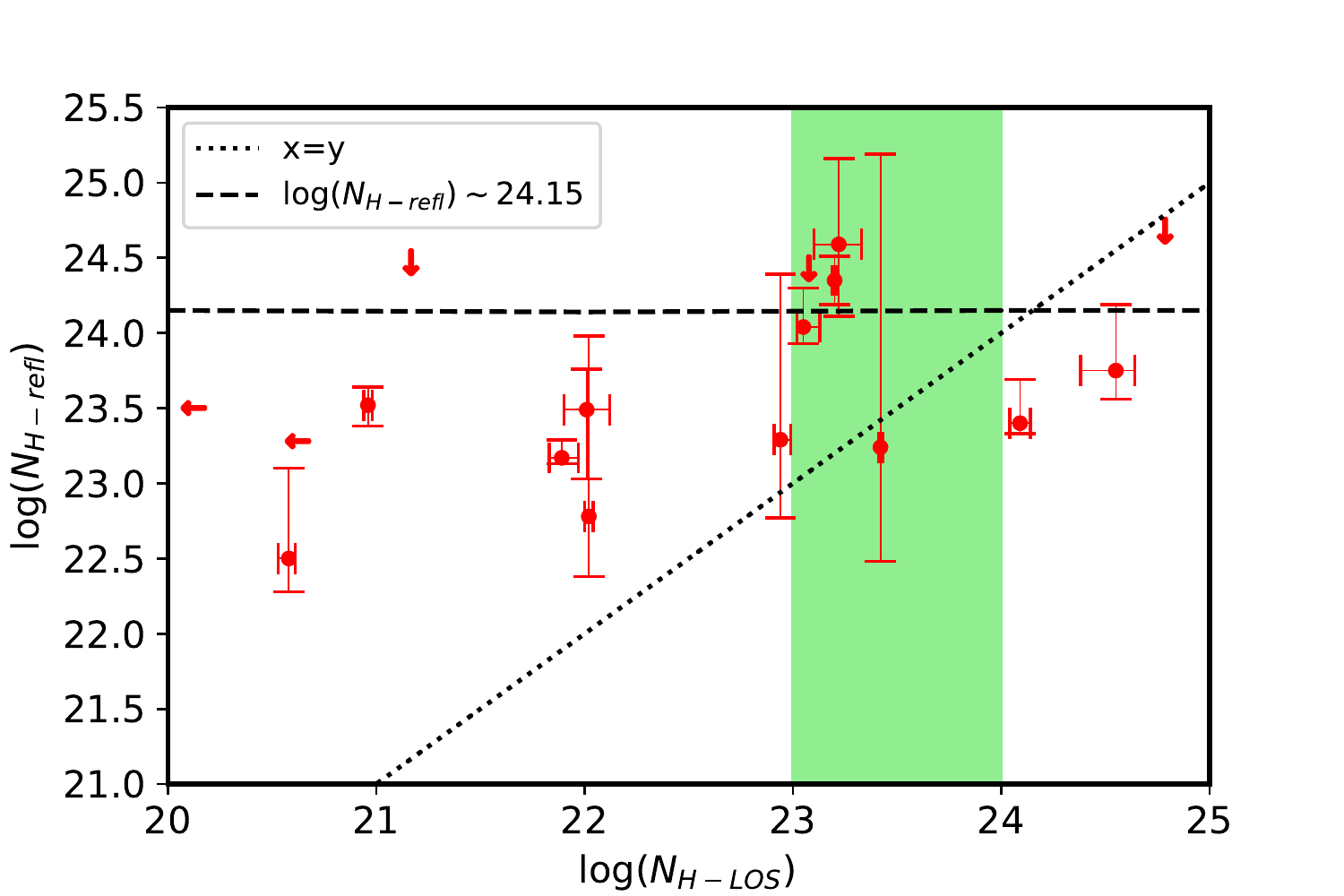}
    
    \caption{Relation between the column density of the torus-like reflector (in log) vs. column density in the line of sight (in log). The red points represent the data point of this work. The black dashed line corresponds to the value obtained by \cite{2021zhao} and the black dotted represents x=y. The green zone is the interval of the column density in the line of sight analyzed in their work. }
    \label{fig:refex_nloss}
\end{figure}

Then, we compare the column density of the reflector and the column density in the line of sight (LOS). \cite{2021zhao}, using all AGN  in the 100-month Palermo \sft/BAT catalog with line-of-sight column density between 10$^{23}$ and 10$^{24}$ cm$^{-2}$ with available \nus data shows that the average torus column density is similar for both Compton thin and CT-AGN, independent of the observing angle, with $\log(N_{\rm H-refl}$ cm$^{-2})$ $\sim$24.15. In Fig. \ref{fig:refex_nloss} we compare the column density of the torus and the absorption in the line of sight of our work. The black dotted line represents the mean value of $\log(N_{\rm H-refl}$ cm$^{-2})$ previously found by \cite{2021zhao}, and the green zone the interval of $\log$(N$_{\rm H-LOS})$ of their work. Note that our data points and their fit are in agreement in the interval of $\log$(L$_{\rm H-LOS}$) of their work, i.e. for the moderately obscured sources in our sample. The majority of galaxies with $\log$(L$_{\rm H-LOS}$)<23.0 in our sample are clearly below the value previously obtained, with a mean value of $\log(N_{\rm H-refl}$ cm$^{-2})$ $\sim$23.36 and $\sigma=0.59$.

In order to explore any correlation between these parameters, we calculate the Kendall $\tau$ correlation coefficient, and we found $\tau$=-0.17 and p-value of 0.51, suggestive of a negative correlation but compatible with a lack of correlation as well between them. Therefore, more data points are necessary to establish any correlation between these parameters. The majority of the objects show a larger $\log$(N$_{\rm H-refl}$) than $\log$(L$_{\rm H-LOS}$), suggesting that the torus is not seen through its densest part,  consistent with what was reported by \cite{2021zhao}.

Regarding the covering factor of the torus-like reflector, we obtain a mean value of CF=0.64 and $\sigma$=0.26. Note that this parameter could be constrained for seven sources. For another seven sources, the best fitting value and an upper limit could be placed, and for an additional three, the best fitting value and lower limit could be placed. In addition, we analyze the correlation between this parameter and the accretion rate by the Kendall $\tau$ correlation test. We find a correlation coefficient $\tau$=-0.06 and p-value of 0.73, suggesting that these parameters are not correlated.

Considering a disk-like reflector, we could constrain R$_{\rm f}$ only for one object (NGC\,7674), while for the others we only obtain lower limits. Also, this model allows us to study the ionization degree of the disk, however, this parameter is also unconstrained with only six galaxies with well-constrained values and one upper limit and five lower limits. 
 
The results presented in this paper suggest that the distribution of the gas in the torus in AGN is a dynamic and very complex structure, showing changes in the physical properties of the torus linked to the luminosity of the AGN, in agreement with what was previously found in the literature in the X-rays and the infrared. Certainly, combining \xmm + \nus is key to exploring the structure and distribution of the reflector and constraining its physical and geometrical parameters, especially in the low luminosity range.

\section{Conclusions}
\label{sect:conclusion}

In this work, we study the reflection of LLAGN by analyzing the broadband X-ray spectra of a BASS/DR2 sample with $\log(\lambda_{Edd})$ < -3 (17 objects) using \xmm+\nus+\sft\ observations and characterizing the reflection features using the \bors model to represent torus reflection and \xill to model accretion disk emission. The goal was to investigate the accretion mechanism by the relation between the spectral index and the accretion rate, as well as constraining the properties of the potential reflector. The main results are summarized below:

   \begin{enumerate}
      \item All objects in our sample are well-fitted with a torus-like reflector. Of these, eight objects are equally well-fitted with a torus and a disk (they are indistinguishable from a statistical point of view and visual inspection). These eight objects have consistent values for the spectral index $\Gamma$ and luminosities when modeled with a torus or a disk reflector. \\
   
      \item In our sample we can classify six objects as unobscured ($\log$(N$_{\rm H}$)<22), nine galaxies as obscured (22<$\log$(N$_{\rm H}$)<24.18) and two as Compton-thick (using as a  threshold N$_H=1.5\times 10^{24}$cm$^{-2}$, 
      according to the torus model). According to the disk case, all the galaxies can be classified as Compton thin. \\
      
      \item Combining \xmm+ \nus  and considering the reflection component in the spectral fitting, the uncertainties on the spectral index and the scatter in the relation between this parameter and the accretion rate are reduced when compared to previous works over similar ranges in accretion rate. \\ 
      
      \item Our work is consistent with the negative slope found in previous works at $\log(\lambda_{Edd})\leq$-3, and also consistent with the change in the $\Gamma-\log(\lambda_{Edd})$ relation at $\log(\lambda_{Edd})\sim$-3, where in high accretion rates sources, the slope is known to be positive. \\
      
      \item We found a tentative correlation between the torus properties (column density) and the accretion rate, suggesting that the torus has a decreasing column density with decreasing accretion rate. Consequently, AGN at $\log(\lambda_{Edd})$<-3 has a lower torus column density compared with more luminous AGN. This column density is derived from reflection as opposed to absorption in the line of sight, so it is representative of the global column density of gas around the X-ray corona. \\
      
      \item All AGN in our sample with a column density in the line of sight, $\log$(N$_{\rm H-LOS}$) < 23.0 have a torus with a column density higher than their $\log$(N$_{\rm H-LOS}$), then the torus could be observed through an underdense region.

   \end{enumerate}

In the future, new X-ray missions such as HEX-P \citep{2019hexp}\footnote{https://hexp.org/} 
facilities will detect a large sample of LLAGN, which could help us further constrain the evolution of the AGN reflection and the accretion physics behind SMBH.

\begin{acknowledgements}

We thank the referee for the valuable comments that improved the manuscript. D.Y. acknowledges financial support from the Doctorate Fellowship program FIB-UV of the Universidad de Valparaíso and the Max Planck Society by a Max Planck partner group. LHG acknowledges funds by ANID – Millennium Science Initiative Program – ICN12$\_$009 awarded to the Millennium Institute of Astrophysics (MAS). ELN acknowledges financial support from ANID Beca 21200718. CR acknowledges support from the Fondecyt Iniciacion grant 11190831 and ANID BASAL project FB210003. MB acknowledges support from the YCAA Prize Postdoctoral Fellowship.  NOC acknowledges support from CONACyT.  J.A.G. acknowledges support from NASA grant 80NSSC21K1567.
     
\end{acknowledgements}

%
%

\bibliographystyle{aa} 
\bibliography{ref}

\begin{appendix}

\section{Notes and comparisons with previous results for individual objects}
\label{app:notes_object}



 \subsection{NGC\,3998}

This object  is  spectroscopically  classified  as  a  LINER \citep{1980Heckman, 1997Ho}, and no significant broad-line polarization \citep{1999barth}. \emph{HST}/WFPC2  optical  images  revealed  an unobscured nucleus and the presence of a bright circumnuclear ionized gas disk \citep{2000Pogge,2018cazzoli}.

In radio, it contains a nucleus \citep{1984hummel} displaying a weak jet-like northern structure \citep{2002filho} and according to \citet{2016frank}, the radio source in NGC\,3998 shows two S-shaped lobes.

\citet{2010pian} found that the spectra of NGC\,3998 is better fitted with a simple absorbed power-law model, allowing the overall normalization of the model fit to each detector to be independent. Later, \citet{2016kawamuro} found the same result using \emph{Suzaku} and \emph{Swift}/BAT and fitting the reflection with \pexr reflection model. They found that the reflection strength is very weak with an upper-limit of R$_{\rm f}$<0.10 and no significant iron K$\alpha$ line emission is detected, either. They suggest that there is little surrounding matter around the nucleus.

Later, \citet{2019younes}, using \emph{XMM-newton} and \emph{NuSTAR} data found that it is an unabsorbed AGN and its spectrum in the energy range 0.5–60 keV using \emph{NuSTAR} data is best fit with a power-law and cutoff energy. They also studied the reflection hump and found only an upper limit for the reflection fraction, with value  R$_{\rm f}$<0.3 at 3$\sigma$.

{In the present work, the data were fitted with \texttt{PEXMON} and the result is consistent with a reflection fraction R$_{\rm f}$<0.15. Using \texttt{borus02} we found that the data is consistent with a Compton thin torus (with $\log(N_{H,refl})$=22.50$_{22.28}^{23.11}$) covering more than 20$\%$ of the sky. The inclination in this model is unconstrained. In case of \texttt{XILLVER} we found a model consistent with a disk with $\log\xi$>3.99 and R$_{\rm f}$>0.26. From a statistical point of view, \texttt{borus02} and \texttt{XILLVER} models are indistinguishable. All the models are consistent with a column density in the line of sight to classify this galaxy as a Compton-thin source with $\log(N_{H, los})\sim 20.5$.}

\subsection{NGC\,3718}

NGC\,3718 was optically classified as a type 1.9 LINER \citep{1997Ho}. \cite{2018cazzoli} showed that its broad H$\alpha$ line is produced in the BLR rather than an outflow. The X-ray spectrum shows a small Fe K$\alpha$ line, indicative of a reflection component \citep{2011younes,2014Lore,2020diaz}. 

Modelling the reflection with \texttt{PEXMON} we found R$_{\rm f}<0.24$. This low reflection signature can be produced by a torus like neutral reflector (\texttt{borus02}), if it is Compton thin ($\log(N_{H,refl})=22.78_{22.38}^{23.98}$), with an unconstrained covering factor (CF<0.99$^{1.00}_{0.05}$). Fitting a disk like reflector instead, using \texttt{XILLVER} requires high levels of ionization with $\log\xi=3.12_{1.35}^{3.61}$ and a R$_{\rm f}>0.86$. A (\texttt{MEKAL}) component was necessary to improve the fit at low energies in all the models. The \texttt{XILLVER} and \texttt{borus02} are indistinguishable.
All the models are consistent with a Compton-thin absorber  in the line of sight to the primary coronal emission  with $\log(N_{H, los})\sim 22$.

\subsection{NGC\,4258*}

NGC\,4258* is a SABbc spiral galaxy,  spectroscopically classified as a 1.9 Seyfert \citep{1997Ho} and as LINER \citep{2014balmaverde}. This source has a highly obscured central X-ray source and is well known for its anomalous arms, discovered on the basis  of  H$\alpha$ imaging \citep{2001wilson}  and for its nuclear H$_{\rm 2}$O megamaser, which traces a dense, edge-on disk on sub-parcsec scales \citep{1994watson, 1995greenhill}. The nucleus of NGC\,4258* also contains a relativistic radio jet \citep{2013doi}.  

\citet{2009reynolds} combined \emph{Suzaku}, \emph{XMM-Newton} and \emph{Swift}, and detected  robust  flux  variability  of  the 6.4 keV iron line and suggested a model in which the line originates from the surface of a warped accretion disk.  Also, during their \emph{Suzaku} observation, they detected high amplitude intraday  variability,  with  fluctuations  on  timescales  as short as 5 ksec. \citet{2005herrnstein} found that the absorption may well arise in the outer layers of the warped geometrically thin accretion disk, further reducing the need for any cold structure other than the accretion disk itself.


{\emph{NuSTAR} and \emph{XMM--Newton} observations of this object are not simultaneous, so we explored possible spectral variability. We found variations in the normalization of the power-law, so we let the normalizations of the power-law free. We then fitted the models. Using \texttt{PEXMON} we found R$_{\rm f}$=0.78$_{0.20}^{3.744}$. Using \texttt{borus02} we found $\log(N_{H,refl})$=23.29$_{22.77}^{24.39}$, so borderline Compton-thin/thick torus. The torus covers less than 81$\%$ of the sky. Fitting with \texttt{XILLVER} we found a disk with $\log\xi$<1.77 and R$_{\rm f}$>3.80. In order to improve the fit in the soft energy band, it was necessary to add three components, modelled with two MEKAL plus a scattered power-law in all the models. Note that in case of \xill, was necessary to use a cut-off power law in the scattered component as in this component, dominate above $\sim$70 keV.} The \texttt{borus02} model  is the best representation of the data.  All the models are consistent with a Compton-thin absorber in the line of sight to the primary coronal emission  with $\log(N_{H, los})\sim 22.94$.

\subsection{ ESO\,253-G003}

This source  is  spectroscopically  catalogued  as  a  Seyfert 2 in \citet{2006veroncetty}. The \xmm and \nus observations of this object are separated by 3 days, so we fit the overlapping 3.0--10.0 keV range of both instruments with a simple absorbed power-law model and checked whether they were consistent. This test showed there was spectral variability between both epochs, with a significant change in the slope $\Gamma$. 
Due to its complexity, the \xmm data was excluded for our analysis. We performed a simultaneous fit of the \nus and \sft/BAT data with \texttt{PEXMON} and significant reflection was detected with  R$_{\rm f}$>1.59. Modelling this reflection with \texttt{borus02} instead of \texttt{PEXMON} results in a marginally Compton thin torus reflector  ($\log(N_{H,refl})$=24.04$_{23.93}^{24.30}$) covering more than 73$\%$ of the sky. Fitting the reflector using \texttt{XILLVER} instead of \texttt{borus02} results in a highly ionized disk with  $\log\xi=2.68_{1.34}^{2.84}$ and R$_{\rm f}>0.85$.  The \texttt{PEXMON} model shows the lowest $\chi_{\nu}^{2}=0.89$ although all models show similar quality fits ($\chi_{\nu}^{2}$=0.905 for \texttt{borus02} and $\chi_{\nu}^{2}$=0.903 for \texttt{XILLVER}). All the models are. indistinguishable, and they are consistent with a Compton-thin absorber in the line of sight to the primary coronal emission  with $\log(N_{H, los})\sim 23$.

\subsection{ NGC 1052}

NGC\,1052 is the brightest elliptical galaxy in the Cetus I group. It was optically classified as a LINER by \citet{1980Heckman}, then it was classified as LINER type 1.9 \citep{1997Ho}.  \citet{2018cazzoli} proposed that this object presents signs of outflowing winds by using optical 2D spectra.

This source is a radio loud galaxy \citep{2007maoz}. At 1.4 GHz \emph{Very Large Array (VLA)} image of NGC 1052 shows a core-dominated radio structure, with only about 15$\%$ of the flux density in extended emission: there are two lobes spanning 3 kpc, possibly with hot spots \citep{1984wrobel}. 

At X-rays, \citet{2020osorio}  presented an extensive study of NGC 1052 using observations from \emph{Chandra}, \emph{XMM-Newton}, \emph{NuSTAR}, and \emph{Suzaku}. They reported variability in the nucleus and found variations both in the intrinsic continuum  flux,  photon  index,  and  in  the  obscuration  along the line of sight. The reflection component is a steady emission both in flux and shape, fully consistent with reflection in a distant structure, perhaps the torus. They argue that NGC 1052 is in the regime of Compton-thin sources, consistent with the fact that the flux of the reflection component is not dominant in the hard band. In addition, \cite{2021mislav} using \emph{NuSTAR}, \emph{XMM-Newton}, \emph{Suzaku} and \emph{BeppoSAX} observations and fitting the \bors model, found covering factor of $\sim$80-100$\%$ and a column density in the range (1-2)$\times$10$^{23}$ cm$^{-2}$ which is well-matched with line-of-sight column density.

In this work, we fit the data with \texttt{PEXMON}, and it is consistent with R$_{\rm f}$=1.26$_{0.79}^{3.05}$. Using \texttt{borus02} model it has ($\log(N_{H,refl})$=24.35$_{24.19}^{24.57}$) so a borderline Compton-thin/thick torus, and CF=0.50$_{0.27}^{0.54}$. Using \texttt{XILLVER} we found $\log\xi$=1.88$_{1.78}^{2.00}$ and R$_{\rm f}$>9.18. In case of \texttt{borus02} it was necessary to add a MEKAL and a scattered power-law to improve the fit. In case of \texttt{XILLVER}, it was necessary to add  two components to improve the fit (an absorber and a \texttt{MEKAL}) and in case of \texttt{PEXMON} three components were necessary (MEKAL, power-law and an absorber). From a visual inspection, we had to add two Gaussian emission lines centered at: S XIV at 2.4 keV and Si XIII at 1.85 keV. The \texttt{borus02} model is the best representation of the data. 
All the models are consistent with a Compton-thin absorber in the line of sight to the primary coronal emission  with $\log(N_{H, los})\sim 23.2$.

\subsection{NGC\,2655}

NGC\,2655 is the brightest member of a group NBGG 12-10 in the Nearby Galaxy Groups catalog of \citet{1988tully}. It is classified as Seyfert 2 in \sft/BAT 70 month catalog \citep{2013baumgartner} and as a Type-2 LINER according to the optical classification done by \citet{1997Ho,2010veroncetti}.

In the X-rays, \citep{2009omaira} found that the 0.2--10 keV \xmm spectrum is well modelled by a MEKAL plus 2 power-law components. \citet{2016kawamuro} fit 2 APEC models to the \emph{Suzaku} spectrum of this galaxy and obtained a line of sight  $\log( N_{H})=2.61^{+0.27}_{-0.17}\times10^{23}$ cm$^{-2}$ and $\Gamma=1.77^{+0.19}_{-0.07}$, which are consistent with earlier fits to \emph{ASCA} data \citep{2002terashima}.  \citet{2016kawamuro} also detected a Fe K$\alpha$ line at 6.4 keV, which was not detected in the previous \emph{ASCA} and \emph{XMM-Newton} observations \citet{2002terashima2, 2009omaira}. 

We fitted the data with \texttt{PEXMON} placing an upper limit on R$_{\rm f}$<0.52. Modeling the reflector with a torus instead, using  \texttt{borus02}, we found a column density of the reflector of  $\log(N_{H,refl})$=23.24$_{22.48}^{25.19}$ and CF>0.11. Modeling the reflection with an ionized disk instead of a torus, using \texttt{XILLVER} results in a highly ionized disk with a large reflection fraction $\log\xi<3.55$ and R$_{\rm f}$=9.84$_{**}^{10}$. In the case of \texttt{borus02} it was necessary to add two MEKAL and scattered power-law to improve the fit. In case of \texttt{XILLVER} it  was necessary to add a MEKAL and an absorber (modelled with zxipcf) components to improve the fit at low energies. In the case of \texttt{PEXMON} an additional \texttt{MEKAL} component was required. The three reflection models showed similar quality fits with $\chi_{\nu}^{2}=0.97$ for \texttt{PEXMON}, $\chi_{\nu}^{2}=1.01$ for both \texttt{borus02} and \texttt{XILLVER}, they are indistinguishable.

All the models are consistent with a heavy but Compton-thin absorber in the line of sight to the primary coronal emission with $\log(N_{H, los})\sim 23.4$. We note that if a torus is producing the reflection, then its average column density is \emph{lower} than the absorber column density in the line of sight. 

\subsection{NGC\,3147*}

NGC\,3147* is an isolated Seyfert 2 galaxy \citep{1997Ho} of Hubble morphological type SA(rs)bc. It was suggested as a true type 2 Seyfert candidate  \citep{2012bianchi}, but later confirmed as an AGN accreting at low rate \citep{2019bianchi}. In the radio, it shows a point-like structure \citep{2001ulvestad,2006krips}.

At X-rays, \citet{2017bianchi} using \emph{NuSTAR} data, shows that spectrum of NGC\,3147* can be simply modelled by a power-law with a standard $\Gamma\sim$1.7 and an iron emission line. These spectral properties, together with significant variability on time-scales as short as weeks, strongly support a line-of-sight free of absorption for this source. They suggested that \emph{NuSTAR} data adds further evidence in favor of an X-ray spectrum completely unaffected by absorption, confirming NGC\,3147 as one of the best cases of true Type 2 Seyfert galaxies, intrinsically characterized by the absence of a BLR. 

{\emph{NuSTAR} and \emph{XMM-Newton} observations of these objects are not simultaneous, then we explored possible spectral variability. We found variations in the normalization of the power-law, so this parameter was left free. We perform a simultaneous fit with this configuration of the data with \texttt{PEXMON} and found R$_{\rm f}$=3.99$_{0.97}^{6.41}$. We also fitted the data using \texttt{borus02} model and found that the data is consistent with a Compton thin torus with density $\log(N_{H,refl})$=23.27$_{22.98}^{23.94}$ covering less than 89$\%$ of the sky. Using the \texttt{XILLVER} model, the fit was consistent with $\log\xi$=0.79$_{0.06}^{1.15}$ and an R$_{\rm f}$>0.91. In case of \texttt{PEXMON}, it was necessary to add one component to improve the fit (a warm absorber modelled with zxipcf). All the models are indistinguishable and consistent with a Compton-thin absorber in the line of sight to the primary coronal emission with $\log(N_{H, los})<20.9$.}

\subsection{NGC\,2110*}

NGC\,2110* is a nearby S0 galaxy, \citep{1991devaucouleur} Seyfert 2 AGN. 
 NGC\,2110* shows a prominent Fe K$\alpha$ line accompanied by variable intrinsic emission \citep{1996hayashi}.  

\citet{2014marinucci} report the X-ray spectral analysis of NGC\,2110* observed by \emph{NuSTAR} in 2012, when the source was at the highest flux level ever observed, and in 2013, when the source had more typical flux levels. They found an upper limit on R$_{\rm f}$ < 0.14, confirming results from past high-energy \emph{BeppoSAX} and \emph{Suzaku} observations \citep{1999malaguti, 2006reeves, 2014rivers}.  Using MyTorus \citep{2009Murphy} to model the reflection, they found a CF of 0.5 with equatorial N$_{\rm H}$ = 2.0 $\pm$ 1.1 $\times$ 10$^{23}$ cm$^{-2}$. 

We re-examine the 2013 \nus observations together with the 2003 \xmm  observation in the overlapping 3--10 keV band, finding a consistent slope and variable normalization. We therefore fit jointly the broad band spectra with untied normalization for the cutoff power-law component between instruments. Modelling the reflection with \texttt{PEXMON} gives R$_{\rm f}$=1.01$_{0.91}^{1.12}$. Modeling the reflection with a torus instead, using \texttt{borus02} we found a Compton thin reflector with $\log(N_{H,refl})$=23.49$_{23.03}^{23.76}$ covering 13$\%$ of the sky, consistent with \cite{2018balokovic}. Modeling the reflection with a disk instead of the torus, with \texttt{XILLVER}, we obtain $\log\xi$=1.99$_{1.82}^{2.01}$ and R$_{\rm f}$>7.80. In the case of \texttt{XILLVER} it was necessary to add two components to improve the fit at low energies: a MEKAL and zxipcf. In case of \bors three components were necessary to improve the fit: a scattered power-law, MEKAL and a zxipcf, and in the other case, four components were necessary: 2 zxipcf + 2*MEKAL. The torus reflector produces the best fit. 

All the models are consistent with a Compton-thin absorber in the line of sight to the primary coronal emission with $\log(N_{H, los})\sim 22.01$. 

\subsection{LEDA\,96373*}

LEDA\,96373 (also 2MASX J07262635–3554214 or IGR J07264–3553) was first observed at high energies during the all-sky hard X-ray IBIS survey \citep{2007krivonos} and reported in the Palermo \sft/BAT survey. This galaxy is classified in NED as a Seyfert 2 and as a Compton thick galaxy \citep{2016koss}.

\citet{2009landi} observed this galaxy with \sft XRT data and reports an excess emission below 2 keV and  found that a double power law model is a good fit. They found an intrinsic column density of $\sim$7 $\times$ 10$^{22}$ cm$^{-2}$ and a photon index $\Gamma\sim$2.5. The source seems to vary by a factor of 2 within a timescale of few days, with an average 2–10 keV flux of $\sim$4.2 $\times$ 10$^{-13}$ erg cm$^{-2}$ s$^{-1}$. 

{Simultaneous \nus and \xmm data are not available for this source, so we explored possible variability. We found variations in the normalization of the power-law, so this parameter was left free. Using \texttt{PEXMON} we found R$_{\rm f}$>9.38. Using \texttt{borus02} we found that the data is consistent with a Compton thin torus ($\log(N_{H,refl})$=23.40$_{23.33}^{23.69}$) covering 59$\%$ of the sky. Then we fitted \texttt{XILLVER} and found $\log\xi$=2.08$_{2.03}^{2.17}$ and R$_{\rm f}$>9.37. In case of \texttt{PEXMON} and \texttt{borus02}, it was necessary to add three components to improve the fit (two MEKAL*zxipcf), and in case of \texttt{XILLVER}, (MEKAL + a scatter power-law)*zxipcf components were necessary. The \texttt{borus02} model is the best representation of the data. 
The fit with \texttt{borus02} requires Compton thick absorption in the line of sight to the coronal emission with $\log(N_{H, los})$= 24.09$^{+24.17}_{-24.06}$.}

\subsection{NGC\,2992}

NGC\,2992 is a spiral galaxy classified as a Seyfert 1.9 in the optical.
This Seyfert galaxy is a changing-look AGN that varies from type 2 to intermediate type sometimes accompanied by extreme X-ray activity and back, over the span of a few years, \citep{2010shu} which is attributed to intrinsic variations of the powerlaw flux and to changes in the absorption in the line of sight \citep{2015parker,2017lore}.

At higher X-ray energies, a combined study with \emph{INTEGRAL}, \emph{Swift} and \emph{BeppoSAX} data published by \citet{2007beckmann} showed that variations in the normalization of the power law were needed when using an absorbed broken power law model to fit the data simultaneously. They found a constant $\Gamma$ and flux variations by a factor of 11 in timescales of months to years.

We used only the \xmm data that overlaps in time with the \nus exposure and perform a simultaneous fit.  Modelling the reflection with \texttt{PEXMON} results in R$_{\rm f}$=0.37$_{0.33}^{0.39}$. Using \texttt{borus02} model for the reflection instead of \texttt{PEXMON} we found a Compton thin torus reflector with $\log(N_{H,refl})$=23.17$_{23.13}^{23.29}$ covering more than 75$\%$ of the sky. Then we fit \texttt{XILLVER} instead of \texttt{borus02} and the fit was consistent with a disk with $\log\xi$<0.08  producing R$_{\rm f}$>8.80. In the case of \texttt{PEXMON} and \texttt{borus02}, it was necessary to add a MEKAL, a scattered power-law and zxicpf components; in the case of \texttt{XILLVER}, two MEKAL components were required. The \texttt{borus02} model is the best representation of the data.

All the models are consistent with a Compton-thin absorber in the line of sight to the primary coronal emission with $\log(N_{H, los})\sim 21.9$. 

\subsection{ M\,51} 

M\,51 (also known as Messier 51a, M\,51a, and NGC 5194) is a very nearby (7.1 Mpc; \citealt{2006takstas}) Compton-thick AGN. It hosts a Seyfert 2 nucleus \citep{1997Ho, 2011dumas} and shows two radio lobes that are filled with hot X-ray gas \citep{2000terashima}  and an outflow of ionised gas \citep{2004bradley}. This is an interacting spiral galaxy and lies in the constellation Canes Venatici.

 \citet{2018brightman} reported a ULX located close enough to the nucleus of M\,51 that \nus cannot resolve it. They modelled this emission with a cut-off power-law model where $\Gamma = -2.21^{+2.26}_{-2.16}$ and E$_{cut} = 1.2_{-1.1}^{+1.4}$ keV. We included the emission from the ULX using the same model setting $\Gamma$ and E$_{\rm cut}$ frozen to the parameters already obtained by them. Here we perform a simultaneous fit of the \xmm, \nus and \sft/BAT data. Using \texttt{PEXMON} reflection model we find R$_{\rm f}>7.86$. Using \texttt{borus02} we found a Compton thick torus reflector ($\log(N_{H,refl})>24.68$) covering less than 57$\%$ of the sky. Then we fit \texttt{XILLVER} and the fit was consistent with an accretion disk with $\log\xi<1.26$ and with R$_{\rm f}> 8.96$. In case of \texttt{PEXMON} and \texttt{XILLVER} one additional component was needed (modelled with zxipcf). The torus model gives a good representation of the data. 
 The fit with this model requires Compton thick absorption in the line of sight to the coronal emission with $\log(N_{H, los})> 24.45$ otherwise with the other models, a Compton thin column density is a good representation of the data.

\subsection{NGC\,2273*}

NGC\,2273* is a Seyfert 2 galaxy \citep{1999risaliti} considered a Compton-thick due to the detection of a strong Fe K$\alpha$ line with an equivalent width $>1$ keV and a low ratio of L$_{X}$/L[OIII] \citet{2005guazaniti}. Observations with  \emph{Suzaku} suggested that the nucleus of NGC\,2273 is obscured by a Compton thick column of  $1.5\times10^{24}$ cm$^{-2}$ \citep{2009awaki}. 
A hidden nucleus is also suggested by the detection of polarized broad lines \citep{2000moran}. 

We explore the spectral variability in this galaxy between the 2003 \xmm observation and the 2014 \nus observation by fitting an absorbed power-law plus 6.4 keV Gaussian model to the overlapping 3--10 keV spectra. we find that allowing only the power-law normalization to vary between instruments/epochs produced an acceptable fit with $\chi_{\nu}^{2}=1.08$ and with consistent values of the normalization, so we will fit these spectra with all the parameters tied. We perform a simultaneous fit of the \xmm, \nus and \sft/BAT data with \texttt{PEXMON} reflection model and the result is consistent with R$_{\rm f}$>9.02. We modeled the reflection using \texttt{borus02} instead of \texttt{PEXMON} finding a Compton thin torus ($\log(N_{H,refl})=23.75_{23.56}^{24.19}$) with a covering factor lower than 59\% of the sky. Modeling the reflection with\texttt{XILLVER} instead produced moderately ionized disk  with $\log\xi=2.15_{1.74}^{2.31}$ and R$_{\rm f}$>8.80. In all the models it was necessary to add two components (MEKAL and absorber). The disk like reflector modelled with \texttt{XILLVER} is the best representation of the data.  The fit with this model requires Compton thick absorption in the line of sight to the coronal emission with $\log(N_{H, los})> 24.55$.

\subsection{HE\,1136-2304}

HE\,1136-2304 changed its optical spectral classification from 1994 (Seyfert 2) to 2014 (Seyfert 1.5) and can be considered an optical changing look AGN \citep{2018zetl}. HE\,1136-2304  has been detected as a variable X-ray source by the \xmm slew survey in 2014 \citep{2016parker}. The 0.2–-2 keV flux increased by a factor of about 30 in comparison to the 1990 \emph{ROSAT} All-Sky Survey flux (RASS; \citet{2000voges}). However, no clear evidence of X-ray absorption variability has been seen.

\cite{2016parker} found an absorbing column density on the X-ray spectrum around $10^{21}$ cm$^{-2}$ in addition to Galactic absorption. This spectrum shows a moderate soft excess and a narrow Fe line at 6.4 keV and a high energy cut-off at $\sim$ 100 keV.

Although the X-ray flux is highly variable in this source, a joint spectral fitting with \emph{XMM-Newton} and \emph{NuSTAR} could be made since the observations were simultaneous. Using \texttt{PEXMON} reflection model we found R$_{\rm f}$=3.99$_{2.85}^{5.10}$. We also fit the data using \texttt{borus02} model instead of \texttt{PEXMON}, and we found that reflection features can be produced by a  Compton thin torus  with $\log(N_{H,refl})$=23.52$_{23.38}^{23.64}$ covering  59$\%$ of the sky. Modelling the reflection with a disk instead, using \texttt{XILLVER} results in a  highly ionized disk with $\log\xi$=3.78$_{3.77}^{3.79}$ and R$_{\rm f}$>1.45. To improve the fit in the low energy range, it was necessary to add one MEKAL component in all the models. The \texttt{borus02} model is the best representation of the data.
All models required a Compton-thin absorber on the line of sight to the primary coronal emission, with $\log(N_{H, los})\sim 20.97$.

\subsection{IGRJ\,11366-6002:}  We did not find previous information about this galaxy in the literature.

Here we perform the fit of the \nus and \sft/BAT and \sft/XRT data. When using \texttt{PEXMON} reflection model, we find R$_{\rm f}$=0.72$_{0.49}^{1.57}$. We also fit the data using \texttt{borus02} model, and we found that the data is consistent with a Compton thick reflector torus ($\log(N_{H,refl})$>24.47) covering $\sim93\%$ of the sky. Using \texttt{XILLVER}, the fit was consistent with $\log\xi$=3.14$_{3.03}^{3.32}$ and  R$_{\rm f}$>1.12. All the models are indistinguishable, and they are consistent with a Compton-thin absorber in the line of sight to the primary coronal emission with $\log(N_{H, los})< 21.2$.
        
\subsection{IC\,4518A}

This galaxy was optically classified as a type 2 Seyfert galaxy \citep{2009zaw} and it is classified as a Compton-thin source \citep{1999bassani, 2008derosa, 2015lore}. 

\citet{2008derosa} presented a 0.2–200 keV broad-band study of this galaxy with \emph{INTEGRAL}, \emph{XMM-Newton}, \emph{Chandra} and \emph{ASCA} to investigate the continuum shape and the absorbing/reflecting medium properties. They fitted \pexr the reflection model and found that in this object the presence of the reflection component above 10 keV is statistically required by the data. However, they found the best fit value larger than 1 that they suggested that could be related with the geometry of the reflector, which should be more complex than that used in their work. 

\citet{2015lore} studied \emph{XMM–Newton} data of this galaxy and found variations in an eight-day period, that correspond to a flux variation of 40$\%$ (41$\%$) in the soft (0.5 - 2 keV) (hard - 5 - 10.0 keV) energy band.  

 for this work, \nus and \xmm data were not simultaneous, thus we explored possible spectral variability. The best fit resulted when the normalization of the power-law is a free parameter. Using \texttt{PEXMON} we found  R$_{\rm f}$=3.41$_{1.10}^{6.17}$. Using \texttt{borus02} we find a Compton-thick reflector torus with $\log(N_{H,refl})$=24.59$_{24.11}^{25.16}$, covering more than 39$\%$ of the sky. Using \texttt{XILLVER} we found a model consistent with $\log\xi$<1.73 and  R$_{\rm f}$>2.94. To improve the fit, three additional components are necessary (MEKAL + scatter power-law)*zxipcf for \texttt{PEXMON},  MEKAL + zxipcf for \texttt{borus02}) and \texttt{XILLVER}). The \texttt{borus02} model is the best representation of the data. In all the models, the absorption in the line of sight is consistent with being Compton-thin ($\log(N_{H, los})\sim$ 23.22).

\subsection{NGC\,7674*} 

NGC\,7674* (Mrk 533) is a nearby luminous infrared galaxy (LIRG). This galaxy is the brightest member of the Hickson 96 interacting galaxy groups and it is a known as a Seyfert 2 galaxy with broad H$\alpha$ and H$\beta$ components in polarized light \citep{1996young}. Later, \citet{2005bianchi} classified it as a changing look AGN in the X-ray range, switching between Compton-thin and Compton thick absorption in the line of sight.

In the X-rays, NGC\,7674* was first reported to be a reflection-dominated AGN by \citet{1998malaguti} from \emph{BeppoSAX} X-ray observations carried out in 1996, with the direct (intrinsic) continuum being fully absorbed by a Compton-thick gas column. \citet{2005bianchi} studied the \xmm spectra of this galaxy, finding a reflection fraction R$_{f}\sim$1.5 and argue that the changes on the CT regimes in this galaxy is associated with material in the line of sight. 
Later, \citet{2017gandhi} present \emph{NuSTAR}, \emph{Suzaku} and \emph{Swift} reporting a flat X-ray spectrum, suggesting that it is obscured by Compton-thick gas. Based upon long-term flux dimming, previous work suggested the alternate possibility that the source is a recently switched-off AGN with the observed X-rays being the lagged echo from the torus. Their data show the source to be reflection-dominated in hard X-rays, but with a relatively weak neutral Fe K$\alpha$ emission line and a strong Fe XXVI ionised line. Also, they construct an updated long term X-ray light curve of NGC\,7674* and find that the observed 2–-10 keV flux has remained constant for the past 20 years. 

As this galaxy is a changing look galaxy and the \xmm and \nus observations are separated by several years, we do not attempt a joint fit and exclude the \xmm observation for our analysis. Using \texttt{PEXMON} we found that the data is consistent with R$_{\rm f}$>2.21. Using \texttt{borus02} to model the reflection instead of \texttt{PEXMON} we found a torus with  $\log(N_{H,refl})$>24.43 covering 70$\%$ of the sky. Modelling the reflection with a disk instead of a torus, using  \texttt{XILLVER} the fit was consistent with $\log\xi$=2.82$_{2.77}^{3.02}$ and R$_{\rm f}$=3.25$_{2.26}^{6.07}$. The \texttt{XILLVER} model is the best representation of the data.

We find heavy but Compton thin absorption in the line of sight to the primary coronal emission with $\log(N_{H, los})\sim 23$ for all the models used.

\subsection{NGC\,5033} 

NGC\,5033 is a nearby spiral galaxy with a low-luminosity Seyfert 1.8 type nucleus \citep{2010veroncetti} with a point-like central X-ray source \citep{1999terashima}. This galaxy has been alternatively classified as a Seyfert of type 1.5 \citep{1997Ho}. This object has a giant neighbor within a short distance: NGC 5005, a Seyfert SAB(rs) galaxy. 

In the radio, it is predominated by a compact core \citep{2001ho}, and showing extended jet-like features to the East-West \citep{2007perestorres}.

In the X-rays, \citet{1999terashima} using ASCA observations found a point-like X-ray source in the 2–10 a keV band. Their X-ray light curve showed variability on a timescale of $\sim$ 10$^{4}$ s with an amplitude of $\sim$20$\%$. 

In this work, we combine the \nus observations and perform a simultaneous fit of the \xmm, \nus and \sft/BAT data. Using \texttt{PEXMON} reflection model results in R$_{\rm f}$>1.0. Using \texttt{borus02}, we found that the data is consistent with a Compton thin torus ($\log(N_{H,refl})$=23.49$_{23.34}^{23.69}$) covering 77$\%$ of the sky. Then we fit \texttt{XILLVER} instead of \texttt{borus02} and the fit was consistent with $\log\xi$<1.0 and R$_{\rm f}$>4.88. To improve the fit, it was necessary to two MEKAL to \texttt{XILLVER}. The \texttt{borus02} model is the best representation of the data. 
In all the models, the required line of sight absorption is compatible with being Compton-thin ($\log(N_{H, los})<$21.0).
\clearpage        

\section{Existence of the reflection component}
\label{apnx:reflex_comp}

The aim of our spectral analysis is to quantify how much reflecting material can exist around the central engine, assuming that there must be material around the SMBH. In this section, we fit the nuclear continuum with and without reflection to probe that this component is usually required by the data. We based our analysis on the statistical significance of the reflection component compared to the model without reflection (i.e. F-test$<$10$^{-3}$), we examine the existence of the reflection component.

We start the analysis by fitting the spectra to the following model: \texttt{constant*phabs*zphabs*cutoffpl}. This model represents a power-law component (\texttt{cutoffpl} - associated with the intrinsic continuum) absorbed by the material along the LOS to the observer (\texttt{phabs}).  The cross normalization constant and the galactic absorption are denoted by \texttt{constant} and \texttt{zphabs} respectively. The column density (N$_{\rm H}$), photon index ($\Gamma$), high energy cut off (E$_{\rm{cut}}$), and normalization of the power-law are the free parameters for this model. In order to avoid including the soft emission in this section of the analysis, we omit the spectra below 3.0 keV.

Then, we use \texttt{constant*phabs*zphabs*(cutoffpl+pexmon)}, to study the case considering the reflection. We will use the  \texttt{pexmon} model to fit the data. This model assumes Compton reflection from neutral X-ray photons in an optically thick material with plane-parallel geometry. The photon index, high energy cut-off, reflection fraction (R$_{\rm f}$), metal and iron abundances, inclination angle, and normalization are all free parameters in \texttt{pexmon}. We have set the photon index of the \texttt{pexmon} to be the same as that of the power-law. The iron abundance was assumed to be solar. The high energy cut off and normalization are free parameters. To take into account the reflection component and exclude the intrinsic power-law continuum in \texttt{pexmon} model, we have set R$_{\rm f}$ to -1. Then, we fit the spectra with this model and test if reflection is required by the data using the F-test tool within \texttt{Xspec}. 

 We present the results of this analysis in Table \ref{table:exitence_reflx}, the addition of a reflection component improves the spectral fit for 14 out of the 17 sources (82$\%$ of the sample). Statistically, the reflection component is not required in three galaxies in our sample. 
NGC\,2655 which has the lowest number of counts, and NGC\,3998 and NGC\,3718 which have the lowest Eddington ratios in our sample. Due to their luminosity, the contribution of the reflection could be very small, and as a consequence we are not able to quantify the reflection component. Because most of the sources do require a reflection component, we will use reflection models to fit all the sources in order to put constraints on this component.

\clearpage

\begin{table*} 
\caption{Power-law and Pexmon model.}           \label{table:exitence_reflx}  
\centering   
\setlength{\tabcolsep}{2.20pt}


\renewcommand{\arraystretch}{1.7} 

\begin{tabular}{ l |c c c |c c c| c | c  } 
\hline       

 & \multicolumn{3}{|c|}{\texttt{PL}} & \multicolumn{3}{|c|}{\texttt{PL+Pexmon} (R$_{f}$=-1)} & \\\hline
 
Name & $\Gamma$ & $\chi^{2}$/d.o.f. & $\chi^{2}_{\nu}$ & $\Gamma$&  $\chi^{2}$/d.o.f. & $\chi^{2}_{\nu}$ &  F-test & Counts 
\\
  & & & & & & & & (3.0-79.0 keV)\\
\hline   

NGC\,3998 & 1.89$_{1.81}^{1.94}$ & 763.2/836 & 0.913 & 1.92$_{1.85}^{1.99}$ & 763.1/835 & 0.914  & 0.741 & 
16468.15\\  

NGC\,3718 & 1.94$_{1.82}^{1.97}$ &  360.2/325 & 1.108 &  1.98$_{1.84}^{2.05}$ & 358.6/324 & 1.107  & 0.230 & 
6998.24\\

NGC\,2655 & 1.84$_{1.52}^{1.98}$ &  118.6/132 & 0.898 & 1.73$_{1.76}^{2.11}$ & 117.9/131 & 0.900 & 0.379 & 
568.62\\

\hline \hline

NGC\,4258* &  1.77$_{1.67}^{1.83}$& 649.4/617 & 1.053 & 1.83$_{1.72}^{1.93}$ & 631.4/615 & 1.027 & 1.763$\times$10$^{-4}$ &  

6593.104 \\

NGC\,3147* & 1.70$_{1.53}^{1.76}$ & 556.2/328 & 1.696 & 1.81$_{1.68}^{2.02}$ &  505.0/326  & 1.549  & 1.458$\times$10$^{-7}$ & 

3054.63 \\

ESO\,253-G003*(Nu) & 1.00$_{**}^{1.21}$ &  124.8/122  & 1.023 & 1.12$_{**}^{1.47}$ & 110.3/121 & 0.912 & 1.143$\times$10$^{-4}$ &
1086.49\\


NGC\,5033 & 1.81$_{1.76}^{1.82}$ &  1247.8/1001  & 1.247  & 1.86$_{1.78}^{1.93}$ & 1114.6/1000  & 1.115 &  2.350$\times$10$^{-26}$ & 

25072.32\\

NGC\,1052 & 1.67$_{1.61}^{1.70}$ & 932.2/822 & 1.134 & 1.93$_{1.86}^{1.99}$ & 895.6/821 & 1.091 & 9.883$\times$10$^{-9}$ &  
11515.75\\

NGC\,2110* & 1.71$_{1.69}^{1.73}$ &  1668.5/1209 & 1.380 & 1.87$_{1.85}^{1.88}$ &  1414.0/1208  & 1.171 & 2.257$\times$10$^{-45}$ & 
44137.44\\

LEDA\,96373* & 1.04$^{+}$ &  768.8/233 & 3.299 & 1.20$^{+}$ & 512.5/232  & 2.209 & 3.340$\times$10$^{-22}$ & 
1407.078\\

NGC\,2992 & 1.72$_{1.70}^{1.73}$ & 2502.1/1566 & 1.598  & 1.79$_{1.77}^{1.81}$ &  1771.0/1565 & 1.132 & 1.345$\times$10$^{-119}$ & 
110668.25\\

M\,51 & 1.0$_{**}^{1.21}$ & 669.1/346 & 1.934 & 1.40$_{**}^{1.48}$ & 609.5/345 & 1.767 & 1.434$\times$10$^{-8}$ & 
1856.124\\

NGC\,2273* & 1.80$^{+}$ & 397.2/165 & 2.407 & 1.71$_{1.40}^{1.79}$ &  303.8/164 & 1.852 & 3.577$\times$10$^{-11}$ & 
930.36\\

HE\,1136-2304 & 1.63$_{1.58}^{1.69}$ & 1027.3/930  & 1.105 & 1.68$_{1.62}^{1.74}$ & 973.6/929  & 1.048 & 1.659$\times$10$^{-12}$ & 
19806.78\\

IGRJ\,11366-6002(Nu) & 1.90$_{1.85}^{1.97}$ &  271.7/242 & 1.122 & 2.10$_{2.06}^{2.22}$ & 255.2/241 & 1.059 & 1.037$\times$10$^{-4}$ & 
2269.164 \\




IC\,4518A &  1.42$_{1.24}^{1.60}$ & 305.2/189  & 1.615 & 1.64$_{1.54}^{1.82}$ &  222.9/188  & 1.186 & 2.26$\times$10$^{-5}$ & 
1238.75\\

NGC\,7674*(Nu) & 1.20$_{**}^{1.39}$ & 277.9/171 & 3.914 & 1.26$_{**}^{1.65}$ &  209.0/170 & 1.229 & 3.659$\times$10$^{-12}$ & 

1640.08 \\

\hline 

\end{tabular} 

\tablefoot{ The symbol $^{+}$ represents the cases where $\chi^{2}$/d.o.f.$>$2.0 , then \texttt{Xspec} is not able to calculate the error. If the F-test throws a probability value below 10$^{-3}$, then the reflection component is required by the data. In bold are marked the cases where, according with the F-test the reflection is not required. (Nu) only NuSTAR data. }

\end{table*}

\clearpage

\section{Effect of the inclusion of the \emph{NuSTAR} data and reflection models in the spectral index estimation}
\label{apnx:scatter_error}

In this work we use \emph{NuSTAR}+\emph{XMM-Newton}+\emph{Swift} observations of a sample of LLAGN from BASS /DR2. We fit all our data with reflection models such as \bors to model a torus-like reflector and \xill to model a disk-like emission. This may affect the best-fit parameters we obtain compared to results in the literature that use data from other instruments and with other reflection models.

\begin{figure}
    \centering
    \includegraphics[width=9.7cm]{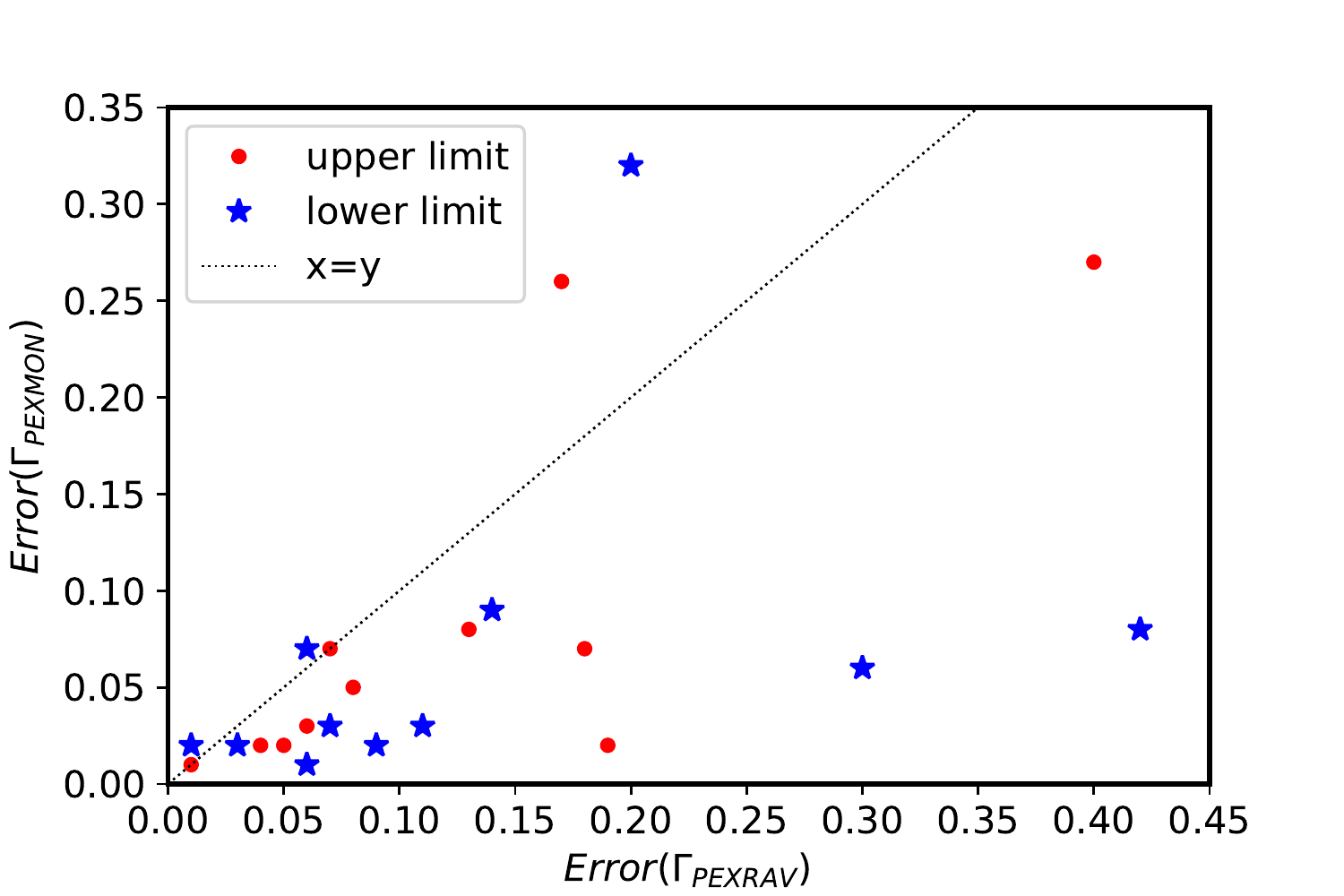}
    
    \caption{Relation between the error between our work using \nus data and \texttt{PEXMON} model and the error values obtained using \texttt{PEXRAV} and observation from \emph{Swift/XRT}, \emph{Swift/BAT}, \emph{ASCA}, \emph{Chandra}, and \emph{Suzaku} by \cite{2017ricciapJS}. The red circles represent the upper limit of the error bar and the blue stars the lower limit. The dotted black line is x=y.   }
    \label{fig:error_DR1_DR2}
\end{figure}

In previous studies, the $\Gamma$ parameter has large error bars (see, for example, \citealt{2018She}). To investigate the improvement in the spectral index estimation, we will compare the error bars of the present work using \texttt{PEXMON} and \emph{NuSTAR}+\emph{XMM-Newton}+\emph{Swift} observations, with those obtained by \cite{2017ricciapJS} using \pexr of the same sample with \emph{Swift/XRT}, \emph{Swift/ BAT }, \emph{ASCA}, \emph{Chandra}, and \emph{Suzaku} observations. In Fig. \ref{fig:error_DR1_DR2} we compare these error bars, with the blue stars representing the lower limits and the red circles the upper limits of the parameter. The black dotted line represents x=y. We found that the errors in this work are smaller compared to \cite{2017ricciapJS}. Then, the uncertainties in the photon index improve by including \nus data and/or using models such as \bors or \xill.

\begin{figure}
    \centering
    \includegraphics[width=9.80cm]{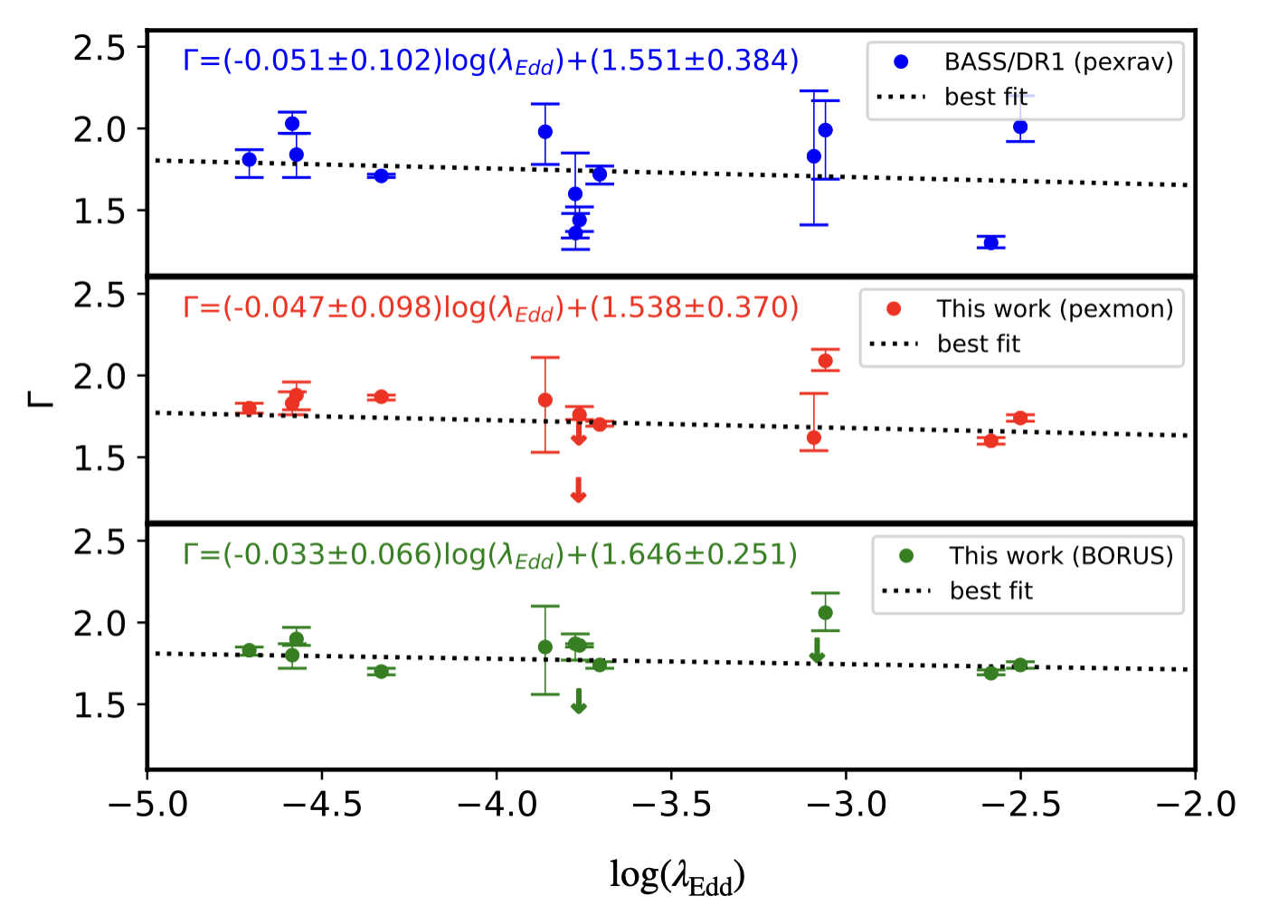}
    
    \caption{Relation between the spectral index, $\Gamma$, and the accretion rate, $\log(\lambda_{Edd})$. The blue points (top panel) represents the values obtained by \citep{2017ricciapJS} using \texttt{PEXRAV} model from the DR1. The middle panel (red points) represents the data points obtained in this work using \texttt{PEXMON} reflection model. In the lower panel (green points) are the values obtained in this work using \bors reflection model. In all the panels, the dotted black lines are the best fit model.  }
    \label{fig:spectr_acc_err}
\end{figure}

Moreover, the $\Gamma$-$\log(\lambda_{Edd})$ relation shows a high scatter (especially in the case of LLAGN), which is still not understood - it could be due to the sensitivity of the measurements or the intrinsic diversity of the nuclei \cite[see][]{2009Gu, 2011younes, yang2015, 2018She}. In this work, we found a significant decrease in the scatter in this correlation. To understand this improvement, the natural question is whether this is related to the inclusion of \nus data or could be an effect of using more physical models than \bors to perform the spectral analysis. In the upper panel in Fig. \ref{fig:spectr_acc_err}, the spectral index is plotted against the accretion rate previously determined by \cite{2017ricciapJS} using \texttt{PEXMON}. In the middle, we plot the relation between the same sample and the values determined in this work with \pexmn including the \nus data. Note that we have not indicated the error bars in the accretion rate for illustrative purposes only. There is an improvement in the scatter and error bars, with smaller uncertainties in the photon index than in the results without including the \nus data. Thus, the inclusion of the \nus data is crucial for the determination of one of the fundamental parameters describing the X-ray emission, like the spectral index $\Gamma$. We also investigate the improvement of the estimate of the parameter by including the \bors model. Comparing the lower panel in Fig. \ref{fig:spectr_acc_err} with the middle panel in the same figure, we see that there is a significant improvement in the scatter of the relation and the uncertainties of the parameter have improved significantly. 

We found that we get lower uncertainties with a more physical model like \bors\. The process of combining \nus data with more physical models like \bors is key to improve the scatter previously observed in this relationship. With our small error bars and improved constraints on the parameters, we can continue the search for the expected correlations in detail.

\section{Tables}
\begin{table*}
\caption{Observational Details. }         
\label{table:observations}      
\centering          
\begin{tabular}{c c c c c c} 
\hline\hline       
Name & Instrument & ObsID & Date & R & Exp. Time \\ 
 & & & & ('') & (ks) \\
(1) & (2) & (3) & (4) & (5) & (6)\\ 
\hline   \hline  \\
NGC\,3998 & \emph{XMM-Newton} & 0790840101 & 2016-10-26 & 49 & 25  \\ \\
& \emph{NuSTAR} &  60201050002 & 2016-10-25 & 49 & 103 \\ \\

\hline  \\

NGC\,3718 & \emph{XMM-Newton} & 0795730101 & 2017-10-24 & 49 & 38  \\ \\
& \emph{NuSTAR} &  60301031002 & 2017-10-24 & 49 & 24\\
\\
\hline \\

NGC\,4258* & \emph{XMM-Newton} & 0400560301 & 2006-11-17 & 49 & 64 \\ \\
& \emph{NuSTAR}& 60101046004 & 2016-01-10 & 49 & 103 \\ \\

\hline \\

ESO\,253-G003* & \emph{XMM-Newton} & 0762920501 & 2015-08-19 & 49 & 27 \\ \\
& \emph{NuSTAR} & 60101014002 & 2015-08-21 & 49 & 22 \\ \\
\hline \\

NGC 1052 & \emph{XMM-Newton} & 0790980101 & 2017-01-17 & 49 & 70 \\ \\
& \emph{NuSTAR} & 60201056002 & 2017-01-17 & 49 & 59 \\
\\
\hline \\

NGC\,2655 & \emph{XMM-Newton} & 0301650301 & 2016-11-10 & 49 & 11 \\ \\
& \emph{NuSTAR} & 60160341004 & 2016-11-10 & 49 & 15 \\
\\
\hline \\

NGC\,3147* & \emph{XMM-Newton} & 0405020601 & 2006-10-06 & 49 & 17 \\ \\
& \emph{NuSTAR} & 60101032002 & 2015-12-27 & 49 & 49 \\
\\
\hline \\
NGC\,2110* & \emph{XMM-Newton} & 0145670101 & 2003-03-05 & 49 & 59 \\ \\
& \emph{NuSTAR} &  60061061002 & 2012-10-05 & 49 & 15 \\
&  & 60061061004 & 2013-02-14 & 49 & 12\\ \\

\hline \\
LEDA\,96373* & \emph{XMM-Newton} & 0674940101 & 2012-04-09 & 49 & 56 \\ \\
& \emph{NuSTAR} & 60061073002 & 2014-07-31 & 49 & 22 \\ \\
\hline \\

NGC\,2992 & \emph{XMM-Newton} & 0840920201 & 2019-05-07 & 49
 & 134 \\
&  &  0840920301 & 2019-05-09 & 49 & 134 \\ \\
& \emph{NuSTAR} & 90501623002 & 2019-05-10 & 49 & 57\\ \\
\hline \\

M\,51 & \emph{XMM-Newton} & 0852030101 & 2019-07-11 & 49 & 77 \\ \\
& \emph{NuSTAR} & 60501023002 & 2019-07-10 & 49 & 169 \\
\\
\hline \\

\end{tabular}
\end{table*}

\begin{table*}
\caption{Cont: Observational Details}   
\label{table:observations_2}      
\centering          
\begin{tabular}{c c c c c c} 
\hline\hline       
Name & Instrument & ObsID & Date & R & Exptime \\ 
 & & & & ('') & (ks) \\
(1) & (2) & (3) & (4) & (5) & (6)\\ 
\hline   \hline  \\

NGC\,2273* & \emph{XMM-Newton} & 0140951001 & 2003-09-05 & 49 & 13 \\ \\
& \emph{NuSTAR} & 60001064002 & 2014-03-23 & 49 & 23 \\ \\
\hline \\

HE\,1136-2304 & \emph{XMM-Newton} & 0741260101 & 2014-07-02 & 49 & 110 \\ \\
& \emph{NuSTAR} & 80002031003 & 2014-07-02 & 49 & 63  \\
& & 80002031002 & 2014-07-02 & 49 & 23 \\ \\
\hline \\

IC\,4518A & \emph{XMM-Newton} & 0401790901 & 2006-08-07 & 11
 & 134 \\
&  &  0406410101 & 2006-08-15 & 49 & 25 \\ \\ & \emph{NuSTAR} & 60061260002 & 2013-08-02 & 49 & 7 \\
\hline \\
NGC\,7674* & \emph{XMM-Newton} & 0200660101 & 2004-06-02 & 49 & 10 \\ \\
& \emph{NuSTAR} & 60001151002 & 2014-09-30 & 49 & 51 \\ \\
\hline \\
IGRJ\,11366-6002  & \emph{NuSTAR} & 60061213002 & 2014-10-29 & 49 & 21 \\
\\
  & \emph{Swift}/XRT & 00080058001 &  2014-10-29 & 25 & 1.6 \\
\\
\hline \\
NGC\,5033 & \emph{XMM-Newton} & 0871020101  & 2020-12-10 & 49 & 21  \\ \\
& \emph{NuSTAR} & 60601023002 & 2020-12-08 & 49 & 104  \\
&  & 60601023004 & 2020-12-12  & 49 & 53 \\

\hline

\end{tabular}
\tablefoot{ Notes: (Col. 1) Name, (Col. 2) instrument, (Col. 3) obsID, (Col. 4) date, (Col. 5) radii of the extraction region and (Col. 5) exposure time (ks). Objects with * symbols are the galaxies with not simultaneous observations.}
\end{table*}

\begin{sidewaystable*}
\caption{Spectral variability of the not simultaneous data between 3.0-10.0 keV.}     

\centering   
\setlength{\tabcolsep}{1.8pt}
\renewcommand{\arraystretch}{1.7} 

\label{Table:not_simul}      
\begin{tabular}{c |c| c | c |c c c c c | c c | c} 
\hline\hline       
Name & Instrument & ObsID & Model & $\Gamma$ & N$_{\rm H}$ & \textbf{Norm PL} & Norm G & Flux & $\chi^{2}$/d.o.f & $\chi^{2}_{\nu}$ & Notes \\

 & & & & & ($10 ^{22}$ cm$^{-2}$) & \textbf{($10 ^{-3}$)}  & ($10 ^{-3}$) & ($10^{-12}$ erg cm$^{-2}$ s$^{-1}$) & &  \\
 (1) & (2) & (3) & (4) & (5) & (6) & \textbf{(7)} & (8) & (9) & (10) & (11) & (12)\\

 \hline \hline

\textbf{NGC\,4258*} & \emph{XMM-Newton} & 0400560301 & PL& $\#$ & $\#$ & \textbf{1.48$_{1.42}^{1.92}$}  & - & 3.84$_{1.14}^{5.35}$ & -  & - & We select \\

 & \emph{NuSTAR} & 60101046004A &  &  1.61$_{1.18}^{1.77}$  & 7.67$_{6.90}^{8.66}$   & \textbf{0.84$_{0.81}^{1.11}$} & - & 2.18$_{0.65}^{3.06}$ & 443.2/387 & 1.145 & PL norm  \\
 &  & 60101046004B &   & $\#$ & $\#$ & \textbf{$\#$}  & - & - &  - & - & variable  \\

\hline

\hline

ESO\,253-G003* & \emph{XMM-Newton} & 0762920501 & PL & \textbf{1.76$_{1.11}^{1.92}$} & $\#$ & $\#$ &  - &0.45$_{**}^{0.51}$ & - & - & $\Gamma$ variable,\\

 & \emph{NuSTAR} & 60101014002A & & \textbf{1.23$_{**}^{1.33}$} & 2.29$_{**}^{5.13}$ & 0.19$_{0.11}^{0.30}$ &-& 1.07$_{**}^{1.24}$ & 119.4/86 & 1.389 & We select only\\
&  & 60101014002B & &$\#$ & $\#$ & $\#$ & - &-& - & - & \emph{NuSTAR}  observations \\

\hline

\textbf{NGC\,3147*}  & \emph{XMM-Newton} &  0405020601 & PL & $\#$ & $\#$ & \textbf{0.53$_{0.38}^{0.78}$} & - & 1.14$_{1.05}^{1.19}$ & - & - & We select \\

 & \emph{NuSTAR} & 60101032002A  &  & 1.79$_{1.69}^{2.02}$ & 2.98$_{1.47}^{4.80}$ & \textbf{1.05$_{0.98}^{1.56}$} & - & 2.19$_{2.03}^{2.25}$  & 243.5/217 & 1.122& PL norm \\

 &  & 60101032002B &   & $\#$ & $\#$ & \textbf{$\#$}  & - & - &  - & - & variable\\

\hline 

  \textbf{NGC  2110*}   & \emph{XMM-Newton} & 0145670101 & PL+G & $\#$ & $\#$ & \textbf{6.78$_{6.34}^{7.46}$} & $\#$ & 25.00$_{21.28}^{30.10}$  & - & - & We select\\
 
 & \emph{NuSTAR} & 60061061004A  & & 1.53$_{1.50}^{1.61}$ & 3.84$_{3.57}^{4.14}$ & \textbf{37.08$_{34.72}^{40.83}$} &  0.049$_{0.045}^{0.53}$ & 134.85$_{112.7}^{163.9}$ & 527.3/453 & 1.164 & PL norm \\
 &  & 60061061004B &  & $\#$ & $\#$ & \textbf{$\#$}  & $\#$ & - &  -  & - &variable\\

\hline

LEDA\,96373* & \emph{XMM-Newton} & 0674940101 & PL+G &  $\#$ & $\#$ &  0.13$_{0.09}^{0.23}$  & $\#$ & 1.14$_{0.16}^{1.67}$ & - & - & We select PL norm tied, they\\

& \emph{NuSTAR} & 60061073002A & &  0.93$_{0.70}^{1.13}$ & 13.59$_{10.32}^{17.14}$ & 0.17$_{0.11}^{0.31}$ & 0.022$_{0.20}^{0.24}$ & 1.40$_{0.16}^{2.19}$  & 245.2/141 & 1.739 & are consistent. The model  \\

& & 60061073002B & & $\#$ & $\#$ & $\#$  & $\#$ & - &  - & - & is not enough to fit the data\\

\hline

NGC\,2273* & \emph{XMM-Newton} & 0140951001 & PL+G &  $\#$ & $\#$ &   0.49$_{0.04}^{1.40}$  & $\#$ & 0.93$_{0.19}^{1.48}$ & - & - & We select  \\

& \emph{NuSTAR} & 60001064002A & & 1.56$_{0.79}^{1.74}$ & 30.04$_{21.21}^{44.99}$ & 0.53$_{0.31}^{1.75}$ & 0.037$_{0.031}^{0.047}$ & 0.98$_{0.19}^{1.47}$ & 109.7/102 & 1.075 & PL norm tied, they \\

& & 60001064002B & & $\#$ & $\#$ & $\#$  & $\#$ & - &  - & - &are consistent \\

\hline

\hline   
\end{tabular}
\tablefoot{(Col. 1) name, (Col. 2) instrument, (Col. 3) obsID, (Col. 4) model, where "PL" represents the power-law and "G" a Gaussian component at 6.4 keV and $\sigma$=0.01 keV. (Col. 5) photon index, (Col. 6) column density, (Col. 7) Power-law normalization, (Col. 8) Normalization of the Gaussian component, (Col. 10 and 11) $\chi^{2}$/d.o.f., and $\chi^{2}_{\nu}$, (Col. 12) Notes about selection of the model. The case where the Normalization of the power-law (Norm PL) is a free parameter between the \xmm and \nus data is marked in bold face. The parameter in this table marked with $\#$ means that is tied to FPMA/\nus. The symbol "$-$" denotes that these components were not require to improve the fit.}
\end{sidewaystable*}

\begin{sidewaystable*}
\caption{Spectral variability not simultaneous data between 3.0-10.0 keV.}     

\centering   
\setlength{\tabcolsep}{1.8pt}
\renewcommand{\arraystretch}{1.7} 

\begin{tabular}{c |c| c | c |c c c c c | c c | c} 
\hline\hline       
Name & Instrument & ObsID & Model & $\Gamma$ & N$_{\rm H}$ & \textbf{Norm PL} & Norm G & Flux & $\chi^{2}$/d.o.f & $\chi^{2}_{\nu}$ & Notes \\

 & & &  & & ($10 ^{22}$ cm$^{-2}$) & \textbf{($10 ^{-3}$)}  & ($10 ^{-3}$) & ($10^{-12}$ erg cm$^{-2}$ s$^{-1}$) & &  \\
 (1) & (2) & (3) & (4) & (5) & (6) & \textbf{(7)} & (8) & (9) & (10) & (11) & (12)\\

 \hline \hline

NGC\,7674* & \emph{XMM-Newton} & 0200660101 & PL+G & $\#$ & $\#$ &  0.20$_{**}^{0.50}$   & $\#$ & 0.052$_{0.029}^{1.243}$ & - & - & Is a changing look, we \\

& \emph{NuSTAR} & 60001151002A &  & 1.71$_{1.34}^{2.17}$ & 18.28$_{13.05}^{26.89}$ & 0.49$_{0.24}^{1.33}$ & 0.0059$_{0.0037}^{0.0079}$ & 0.79$_{0.03}^{1.43}$ & 12.4/76 & 1.479 & did not combine the data. \\

& & 60001151002B & &$\#$ & $\#$ & $\#$  & $\#$ & - &  - &  \\

\hline 

\textbf{IC 451A*} & \emph{XMM-Newton} & 0406410101 & PL+G & $\#$ & $\#$ &   \textbf{0.97$_{0.56}^{1.06}$}  & $\#$ & 1.64$_{0.09}^{3.46}$ &- & - & We select  \\

 & \emph{NuSTAR} & 60061260002A  & & 1.70$_{1.36}^{1.83}$ & 21.57$_{18.30}^{24.35}$ & \textbf{2.64$_{1.44}^{2.89}$} & 0.015$_{0.012}^{0.017}$ & 4.29$_{0.09}^{9.22}$ & 143.4/123 & 1.166 & PL norm \\

 &  & 60061260002B & &$\#$ & $\#$ & \textbf{$\#$}  & $\#$ & - &  - & - & variable \\
\hline \hline

\hline   
\end{tabular}
\end{sidewaystable*}


\begin{sidewaystable*}

\caption{ Final compilation of the coronal parameters of the best-fit models for the sample.  }             
\centering   
\begingroup
\setlength{\tabcolsep}{3.5pt}
\renewcommand{\arraystretch}{1.7} 
\label{table:corona}

\begin{tabular}{ l |c c c c |c c c c | c c c c | c  } 
\hline       

 & \multicolumn{4}{|c|}{\pexmn} & \multicolumn{4}{|c|}{\bors} & \multicolumn{4}{|c|}{\xill} & \\\hline
 
 
Name & $\Gamma$  & $E_{\rm cut}$ (keV)  & $\chi^{2}$/d.o.f  & $\chi^{2}_{\nu}$  & $\Gamma$  & $E_{\rm cut}$ (keV)  & $\chi^{2}$/d.o.f & $\chi^{2}_{\nu}$ & $\Gamma$  &  $E_{\rm cut}$ (keV)  & $\chi^{2}$/d.o.f & $\chi^{2}_{\nu}$ &  BM \\
 
 \hline 
 NGC\,3998 & 1.80$_{1.77}^{1.83}$ & 81.78$_{61.07}^{123.99}$ & 840.7/896 & 0.938 & 1.83$_{1.80}^{1.85}$ & 119.02$_{92.00}^{173.06}$ & 837.6/895 & 0.936 &  1.79$_{1.76}^{1.83}$ &  568.75$_{108.96}^{**}$ & 838.0/895 & 0.936 & T/D\\
 
NGC\,3718 & 1.88$_{1.79}^{1.96}$ & 111.57$_{55.39}^{994.30}$ & 408.2/371 & 1.100 & 1.90$_{1.86}^{1.97}$ & 252.99$_{80.04}^{**}$ & 405.9/369 & 1.100 & 1.86$_{1.77}^{1.95}$ & 142.89$_{64.63}^{1140.6}$ & 404.8/369 & 1.097 &  T/D\\

NGC\,4258* & 1.83$_{1.76}^{1.90}$ & 142.15$_{93.54}^{396.60}$ & 740.0/666 & 1.111 & 1.80$_{1.72}^{1.87}$ & 371.07$_{120.95}^{**}$ & 731.5/665 & 1.100 & 1.86$_{1.78}^{1.93}$ & 208.36$_{91.48}^{**}$ & 739.3/668 & 1.187 & Torus \\

ESO\,253-G003* & 1.30$_{**}^{1.65}$ & 79.45$_{**}^{2000.0}$ &  107.1/121& 0.885 & 1.40$_{**}^{1.52}$ & 90.08$_{38.99}^{187.57}$ & 108.6/120 & 0.905 &  1.41$_{**}^{1.62}$ & 228.04$_{89.35}^{**}$  & 108.37/120 & 0.903 &  T/D\\

NGC 1052 & 1.76$_{1.73}^{1.81}$ & 134.55$_{95.96}^{223.48}$ & 956.7/872 & 1.097 & 1.86$_{1.85}^{1.87}$ & 2000.0$_{374.77}^{**}$ & 949.9/872 & 1.089 & 1.45$_{1.40}^{1.57}$ & 40.39$_{38.09}^{42.05}$ & 1286.0/872 & 1.4747 & Torus \\

NGC\,2655 &  1.85$_{1.53}^{2.11}$ & 209.70$_{59.54}^{**}$ & 171.8/168 &  1.022 & 1.85$_{1.56}^{2.10}$ & 221.28$_{**}^{2000.0}$ & 171.2/167 & 1.025 & 1.90$_{1.78}^{2.23}$ & 230.86$_{79.97}^{1183.5}$ &  178.1/169 & 1.054 &  T/D\\

NGC\,3147* &  1.71$_{1.66}^{1.75}$ & 65.65$_{**}^{138.94}$ & 394.3/378 & 1.043  & 1.76$_{1.74}^{1.79}$ & 2000$_{177.53}^{**}$  & 379.3/377 & 1.006 & 1.72$_{1.64}^{1.80}$ & 71.68$_{44.69}^{191.61}$ & 384.7/377 & 1.020 &  T/D \\

NGC\,2110* & 1.87$_{1.85}^{1.88}$ & 3000.0$_{1419.73}^{**}$ & 1600.4/1261 & 1.269 & 1.70$_{1.68}^{1.72}$ & 354.86$_{303.67}^{465.42}$ & 1293.2/1260 & 1.026 & 1.82$_{1.81}^{1.83}$ & 1870.56$_{764.82}^{**}$ & 1564.3/1263 & 1.239 & Torus\\

LEDA\,96373* & 1.20$_{**}^{1.30}$ & 38.02$_{33.40}^{44.08}$ & 403.1/276 & 1.460 & 1.87$_{1.77}^{1.93}$ & 2000.0$_{596.45}^{**}$  & 309.9/275 & 1.127 & 1.77$_{1.69}^{1.84}$ & 114.47$_{53.53}^{184.62}$ & 461.0/276 & 1.670 &  Torus\\

NGC\,2992 & 1.60$_{1.58}^{1.62}$& 55.28$_{51.78}^{58.83}$ & 2092.1/1619 & 1.2922 & 1.69$_{1.68}^{1.71}$ & 259.51$_{183.62}^{376.14}$ &  1738.7/1618 & 1.075 & 1.65$_{1.60}^{1.68}$ & 64.80$_{61.66}^{69.13}$ & 2140.5/1620 & 1.3213 & Torus \\

M\,51 & 1.90$_{1.76}^{2.06}$ & 3000$_{381.74}^{**}$ & 482.9/341 & 1.416 & 1.56$_{**}^{1.76}$ & 2000$_{70.89}^{**}$ & 470.5/343 & 1.372 & 1.90$_{1.72}^{2.24}$  & 3000$_{395.18}^{**}$ & 493.8/341 & 1.448 & Torus \\

NGC\,2273* & 1.80$_{1.51}^{2.10}$ & 57.86$_{**}^{313.43}$ & 226.8/197 & 1.151 & 1.44$_{**}^{2.00}$ & 60.41$_{36.33}^{**}$ & 173.0/196  & 0.882 & 2.14$_{1.99}^{2.27}$ & 575.64$_{298.54}^{**}$ & 246.3/196 & 1.257 &  D \\

HE\,1136-2304 &  1.74$_{1.72}^{1.76}$ & 97.20$_{77.12}^{136.22}$ & 1066.0/987 & 1.080 & 1.74$_{1.72}^{1.76}$ & 394.08$_{172.4}^{**}$ & 1054.5/986 & 1.069 & 1.567$_{1.553}^{1.598}$ & 1014.66$_{554.79}^{**}$ & 1126.3/986 & 1.142 & Torus \\

IGRJ\,11366-6002  & 2.10$_{2.03}^{2.16}$ &   2999.36$_{375.28}^{**}$ & 261.9/245 & 1.069 & 2.06$_{1.91}^{2.16}$ & 206.76$_{100.57}^{**}$ & 252.3/244 & 1.034 & 1.92$_{1.85}^{2.00}$ & 2000$_{395.73}^{**}$ & 250.49/244 & 1.026 &  T/D\\

IC\,4518A* & 1.62$_{1.54}^{1.89}$& 50.0$_{**}^{85.95}$ & 245.7/215 & 1.143 & 1.57$_{**}^{1.83}$ & 34.19$_{**}^{75.18}$ & 222.6/217 & 1.026 &  1.76$_{1.66}^{1.92}$ & 73.03$_{44.59}^{173.61}$ & 250.4/217 & 1.154 & Torus \\

NGC\,7674* & 1.82$_{1.45}^{2.02}$ & 180.88$_{107.29}^{**}$ & 189.0/169 & 1.118 & 1.47$_{**}^{1.58}$ & 30$_{**}^{40.22}$ & 190.7/168 & 1.135 & 1.42$_{**}^{1.84}$ & 95.68$_{67.58}^{116.13}$ & 179.4/168 & 1.068 &  D \\

NGC\,5033 & 1.70$_{1.69}^{1.72}$ & 50$_{**}^{55.13}$ & 1215.4/1059 & 1.148 & 1.74$_{1.72}^{1.76}$ & 142.95$_{102.90}^{200.67}$ & 1177.4/1058 & 1.113  & 1.77$_{1.72}^{1.81}$  & 61.35$_{48.78}^{74.93}$ & 1193.8/1054 & 1.133 & Torus \\

   \hline 
 \end{tabular}
 
 \endgroup
\tablefoot{Notes: photon index ($\Gamma$), the high energy cut-off (E$_{\rm cut}$) in keV, $\chi^{2}$/d.o.f and $\chi^{2}_{\nu}$ for \pexmn, \bors (B) and \xill (X). BM shows
the best model according with Akaike criterion. The symbol T/D represents that the models are indistinguishable. Objects with * symbols are the galaxies with not simultaneous observations. The quantities in this table marked with ** are unconstrained errors.}
\end{sidewaystable*}


\begin{sidewaystable*}

\caption{Final compilation of the reflector parameters of the best-fit models for the sample.  }           
\centering  

\setlength{\tabcolsep}{2.3pt}
\renewcommand{\arraystretch}{1.5} 
\label{table:reflection_parameters}

\begin{tabular}{ l |c c c c|c c c c c| c c c c c| c  } 
\hline       
 & \multicolumn{4}{|c|}{\pexmn} & \multicolumn{5}{|c|}{\bors} & \multicolumn{5}{|c|}{\xill} &\\
\hline

Name & R$_{\rm f}$ & Incl & $\chi^{2}$/d.o.f. & $\chi^{2}_{\nu}$ & $\log(\rm{N_{H-refl}})$  & CF & Cos($\theta_{\rm incl}$) & $\chi^{2}$/d.o.f. & $\chi^{2}_{\nu}$ & $\log\xi$ & R$_{\rm f}$ & Incl (deg) & $\chi^{2}$/d.o.f. & $\chi^{2}_{\nu}$ & BM \\
 
 \hline 
 NGC\,3998 & 0.07$_{**}^{0.15}$ & 0.01$_{**}^{90.0}$ & 840.7/896 & 0.938 & 22.50$_{22.28}^{23.11}$ & 1.00$_{0.20}^{1.00}$ & 0.18$_{0.05}^{**}$ & 837.6/895 & 0.936 & 4.07$^{**}_{3.99}$ & 4.1$_{0.26}^{**}$ & 18.33$_{0.94}^{58.68}$  & 837.95/895 & 0.936 & T/D\\
 
NGC\,3718 & 0.01$_{**}^{0.24}$ & 0.02$_{**}^{90}$ & 408.2/371 & 1.100 & 22.78$_{22.38}^{23.98}$ & 0.99$_{0.05}^{1.0}$ & 0.30$_{**}^{0.95}$ & 405.9/369 & 1.100 & 3.12$_{1.35}^{3.61}$ & 4.91$_{0.86}^{**}$ & 88.74$_{81.30}^{89.93}$ & 404.8/369 & 1.097 & T/D \\

NGC\,4258* & 0.78$_{0.20}^{3.44}$ & 80.14$_{**}^{87.67}$ & 740.0/666 & 1.111 & 23.29$_{22.77}^{24.39}$ & 0.38$_{0.05}^{0.81}$ & 0.95$_{0.27}^{**}$ & 731.5/665 & 1.100 & 0.48$_{**}^{2.06}$ & 5.39$_{1.54}^{**}$ & 87.97$_{67.03}^{89.32}$ & 739.3/668 & 1.187 &  Torus\\

ESO\,253-G003 & 10.00$_{1.59}^{**}$ & 86.60$_{69.93}^{89.03}$  & 107.1/121 & 0.885 & 24.04$_{23.93}^{24.30}$  &  0.73$_{0.05}^{1.00}$ & 0.76$_{**}^{0.94}$ & 108.6/120 & 0.905 & 2.68$_{1.34}^{2.84}$ &  2.08$_{0.85}^{**}$ & 65.99$_{**}^{83.64}$ & 108.3/120 & 0.903 & T/D\\

NGC 1052 & 1.26$_{0.79}^{3.05}$ & 74.71$_{61.54}^{84.16}$ & 956.7/872 & 1.097 & 24.35$_{24.19}^{24.51}$ & 0.50$_{0.27}^{0.54}$ & 0.55$_{0.45}^{0.73}$ & 949.9/872 & 1.089 & 1.88$_{1.78}^{2.00}$ & 10.00$_{9.18}^{**}$ & 88.12$_{86.06}^{88.29}$ & 1286.0/872 & 1.4747 &  Torus \\

NGC\,2655 & 0.07$_{**}^{0.51}$ & 0.01$_{**}^{80.10}$ & 171.8/168 & 1.022 & 23.24$_{22.48}^{25.19}$ & 0.83$_{0.11}^{1.00}$ & 0.85$_{0.05}^{**}$ & 159.4/167 & 0.954 & 1.85$_{**}^{3.55}$ & 9.84$_{**}^{10}$ & 89.09$_{24.98}^{89.51}$ & 178.1/169 & 1.054 & T/D \\

NGC\,3147* & 3.99$_{0.97}^{6.41}$ & 87.11$_{**}^{88.19}$ & 394.3/378 & 1.043 & 23.27$_{22.98}^{23.94}$ &  0.89$_{0.26}^{1.00}$ & 0.90$_{**}^{0.05}$ & 379.3/377 & 1.006 & 0.79$_{0.06}^{1.15}$ & 9.48$_{0.91}^{**}$ & 87.61$_{62.87}^{88.14}$ & 384.7/377 & 1.020 & T/D\\

NGC\,2110* & 1.01$_{0.91}^{1.12}$ & 79.58$_{79.27}^{80.68}$ & 1600.4/1261 & 1.2691 & 23.49$_{23.03}^{23.76}$ & 0.13$_{0.11}^{0.30}$ & 0.45$_{0.23}^{0.73}$ &  1293.2/1260 & 1.026 & 1.99$_{1.82}^{2.01}$ & 10.00$_{7.80}^{**}$ & 88.94$_{88.47}^{89.05}$ & 1564.3/1263 & 1.236 &  Torus \\

LEDA\,96373* & 10.00$_{9.38}^{**}$ & 65.49$_{50.08}^{69.11}$ & 403.1/276 & 1.460 & 23.40$_{23.33}^{23.69}$ & 0.59$_{0.41}^{0.76}$& 0.95$_{0.59}^{**}$ & 309.9/275  & 1.127 & 2.08$_{2.03}^{2.17}$  & 10.00$_{9.37}^{**}$ & 19.10$_{1.51}^{43.32}$ & 461.0/276 & 1.670 &  Torus  \\

NGC\,2992 & 0.37$_{0.33}^{0.39}$ &  0$_{**}^{11.76}$ & 2092.1/1619 & 1.292 & 23.17$_{23.13}^{23.29}$ & 0.75$_{0.05}^{1.00}$ & 0.85$_{0.63}^{**}$ & 1738.7/1618  & 1.075 & 0.006$_{**}^{0.08}$ & 10.00$_{8.80}^{**}$ & 88.41$_{88.11}^{88.55}$ & 2140.5/1620 & 1.3213 &  Torus\\

M\,51 & 10.00$_{7.86}^{**}$ &  0.06$_{**}^{52.70}$ & 482.9/341 & 1.416 & 25.40$_{24.68}^{**}$ & 0.34$_{0.05}^{0.57}$ & 0.95$_{0.80}^{**}$ & 470.5/343 &  1.372 & 0.05$_{**}^{1.26}$ & 10.00$_{8.96}^{**}$ & 18.00$_{**}^{47.80}$ & 493.8/341 &  1.448 &  Torus\\

NGC\,2273* & 10.00$_{9.02}^{**}$  & 0.09$_{**}^{52.65}$ & 226.8/197 & 1.151 & 23.75$_{23.56}^{24.19}$ & 0.15$_{0.05}^{0.59}$ & 0.94$_{0.72}^{**}$ & 173.5/196 & 0.882 &  2.14$_{2.02}^{2.31}$ &  9.99$_{8.80}^{**}$ & 8.90$_{**}^{45.34}$ & 246.3/196 & 1.257 & D\\

HE\,1136-2304 & 3.99$_{2.85}^{5.10}$ & 87.77$_{86.75}^{88.41}$ & 1066.0/987 & 1.080 & 23.52$_{23.38}^{23.64}$ & 0.59$_{0.50}^{0.68}$ & 0.05$_{**}^{0.39}$ &  1054.5/986& 1.069 &  3.78$_{3.77}^{3.79}$ & 9.93$_{1.45}^{**}$ & 0.05$_{**}^{26.84}$ & 1126.3/986 & 1.142 &  Torus \\

IGRJ\,11366-6002  & 0.68$_{0.42}^{2.23}$ & 18.37$_{**}^{77.64}$ & 261.9/245& 1.069 & 24.75$_{24.48}^{**}$ & 0.93$_{0.77}^{0.97}$ & 0.94$_{0.67}^{ 0.95}$ & 252.3/244 & 1.034 & 3.19$_{3.08}^{3.40}$  & 3.68$_{1.29}^{**}$ & 63.99$_{37.40}^{85.73}$ & 250.49/244 & 1.026 & T/D \\

IC\,4518A* &  3.41$_{1.10}^{6.17}$ & 75.58$_{**}^{83.72}$ & 245.7/215 & 1.143 & 24.59$_{24.11}^{25.16}$ & 0.56$_{0.39}^{1.00}$ & 0.55$_{0.30}^{0.72}$ & 222.6/217 & 1.026 & 1.35$_{**}^{1.73}$ & 9.99$_{2.94}^{**}$ & 82.47$_{59.56}^{84.84}$ & 250.4/217 & 1.154 & Torus \\

NGC\,7674* & 3.22$_{2.21}^{**}$ & 18.17$_{**}^{80.86}$ & 189.0/169 & 1.118 & 25.40$_{24.43}^{**}$ & 0.70$_{0.52}^{0.90}$ & 0.95$_{0.84}^{**}$ & 190.7/168 & 1.135 & 2.82$_{2.77}^{3.02}$ &  3.25$_{2.26}^{6.07}$ &  0.0001$_{**}^{71.42}$ & 179.4/168 & 1.068 & D\\

NGC\,5033 & 1.35$_{1.00}^{**}$ & 78.47$_{56.20}^{87.66}$ & 1232.9/1058 & 1.165 &  23.49$_{23.34}^{23.69}$ & 0.77$_{0.53}^{0.92}$ & 0.95$_{0.76}^{**}$  & 1177.4/1058 & 1.113 & 0.45$_{**}^{1.00}$ & 10.00$_{4.88}^{**}$ & 87.41$_{84.68}^{87.83}$ & 1193.8/1054 & 1.133 &  Torus \\

   \hline 
 \end{tabular}
\tablefoot{
Notes: Reflection fraction (R$_{\rm f}$), the inclination (Incl) and the ionization degree (in log) of \pexmn and \xill models. The column density of the torus like reflector ($\log(\rm{N_{H,refl}})$), the covering factor (CF) and the inclination (Cos($\theta_{\rm incl}$)) for \bors. BM shows
the best model according with Akaike criterion. The symbol T/D represents that the models are indistinguishable. Objects with * symbols are the galaxies with not simultaneous observations.  The quantities in this table marked with ** are unconstrained errors except in the case of the CF, where we report the cases the error is limited by the model.}

\end{sidewaystable*}

\begin{sidewaystable*}


\caption{Summary of other parameters of the best fit model.}              
\centering     
\setlength{\tabcolsep}{2.20pt}
\renewcommand{\arraystretch}{1.8} 

\small
\label{table:soft}

\begin{tabular}{ l |c c c c c|c c c c c| c c c c c  } 
\hline       
& \multicolumn{5}{|c|}{\pexmn} & \multicolumn{5}{|c|}{\bors} & \multicolumn{5}{|c}{\xill}\\
\hline

Name &  $\log(N_{{\rm H},los})$ &  $\log(N_{{\rm H},ext})$ & kT$_{\rm 1}$  & kT$_{\rm 2}$  & BM&   $\log(N_{{\rm H},los})$ &  $\log(N_{{\rm H},ext})$ & kT$_{\rm 1}$  & kT$_{\rm 2}$  & BM & $\log(N_{{\rm H},los})$ &  $\log(N_{{\rm H},ext})$ & kT$_{\rm 1}$  & kT$_{\rm 2}$  & BM \\
 
 \hline 
 NGC\,3998 & 20.54$_{20.46}^{20.62}$ & - & - & - & -& 20.58$_{20.53}^{20.61}$ & - &-& -& - & 20.51$_{20.40}^{20.62}$& - & -& -& - \\   
 
NGC\,3718 & 22.03$_{21.98}^{22.07}$ & - & 0.89$_{0.69}^{**}$ & - & ME & 22.02$_{21.99}^{22.07}$ & - & 0.88$_{0.67}^{**}$ & - & ME & 22.03$_{21.98}^{22.08}$ & - & 0.89$_{0.69}^{**}$ & - & ME\\

NGC\,4258* & 22.94$_{22.92}^{22.95}$ & 20.08$_{**}^{20.52}$ & 0.58$_{0.57}^{0.59}$ &  0.20$_{0.19}^{0.21}$ & 2ME+PL & 22.94$_{22.92}^{22.96}$ & 20.08$_{**}^20.63$ & 0.58$_{0.57}^{0.59}$ &  0.20$_{0.19}^{0.21}$ & 2ME+PL & 22.94$_{22.92}^{22.96}$ & 20.08$_{**}^{20.75}$ & 0.58$_{0.57}^{0.59}$ &  0.20$_{0.19}^{0.22}$ & 2ME+PL \\

ESO\,253-G003* & 22.78$_{22.15}^{23.09}$ & - & - & - & - & 23.05$_{22.94}^{23.16}$  & - & - & - & - & 22.94$_{22.74}^{23.12}$ & - & - & - & -\\

NGC 1052 & 23.16$_{23.15}^{23.18}$ & 20.53$_{**}^{20.70}$ & 0.65$_{0.61}^{0.68}$ & - & ab*(ME+PL) & 23.20$_{23.19}^{23.21}$ & 20.71$_{20.60}^{20.82}$ & 0.63$_{0.59}^{0.67}$ & - & ME+PL & 23.83$_{23.76}^{23.87}$ & - &  0.86$_{0.85}^{0.87}$ & -  & ab*ME\\

NGC\,2655 &  23.31$_{23.19}^{23.36}$ & - & 0.65$_{0.61}^{0.69}$ & 0.09$_{**}^{0.43}$ & ab*2ME &  23.42$_{23.25}^{23.51}$ & 20.32$_{**}^{21.79}$ & 0.62$_{0.58}^{0.70}$ & 0.15$_{**}^{0.30}$ & 2ME+PL & 23.29$_{23.22}^{23.37}$ & - & 0.66$_{0.62}^{0.69}$ & - & ME*ab\\

NGC\,3147* & 20.54$_{**}^{20.61}$ & - & - & - & ab & 20.54$_{**}^{20.62}$ & - & - & - & - & 22.78$_{**}^{20.98}$ & - & - & - & - \\

NGC\,2110* & 21.76$_{21.72}^{21.85}$ & - &  0.65$_{0.64}^{0.67}$ & 0.09$_{**}^{0.39}$ & 2ME*2ab &  22.01$_{21.96}^{22.06}$ & 21.27$_{**}^{21.82}$ & 0.71$_{0.69}^{0.72}$ & - & (ME+PL)*ab & 21.88$_{21.83}^{21.93}$ &  - & 0.70$_{0.69}^{0.71}$ & - & ME*ab \\

LEDA\,96373* & 22.67$_{22.57}^{22.76}$ & - & 0.63$_{0.55}^{0.76}$ &  0.22$_{0.20}^{0.23}$ & 2ME*ab & 24.09$_{24.06}^{24.17}$ & -  & 0.59$_{0.41}^{0.76}$ & 0.22$_{0.19}^{0.23}$ & 2ME*ab & 23.56$_{23.49}^{23.62}$ & 23.11$_{23.03}^{23.18}$ &  0.79$_{0.76}^{0.80}$ & - & (ME+PL)*ab \\

NGC\,2992 & 21.89$_{21.88}^{21.91}$ & 21.20$_{21.11}^{21.26}$ & 0.81$_{0.78}^{0.84}$ & - & (PL+ME)*ab & 21.89$_{21.88}^{21.90}$ & 20.00$_{**}^{21.25}$ & 0.79$_{0.76}^{0.82}$ &- & (PL+ME)*ab & 21.90$_{21.88}^{21.89}$ & - & 0.63$_{0.62}^{0.65}$ & 0.19$_{0.18}^{0.21}$ & 2ME\\

M\,51 & 22.36$_{**}^{22.88}$ & - & - & - & ab & 24.77$_{24.45}^{**}$ & - & - & - & - &  21.93$_{**}^{22.69}$ & - & - & - & ab\\

NGC\,2273* & 23.15$_{22.75}^{23.31}$ & - & 0.71$_{0.52}^{0.79}$ & - & ME*ab & 24.55$_{24.43}^{24.66}$ & - & 0.26$_{0.22}^{0.34}$ & - & ME*ab &  23.36$_{23.27}^{23.44}$ & - & 0.71$_{0.50}^{0.80}$ & - & ME*ab \\

HE\,1136-2304 & 20.97$_{20.95}^{20.99}$ & - & 0.59$_{0.50}^{0.67}$ & - & ME & 20.96$_{20.94}^{20.98}$ &  - &  0.59$_{0.51}^{0.67}$ & - & ME & 20.96$_{20.94}^{20.99}$ & - &  0.65$_{0.54}^{0.77}$ & - & ME \\

IGRJ\,11366-6002  & 21.81$_{**}^{21.90}$ & - & - & - & -& 21.99$_{**}^{22.24}$  & - & - & - &-& 21.81$_{**}^{21.89}$ & - & - & - & -\\

IC\,4518A* &  23.22$_{23.17}^{23.32}$& 20.96$_{**}^{23.19}$ & 0.70$_{0.67}^{0.72}$ & - & (ME+PL)*ab & 23.22$_{23.16}^{23.30}$ & - & 0.71$_{0.59}^{0.75}$ & - & ME*ab & 23.21$_{23.15}^{23.28}$ & - & 0.70$_{0.68}^{0.74}$ & - & ME*ab\\

NGC\,7674* & 23.09$_{22.96}^{23.22}$ & - & - & - & - & 23.06$_{22.98}^{23.21}$ & - &  - &  - & - & 22.99$_{22.75}^{23.08}$ & - & - & - & -\\

NGC\,5033 & 20$_{**}^{20.09}$ & - & - & - & - & 20$_{**}^{20.12}$ & - & - & - & - & 20.99$_{20.75}^{21.13}$ & - & 0.82$_{0.59}^{**}$  & 0.23$_{0.18}^{0.54}$ & 2ME \\
 \hline 
 \end{tabular}
\tablefoot{Name, column density in the hard energy band ($\log(N_{{\rm H},los})$), column density in the soft energy band ($\log(N_{{\rm H},ext})$),  temperature of the thermal components (kT$_{\rm 1}$ and kT$_{\rm 2}$).  BM means the best model added to the reflection and the central emission to improve the fit. ME means MEKAL, PL power-law in the scattered component and ab the warm absorption modelled with \texttt{zxipcf} in \texttt{xspec}. - denotes that these were not require to improve the fit. Objects with * symbols are the galaxies with not simultaneous observations.}

\end{sidewaystable*}

\begin{sidewaystable*}
    \caption{Summary of other parameters of the best fit models.}             
\label{table:other}      
\centering    
\setlength{\tabcolsep}{1.6pt}
\renewcommand{\arraystretch}{1.8}

\begin{tabular}{ l |c c c c |c c c c | c c c c } 
\hline  
& \multicolumn{4}{|c|}{\pexmn} & \multicolumn{4}{|c|}{\bors} & \multicolumn{4}{|c}{\xill}\\
\hline
Name & N$_{\rm {H,W}}$ & $\log(\xi_{W})$ & Norm PL$_{\rm XMM}$  & Norm PL$_{\rm Nu}$  &N$_{\rm {H,W}}$ & $\log(\xi_{W})$ & Norm PL$_{\rm XMM}$  & Norm PL$_{\rm Nu}$ & N$_{\rm {H,W}}$ & $\log(\xi_{W})$ & Norm PL$_{\rm XMM}$  & Norm PL$_{\rm Nu}$ \\

&  ($10 ^{22}$ cm$^{-2}$) & &($10 ^{-3}$) &($10 ^{-3}$) & ($10 ^{22}$ cm$^{-2}$) & &($10 ^{-3}$) &($10 ^{-3}$) & ($10 ^{22}$ cm$^{-2}$) & &($10 ^{-3}$) &($10 ^{-3}$)     \\

 \hline 
 NGC\,3998 & - & - & $\#$ & 2.33$_{2.23}^{2.43}$ & - & - & $\#$ & 2.35$_{2.26}^{2.44}$ & - & - & $\#$ & 2.05$_{**}^{3.41}$  \\
 
NGC\,3718 & - & - & $\#$ & 0.53$_{0.47}^{0.59}$ & - & - & $\#$ & 0.52$_{0.47}^{0.58}$ & - & - & $\#$ & 0.08$_{0.05}^{0.09}$\\

NGC\,4258* & - & - & 1.70$_{1.50}^{1.83}$ & 1.13$_{0.99}^{1.24}$ & - & - & 1.60$_{1.42}^{1.74}$ & 1.04$_{0.92}^{1.25}$ & - & - & 0.032$_{0.028}^{0.051}$ & 0.021$_{0.017}^{0.028}$\\

ESO\,253-G003* & - & - & - & 0.17$_{0.11}^{0.36}$ & - & - & - & 0.30$_{0.26}^{0.46}$ & - & - & - & 0.02$_{0.01}^{0.03}$ \\

NGC 1052 &  0.05$_{**}^{45.54}$ & 6$_{2.37}^{**}$ & $\#$ & 2.79$_{2.66}^{3.03}$ & - & - & $\#$ & 3.46$_{**}^{3.51}$ & 19.25$_{19.00}^{19.51}$ & 2.032$_{2.029}^{2.043}$ & $\#$ & 0.043$_{0.042}^{0.044}$ \\

NGC\,2655 & 2.11$_{1.23}^{2.81}$ & 1.32$_{1.15}^{1.55}$ & $\#$ & 0.91$_{0.42}^{1.22}$ & -&- & $\#$ & 0.80$_{0.54}^{1.39}$ & 2.57$_{1.71}^{3.77}$ & 1.36$_{1.23}^{1.64}$ & $\#$ & 0.02$_{0.01}^{0.03}$\\

NGC\,3147* & - & - & 1.89$_{0.92}^{6.21}$ & 0.68$_{0.64}^{0.72}$ & - & - & 0.34$_{0.30}^{0.38}$ & 0.71$_{0.66}^{0.75}$ & - & - & 0.0066$_{0.0056}^{0.0071}$ & 0.012$_{0.010}^{0.016}$\\

NGC\,2110*$^{\rm w}$ & 0.17$_{0.13}^{0.24}$ & -0.55$_{-0.83}^{-0.43}$ & 10.80$_{9.70}^{13.10}$ & 59.40$_{58.50}^{60.50}$ & 1.53$_{1.39}^{1.66}$ & 4.30$_{4.03}^{4.51}$ & 8.38$_{7.51}^{8.79}$ & 4.44$_{4.22}^{4.61}$ & 5.23$_{5.07}^{5.37}$ & 1.36$_{1.29}^{1.43}$ & 0.30$_{0.28}^{0.59}$ & 1.54$_{1.37}^{1.57}$ \\

LEDA\,96373* & 0.79$_{0.67}^{0.87}$ & -0.93$_{-1.07}^{-0.65}$ & $\#$ &  0.12$_{0.11}^{0.14}$ & 0.80$_{0.68}^{0.83}$  & -0.55$_{-1.50}^{-0.33}$& $\#$ & 5.93$_{4.52}^{9.06}$ & 7.94$_{7.28}^{8.85}$ & 2.01$_{1.97}^{2.04}$ & $\#$ &  0.005$_{0.004}^{0.006}$ \\

NGC\,2992 & 0.49$_{0.43}^{0.63}$ & 2.53$_{2.39}^{2.69}$ & $\#$ & 18.58$_{18.16}^{19.06}$ &  0.64$_{0.52}^{0.79}$ & 2.48$_{2.36}^{2.59}$ & $\#$ &  20.80$_{20.30}^{21.40}$ & -& - & $\#$ & 0.40$_{0.39}^{0.41}$ \\

M\,51 &  499.96$_{224.16}^{**}$ & 4.25$_{3.36}^{4.53}$ & $\#$ & 0.056$_{0.036}^{0.099}$ & -& - & $\#$ & 0.36$_{0.22}^{1.18}$ & 488.99$_{278.02}^{**}$ & 4.18$_{3.88}^{4.38}$ & $\#$ & 0.0012$_{0.0011}^{0.0016}$   \\

NGC\,2273* &  2.95$_{1.78}^{5.33}$ & 1.24$_{0.26}^{1.49}$ & $\#$ &  0.29$_{0.08}^{0.57}$ & 0.94$_{0.24}^{1.24}$ & -1.14$_{-1.31}^{0.35}$ & $\#$ & 5.14$_{1.60}^{10.52}$  & 2.97$_{2.43}^{5.72}$ & 1.23$_{0.45}^{1.39}$ & $\#$ &  0.006$_{0.005}^{0.010}$\\

HE\,1136-2304 & - & - & $\#$ & 2.53$_{2.47}^{2.59}$ & - & - & $\#$ & 2.53$_{2.48}^{2.59}$ & - & - & $\#$ & 0.0042$_{0.0041}^{0.069}$\\

IGRJ\,11366-6002  & - & - & - & 2.42$_{2.14}^{2.79}$ & - & - & - & 2.26$_{1.87}^{2.50}$ & - & - & - & 1.43$_{0.97}^{2.74}$ \\

IC\,4518A* & 3.79$_{1.91}^{6.03}$ & 1.37$_{1.09}^{1.50}$ & 0.77$_{0.60}^{1.46}$ & 1.65$_{1.41}^{1.98}$ & 4.41$_{1.95}^{5.77}$ & 1.44$_{1.22}^{1.64}$ & 0.65$_{0.45}^{0.95}$ & 2.13$_{1.59}^{3.00}$ &  3.97$_{2.23}^{5.79}$ & 1.39$_{1.31}^{1.57}$ & 0.015$_{0.010}^{0.199}$ & 0.032$_{0.026}^{0.041}$ \\

NGC\,7674* &- &- & - &  0.36$_{0.14}^{0.47}$ & - & - & - & 0.24$_{0.19}^{0.32}$ & - & - & - & 0.003$_{0.002}^{0.005}$ \\

NGC\,5033 & - & - & $\#$ & 1.93$_{1.90}^{1.98}$ & - & - & $\#$ & 1.90$_{1.86}^{1.95}$ & - & - & $\#$ & 0.032$_{0.029}^{0.033}$\\

   \hline 
 \end{tabular}
\tablefoot{Name, Column density and ionization degree of the warm absorber (N$_{H,W}$ and $\log(\xi_{W})$) and the normalization of the power law emission of \xmm and \nus observations (Norm PL$_{XMM}$  and Norm PL$_{Nu}$). The parameter in this table marked with $\#$ (power-law normalization of \xmm) is tied to \nus, "$-$" denotes that these were not require to improve the fit. Objects with * symbols are the galaxies with not simultaneous observations. NGC\,2110 (denoted as NGC\,2110$^{\rm w}$) need two warm absorbers to improve the \texttt{pexmon} fit. In addition to the absorber presented in this table, the other correspond to: N$_{H,W}$=5.38$_{5.23}^{5.52}$ and $\log(\xi_{W})$=1.31$_{1.27}^{1.35}$.}
\end{sidewaystable*}

\begin{table*}
    \caption{Summary of the Cross normalization constants between the instruments. }             
\label{table:cab}      
\centering    
\setlength{\tabcolsep}{2.5pt}
\renewcommand{\arraystretch}{1.8}

\begin{tabular}{ l |c c c |c c c | c c c  } 
\hline       
 & & \rm \pexmn & & & \rm \bors  & & & \rm \xill &     \\ \hline 
Name &  $C_{\rm {A/B}}$  & $C_{\rm {XMM}}$  & $C_{\rm {BAT}}$  & $C_{\rm {A/B}}$  & $C_{\rm {XMM}}$  & $C_{\rm {BAT}}$  & $C_{\rm {A/B}}$  & $C_{\rm {XMM}}$  & $C_{\rm {BAT}}$   \\
 
 \hline 
 NGC\,3998 & 1.04$_{1.02}^{1.06}$ &0.91$_{0.88}^{0.94}$ & 0.98$_{0.80}^{1.17}$ & 1.04$_{1.02}^{1.06}$ & 0.91$_{0.90}^{0.94}$ & 0.96$_{0.79}^{1.13}$ & 1.04$_{1.02}^{1.06}$ & 0.92$_{0.89}^{0.94}$ & 0.96$_{0.79}^{1.14}$ \\
 
NGC\,3718 & 0.99$_{0.96}^{1.03}$ & 0.98$_{0.94}^{1.03}$ & 4.53$_{3.60}^{5.53}$ & 1.00$_{0.96}^{1.02}$ & 0.99$_{0.94}^{1.03}$ & 4.42$_{3.52}^{5.26}$ & 0.99$_{0.96}^{1.03}$ & 0.98$_{0.94}^{1.03}$ & 4.46$_{3.57}^{5.41}$\\

\textbf{NGC\,4258*} & 0.99$_{0.96}^{1.02}$ & \textbf{1.09$_{1.03}^{**}$} & 2.64$_{2.31}^{2.98}$ & 0.99$_{0.96}^{1.02}$ & \textbf{1.07$_{0.96}^{**}$} & 2.56$_{2.24}^{2.93}$ &  0.99$_{0.97}^{1.02}$ & \textbf{1.09$_{**}^{1.10}$} & 2.60$_{2.28}^{2.97}$ \\

ESO\,253-G003* & 1.14$_{1.06}^{1.23}$ & -& 0.89$_{0.66}^{1.14}$ & 1.14$_{1.06}^{ 1.23}$ & - & 0.88$_{0.67}^{1.13}$ &  1.14$_{1.06}^{1.23}$ & - &  0.85$_{0.65}^{1.08}$ \\

NGC 1052 & 1.00$_{0.98}^{1.03}$ & 0.89$_{0.87}^{0.91}$ & 0.86$_{0.76}^{0.96}$ & 1.00$_{0.97}^{1.02}$ & 0.87$_{0.85}^{0.89}$ & 0.83$_{0.74}^{0.93}$ &  1.00$_{0.99}^{1.02}$ & 0.89$_{0.87}^{0.90}$ & 0.90$_{0.79}^{1.00}$  \\

NGC\,2655 & 1.16$_{1.05}^{1.05}$ & 1.09$_{0.93}^{1.27}$ & 2.21$_{1.63}^{2.90}$ & 1.15$_{1.03}^{1.27}$ & 1.07$_{0.91}^{1.25}$ & 2.22$_{1.63}^{2.89}$ & 1.15$_{1.04}^{1.28}$ & 1.07$_{0.92}^{1.16}$ & 2.18$_{1.62}^{2.78}$ \\

\textbf{NGC\,3147*} &  1.04$_{0.99}^{1.09}$ & \textbf{0.95$_{**}^{0.97}$} & 1.08$_{0.71}^{1.47}$ & 1.04$_{0.99}^{1.09}$ & \textbf{0.99}$_{0.90}^{**}$ & 1.00$_{0.64}^{1.34}$ & 1.04$_{0.90}^{1.09}$ & \textbf{0.92}$_{**}^{0.99}$  & 1.08$_{0.70}^{1.47}$\\

\textbf{NGC\,2110*} &  1.03$_{1.02}^{1.05}$ & \textbf{1.03$_{**}^{1.10}$} & 0.58$_{0.57}^{0.59}$ & 1.04$_{1.02}^{1.05}$ & \textbf{0.90$_{**}^{0.97}$} & 0.60$_{0.59}^{0.61}$ & 1.03$_{1.02}^{1.05}$ & \textbf{0.97$_{**}^{1.08}$}  & 0.58$_{0.57}^{0.59}$ \\

LEDA\,96373* & 1.12$_{1.05}^{1.20}$ & 0.84$_{0.79}^{0.90}$ & 0.87$_{0.71}^{1.06}$ & 1.13$_{1.05}^{1.21}$& 0.87$_{0.81}^{0.94}$ & 0.85$_{0.70}^{1.03}$ & 1.12$_{1.05}^{1.20}$ & 0.86$_{0.80}^{0.92}$  & 0.88$_{0.71}^{1.06}$ \\

NGC\,2992 & 1.027$_{1.019}^{1.033}$ & 0.91$_{0.90}^{0.92}$ & 0.127$_{0.119}^{0.133}$ &  1.026$_{1.020}^{1.033}$ & 0.909$_{0.903}^{0.915}$ & 0.122$_{0.110}^{0.136}$ & 1.026$_{1.019}^{1.033}$ & 0.915$_{0.909}^{0.920}$  & 0.13$_{0.12}^{0.15}$ \\

M\,51 &  0.99$_{0.93}^{1.06}$ & 1.06$_{0.97}^{1.16}$ & 2.67$_{1.95}^{3.46}$ & 1.00$_{0.94}^{1.07}$ & 1.06$_{0.98}^{1.16}$ & 2.08$_{1.54}^{2.77}$ & 0.99$_{0.92}^{1.06}$ & 1.07$_{0.98}^{1.17}$ & 2.67$_{1.91}^{3.52}$ \\

NGC\,2273* &  1.09$_{0.99}^{1.18}$& 1.15$_{0.99}^{1.33}$ & 0.84$_{0.54}^{1.17}$ & 1.08$_{0.99}^{1.17}$ & 1.04$_{0.90}^{1.19}$ & 0.65$_{0.43}^{0.91}$ &  1.09$_{1.00}^{1.18}$& 1.09$_{0.94}^{1.97}$& 0.84$_{0.55}^{1.14}$ \\

HE\,1136-2304 & 1.07$_{1.05}^{1.08}$ & 1.03$_{1.01}^{1.04}$ & 0.63$_{0.50}^{0.77}$ & 1.064$_{1.046}^{1.082}$ & 1.03$_{1.01}^{1.04}$ & 0.58$_{0.45}^{0.71}$ &  1.065$_{1.047}^{1.081}$ & 1.023$_{1.015}^{1.039}$  & 0.62$_{0.49}^{0.75}$\\

IGRJ\,11366-6002  & 1.15$_{1.09}^{1.22}$ & - & 1.46$_{1.21}^{1.81}$ & 1.15$_{ 1.09}^{ 1.22}$ & - & 1.42$_{ 1.14}^{ 1.71}$ & 1.15$_{1.09}^{1.22}$ & - & 1.46$_{1.16}^{1.78}$\\

\textbf{IC\,4518A*} & 1.12$_{1.04}^{1.20}$ & \textbf{0.97$_{**}^{0.99}$} & 1.11$_{0.95}^{1.25}$ & 1.11$_{1.03}^{1.19}$ & \textbf{1.10$_{0.90}^{**}$} & 1.09$_{0.93}^{1.28}$ & 1.11$_{1.04}^{1.19}$& \textbf{0.90$_{**}^{1.00}$} & 1.09$_{0.94}^{1.27}$ \\

NGC\,7674* & 1.02$_{0.95}^{1.09}$ & - & 1.58$_{1.15}^{2.03}$ & 1.02$_{0.96}^{1.09}$ & - & 1.60$_{1.17}^{2.07}$ & 1.02$_{0.95}^{1.09}$ & - & 1.57$_{1.16}^{2.00}$\\

NGC\,5033 & 1.05$_{1.03}^{1.07}$ & 0.75$_{0.73}^{0.76}$ & 0.28$_{0.13}^{0.42}$ & 1.05$_{1.04}^{1.07}$ & 0.76$_{0.74}^{0.78}$ & 0.27$_{0.14}^{0.41}$ &  1.05$_{1.04}^{1.07}$ & 0.76$_{0.74}^{0.78}$ & 0.28$_{0.13}^{0.42}$\\

   \hline 
 \end{tabular}
\tablefoot{Name, cross normalization constant between the detector FPMA and FPMB of \nus (C$_{\rm A/B}$, \nus and \xmm (C$_{\rm XMM}$) and \nus and \sft (C$_{\rm BAT}$). In case of the not simultaneous data \xmm+\nus observations, where the normalization of the power-law is a free parameter (objects denoted in bold face), we put the restriction of C$_{\rm XMM}$=[0.9,1.10]. For IGRJ\,11366-6002 we also obtain $C_{XRT}$=1.30$_{1.09}^{1.50}$ (\texttt{PEXMON}), $C_{XRT}$=1.13$_{1.09}^{1.22}$ (\texttt{borus02}) and $C_{XRT}$=1.32$_{1.11}^{1.54}$ (\texttt{XILLVER}). The error of this parameters is in the normalization of the power-law then we reported here only the best fit value. The quantities in this table marked with ** are unconstrained errors.  }
\end{table*}

\begin{table*}
    \caption{Soft (2-10 keV) and hard (10–79 keV) intrinsic and observed luminosities for all the sample. }             
\label{table:luminos}      
\centering    
\setlength{\tabcolsep}{3.5pt}
\renewcommand{\arraystretch}{2.0}

\begin{tabular}{ l |c c  |c c |c} 
\hline 
& \multicolumn{2}{|c|}{Observed} & \multicolumn{2}{|c|}{Intrinsic} & \\
 \hline 
Name & $\log(L_{2.0-10.0})$  & $\log(L_{10.0-79.0})$ & $\log(L_{2.0-10.0})$  & $\log(L_{10.0-79.0})$ & Model\\


 
 \hline 
 NGC\,3998 &  41.14$_{41.12}^{41.15}$ & 41.33$_{41.27}^{41.34}$ & 41.15$^{41.16}_{41.14}$ & 40.67$^{40.71}_{40.62}$     & \bors  \\
 &  41.28$_{41.12}^{41.40}$ & 41.46$_{41.16}^{41.62}$ &  41.28$_{41.12}^{41.40}$ &  41.16$_{40.88}^{41.33}$    & \xill  \\
 \hline

NGC\,3718 &  40.45$_{40.39}^{40.46}$ & 40.64$_{40.45}^{40.65}$ & 40.49$_{40.46}^{40.52}$ & 40.64$_{40.55}^{40.72}$ &\bors \\
&   40.45$_{40.10}^{40.61}$ & 40.63$_{40.03}^{40.71}$ & 40.49$^{40.67}_{40.20}$ & 40.63$^{40.79}_{40.36}$ & \xill  \\
\hline

NGC\,4258* & 39.93$^{37.90}_{40.65}$ & 40.14$_{40.07}^{40.22}$ & 40.15$^{40.05}_{40.70}$ & 40.15$^{40.22}_{40.07}$ &\bors \\
\hline

ESO\,253-G003$^{*}_{Nu}$ & 42.66$_{42.10}^{42.69}$ & 43.56$_{43.11}^{43.67}$ & 42.92$_{42.05}^{43.10}$ &  43.59$_{43.00}^{43.81}$ & \bors \\
& 42.70$_{42.52}^{43.04}$  & 43.62$_{43.15}^{43.65}$  & 42.89$_{42.23}^{43.15}$ & 
 43.62$_{43.42}^{43.76}$      & \xill  \\
\hline

NGC 1052 &   41.44$_{41.36}^{41.48}$ & 42.13$_{41.83}^{42.18}$ & 41.76$^{41.70}_{41.81}$ & 42.15$^{42.23}_{42.04}$ &\bors\\
\hline

NGC\,2655 & 40.61$_{39.90}^{40.67}$ & 41.18$_{40.76}^{41.21}$ & 41.03$^{41.11}_{40.95}$  & 41.19$^{41.26}_{41.08}$ &\bors \\
& 40.60$_{40.19}^{40.74}$  & 40.19$_{40.09}^{40.32}$  & 40.96$_{40.66}^{41.14}$ &   41.20$_{39.85}^{41.51}$   & \xill  \\
\hline

NGC\,3147* & 41.40$_{40.08}^{41.60}$ & 42.01$_{41.69}^{42.04}$ & 41.40$^{41.56}_{41.16}$ &
42.01$^{42.12}_{41.87}$  &\bors\\
& 40.39$_{40.18}^{41.77}$  & 41.95$_{41.93}^{42.07}$ & 41.39$_{40.62}^{41.73}$  &    41.95$_{41.86}^{42.02}$   & \xill  \\
\hline

NGC\,2110* &   42.44$_{42.25}^{42.58}$ & 43.61$_{43.60}^{43.62}$ & 42.52$^{42.65}_{42.34}$   & 43.61$^{43.62}_{43.60}$ & \bors \\
\hline

LEDA\,96373* &  42.34$_{42.10}^{42.34}$ & 43.41$_{42.98}^{43.42}$ & 43.48$^{43.56}_{43.38}$ & 43.80$^{43.92}_{43.63}$ & \bors \\
\hline

NGC\,2992 &  42.911$_{42.907}^{42.914}$ & 43.318$_{43.290}^{43.322}$ & 42.95$^{42.95}_{42.94}$  & 43.32$^{43.34}_{43.30}$ & \bors\\
\hline

M\,51 &  38.94$_{38.72}^{40.11}$ & 39.81$_{37.38}^{40.41}$ & 39.68$^{39.80}_{39.51}$ & 40.25$^{40.75}_{40.32}$  &\bors\\
\hline


NGC\,2273* & 40.89$_{39.91}^{41.00}$ & 41.76$_{39.70}^{40.95}$ & 42.33$_{40.94}^{42.54}$ & 41.99$_{41.88}^{42.20}$ &\xill \\
\hline

HE\,1136-2304 & 43.23$_{43.19}^{43.24}$ & 43.54$_{43.36}^{43.55}$ & 43.23$^{43.25}_{43.21}$ & 43.54$^{43.61}_{43.46}$ &\bors \\
\hline

IGRJ\,11366-6002  &  42.35$_{42.28}^{42.35}$ & 42.59$_{42.14}^{42.52}$  & 42.39$^{42.59}_{42.02}$ & 42.59$^{42.46}_{42.69}$ &\bors \\
& 42.33$_{42.14}^{42.57}$  & 42.58$_{42.10}^{42.70}$ & 42.39$_{42.05}^{42.57}$ &  
 42.58$_{42.26}^{42.76}$    & \xill  \\
\hline

IC\,4518A* & 41.97$_{41.37}^{42.10}$ & 43.08$_{42.88}^{43.13}$ & 42.38$^{42.42}_{42.35}$ & 
 43.10$^{43.16}_{43.02}$ & \bors \\
\hline

NGC\,7674$_{nu}^{*}$ & 42.17$_{41.52}^{42.98}$ & 43.03$_{42.03}^{43.58}$ & 42.42$^{42.31}_{42.50}$ &
 43.07$^{4300}_{43.12}$ & \xill
  \\

\hline 
 NGC\,5033 & 41.04$_{40.01}^{40.05}$  & 41.43$_{41.34}^{41.44}$ & 41.04$_{41.02}^{41.06}$ & 41.43$_{41.38}^{41.47}$ & \bors  \\

   \hline 
\end{tabular}
\tablefoot{Name, soft and hard luminosities (observed and intrinsic) and the best fit model.  Objects with * symbols are the galaxies with not simultaneous observations. 
$_{nu}$ mark objects with only \nus data. 
}
\end{table*}

\begin{table*}
\caption{Bolometric correction, bolometric luminosity and accretion rate for the sample of galaxies.}             
\label{table:bolometric_info}      
\centering   
\setlength{\tabcolsep}{2.0pt}
\renewcommand{\arraystretch}{1.9} 
\begin{tabular}{l |c c c} 
\hline\hline       
Name &k$_{\rm 2.0-10.0}$ & L$_{\rm Bol}$ & $\log(\lambda_{\rm Edd})$\\
 (1) & (2) & (3) & (4) \\ 
\hline   

NGC\,3998	& 15.35$^{15.41}_{15.29}$ &	42.33$^{42.34}_{42.32}$ &	-4.71$^{-5.40}_{-4.02}$ \\

NGC\,3718	& 15.33$^{15.40}_{15.27}$ & 	41.68$^{41.85}_{41.38}$ &	-4.67$^{-5.10}_{-4.17}$ \\

NGC\,4258* &	15.33$^{15.39}_{15.27}$ &	41.08$^{41.11}_{41.05}$ &	-4.59$^{-5.26}_{-3.92}$ \\

NGC\,5033	& 15.34$^{15.40}_{15.28}$ &	42.02$^{42.03}_{42.00}$	& -4.00$^{-4.46}_{-3.09}$ \\

ESO\,253-G003*	& 15.82$^{16.11}_{15.57}$	&44.09$^{44.34}_{43.77}$ &	-3.76$^{-4.21}_{-3.38}$ \\

NGC\,1052	&15.39$^{15.45}_{15.33}$ &	42.92$^{42.95}_{42.89}$ &	-3.86$^{-4.53}_{-3.19}$ \\

NGC\,2655	&15.34$^{15.40}_{15.28}$ &	42.22$^{42.30}_{42.13}$ &	-4.09$^{-4.71}_{-3.48}$ \\

NGC\,3147* &	15.36$^{15.43}_{15.29}$ &	42.59$^{42.75}_{42.35}$ &	-4.33$^{-4.87}_{-3.87}$ \\

NGC\,2110* &	15.59$^{15.70}_{15.47}$ &	43.71$^{43.84}_{43.53}$ &	-3.78$^{-4.35}_{-3.26}$ \\

LEDA\,96373* &	16.74$^{16.99}_{16.46}$ &	44.73$^{44.84}_{44.60}$ &	-2.59$^{-3.18}_{-2.02}$\\

NGC\,2992	 & 15.88$^{15.92}_{15.83}$ &	44.15$^{44.15}_{44.14}$ &	-2.29$^{-2.99}_{-1.60}$ \\

M\,51 &	15.33$^{15.39}_{15.27}$ & 	40.87$^{40.99}_{40.69}$ &	-3.83$^{-4.41}_{-3.31}$ \\

NGC\,2273* &	15.54$^{15.69}_{15.35}$ &	43.60$^{43.84}_{43.06}$	& -2.50$^{-2.96}_{-2.34}$ \\

HE\,1136-2304	&16.22$^{16.27}_{16.16}$ &	44.44$^{44.46}_{44.42}$ &	-3.06$^{-3.74}_{-2.38}$ \\

IGRJ11366 &	15.53$^{15.66}_{15.38}$ &	43.58$^{43.78}_{43.20}$ &	-3.09$^{-3.59}_{-2.77}$ \\

IC\,4518A	&15.51$^{15.61}_{15.39}$ &	43.51$^{43.67}_{43.26}$ &	-3.39$^{-3.93}_{-2.94}$ \\

NGC\,7674*&	15.53$^{15.72}_{15.35}$ &	43.59$^{43.89}_{43.03}$	& -3.70$^{-4.10}_{-3.56}$ \\

\hline 
\end{tabular}
\tablefoot{(Col. 1) name, (Col. 2) bolometric correction, (Col. 3) bolometric luminosity and (Col. 4) accretion rate. All these values were estimated using \cite{2020duras}.  
}
\label{table:lumi_kcorrect}
\end{table*}

        
\clearpage        

\section{Spectral models}       
\label{app:plots}

\begin{figure*}
\centering
    \includegraphics[width=0.850\textwidth]{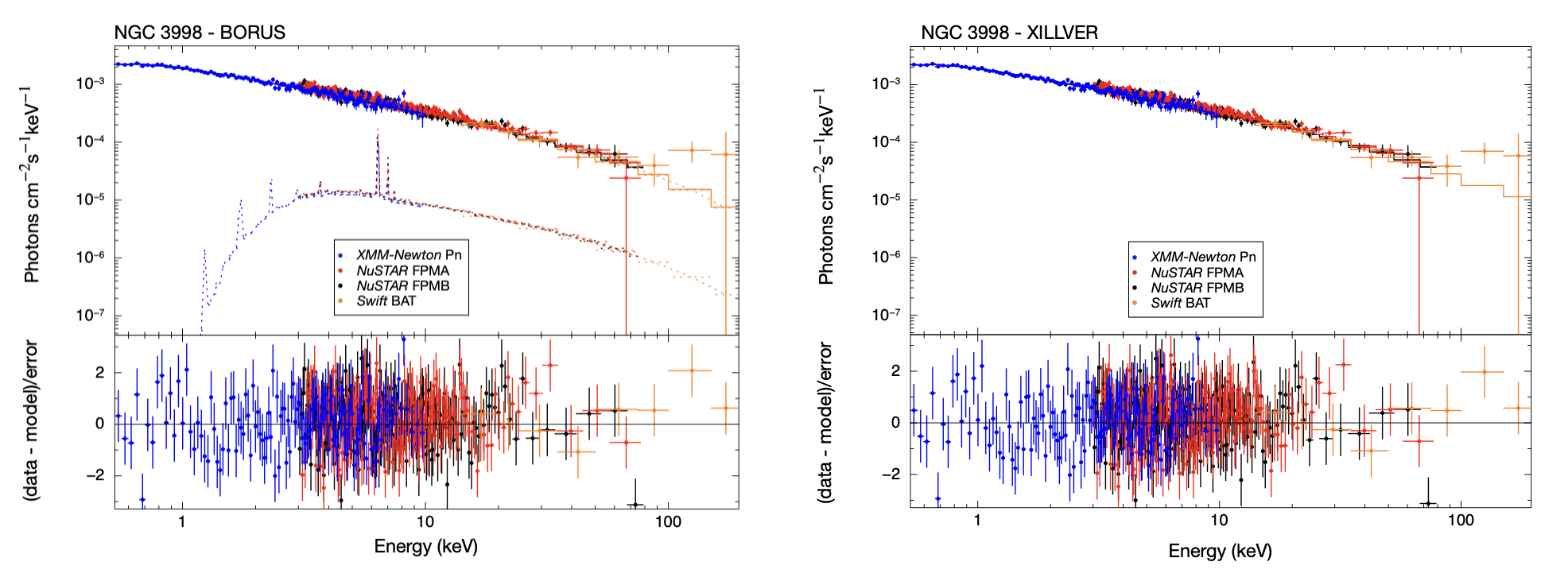}
    \includegraphics[width=0.850\textwidth]{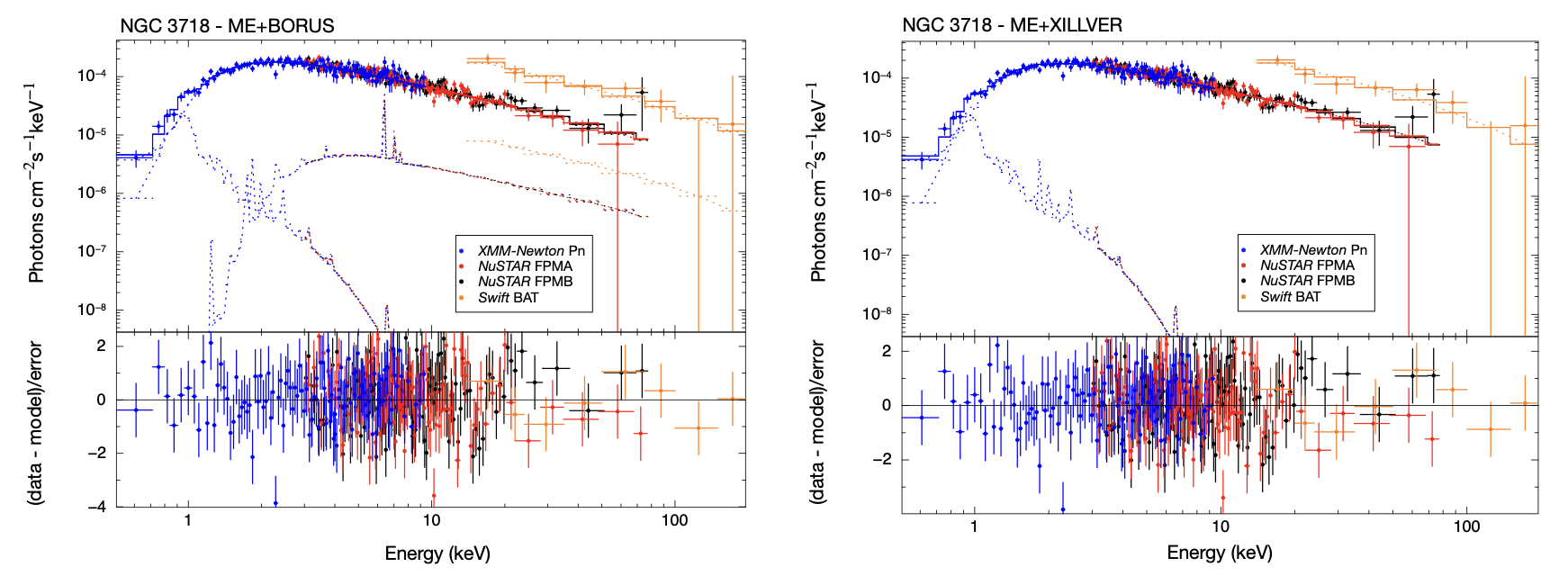}
    \includegraphics[width=0.850\textwidth]{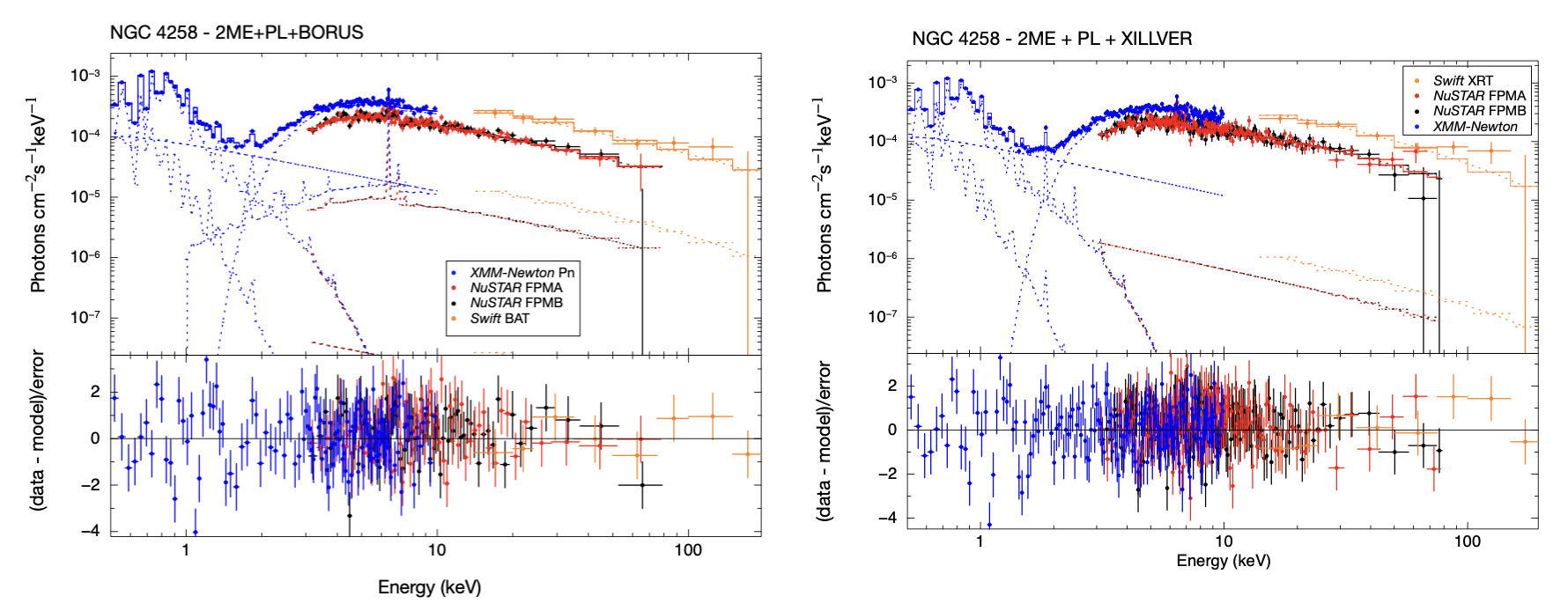}
    \includegraphics[width=0.850\textwidth]{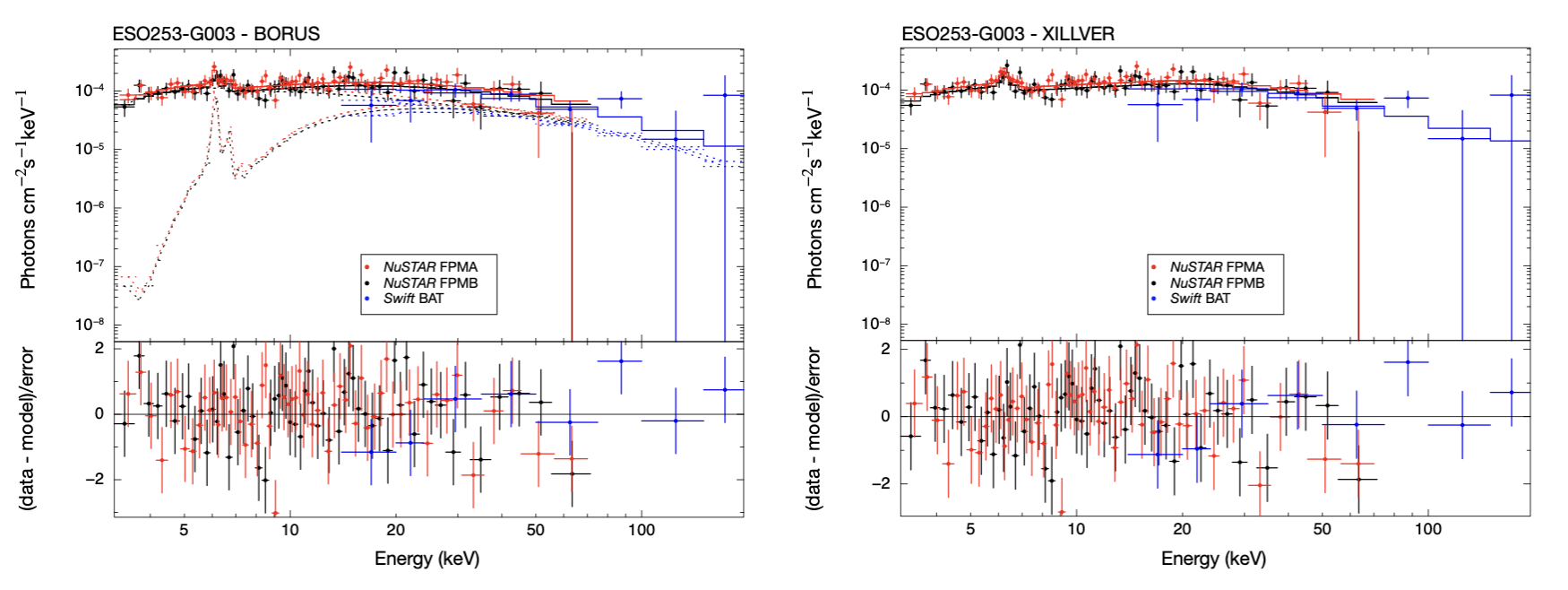}
    \caption{Spectral modelling of NGC\,3998, NGC\,3718, NGC\,4258* and ESO\,253-G003*. The plots correspond to \texttt{borus02} (left) and \texttt{XILLVER} (right). }
     \label{fig:spectr_1}
\end{figure*}


\begin{figure*}
\centering
    \includegraphics[width=0.850\textwidth]{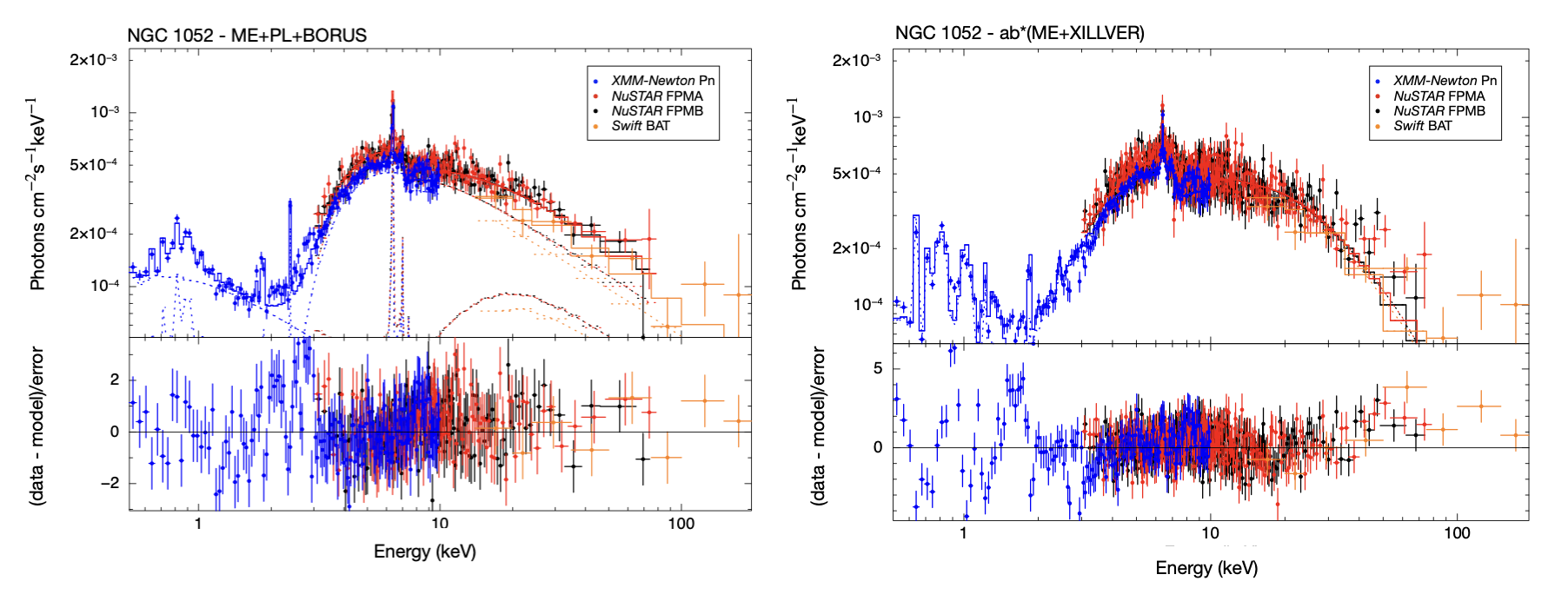}
    \includegraphics[width=0.850\textwidth]{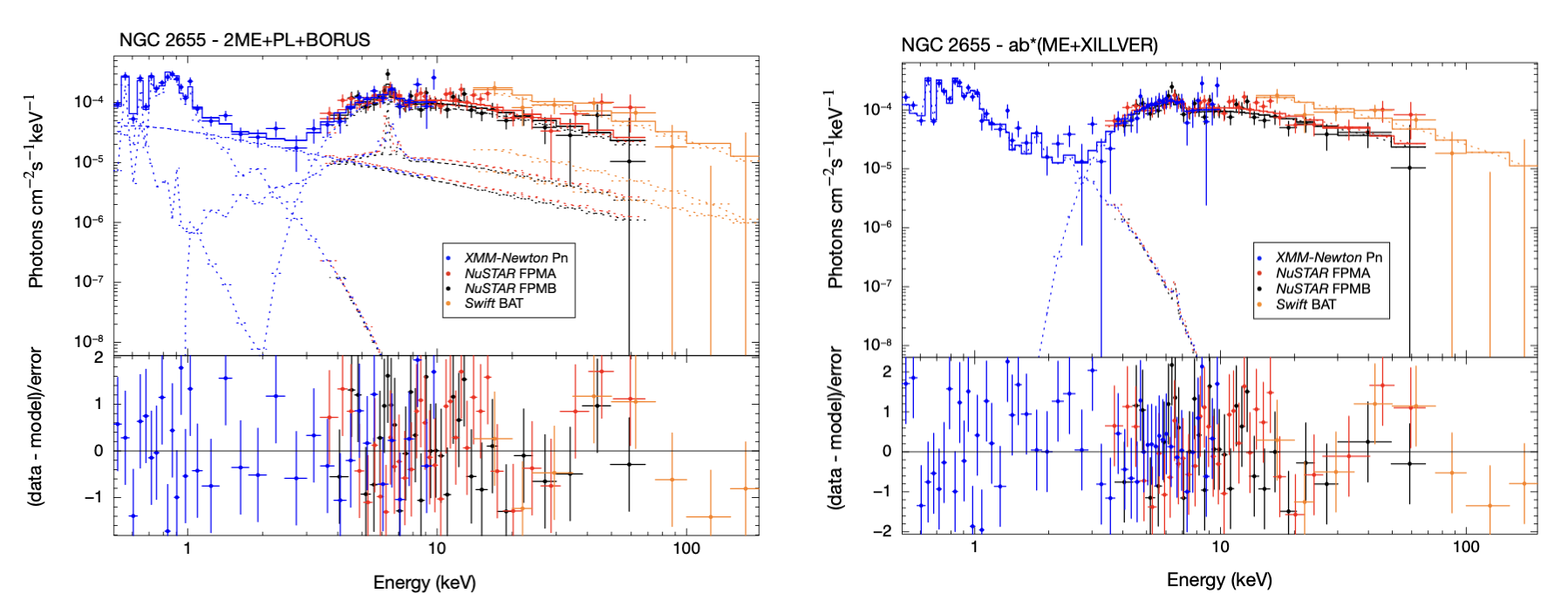}
    \includegraphics[width=0.850\textwidth]{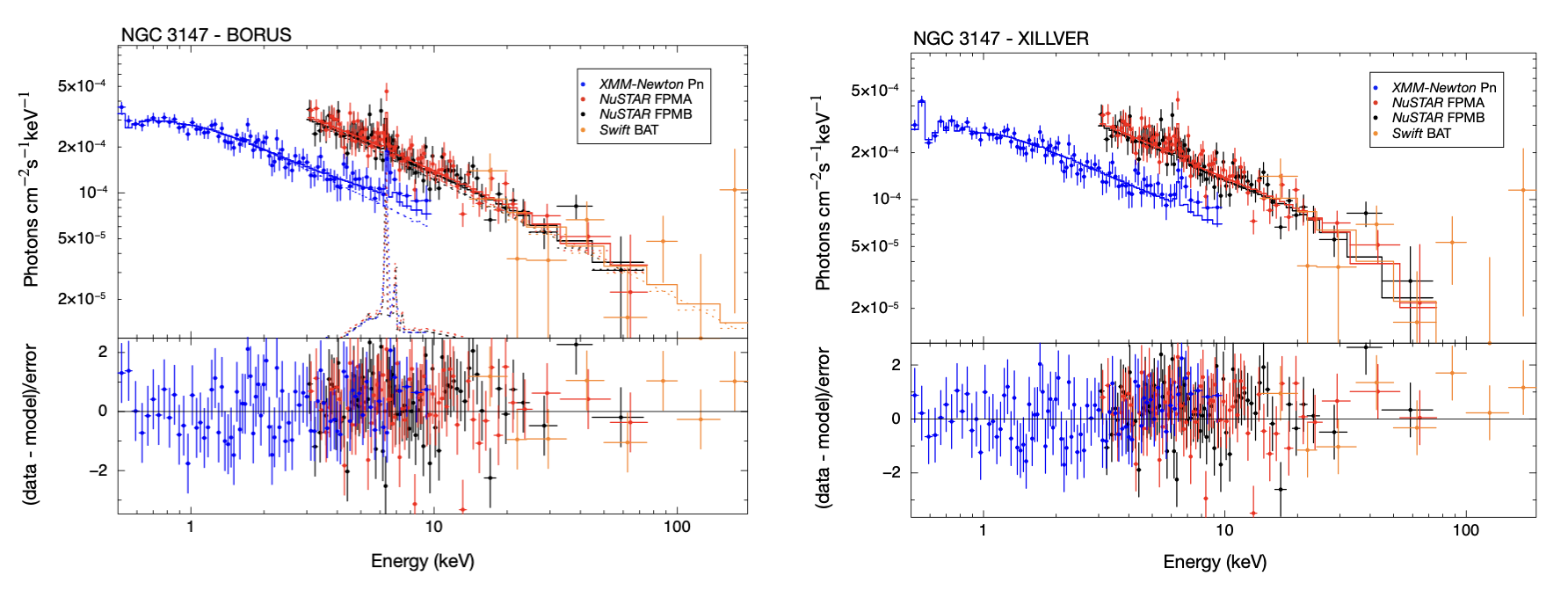}
    \includegraphics[width=0.850\textwidth]{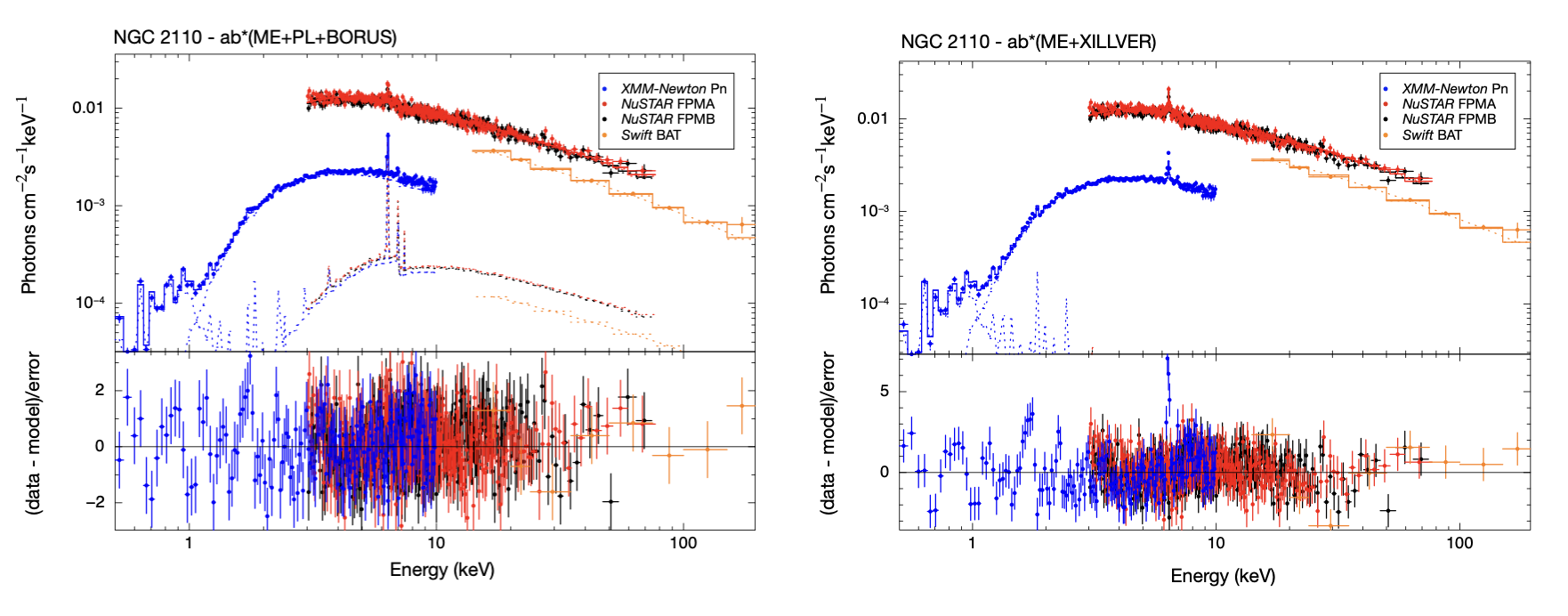}
    \caption{Spectral modelling of NGC 1052, NGC\,2655, NGC\,3147 and NGC\,2110. The plots correspond to \texttt{borus02} (left) and \texttt{XILLVER} (right). }
     \label{fig:spectr_2}
\end{figure*}


\begin{figure*}
\centering
    \includegraphics[width=0.850\textwidth]{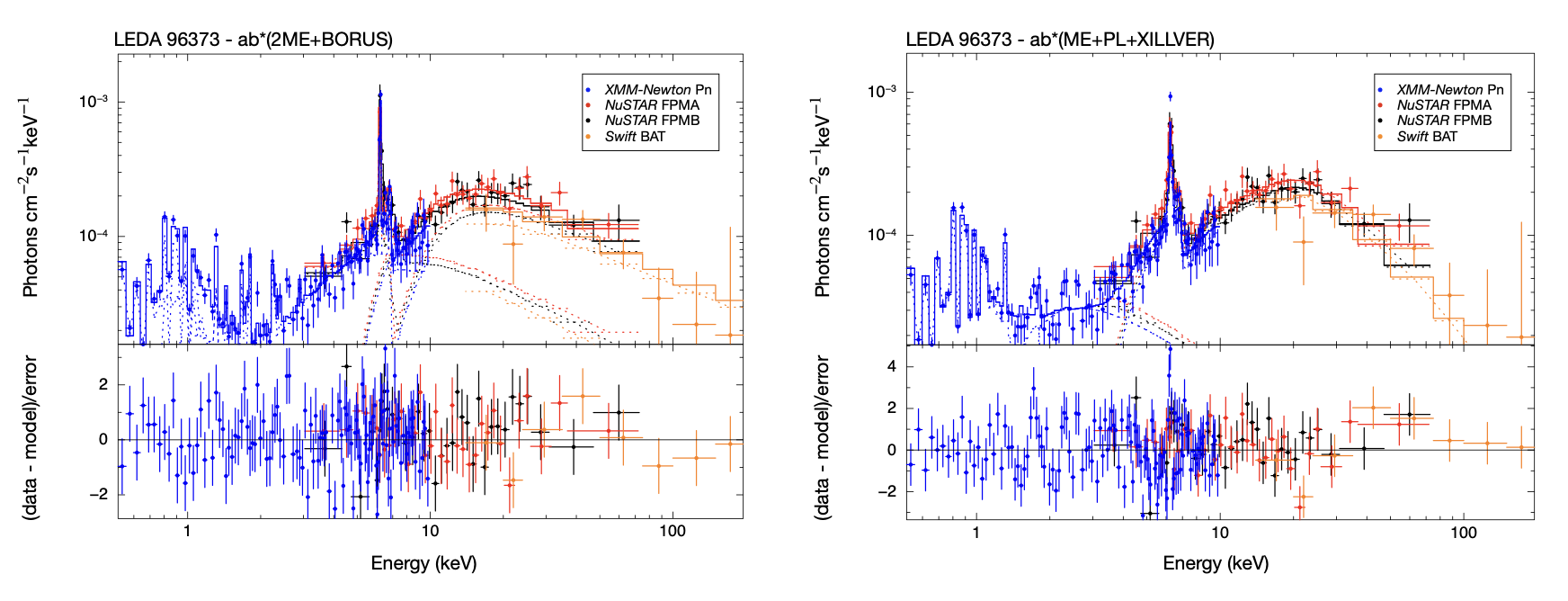}
    \includegraphics[width=0.850\textwidth]{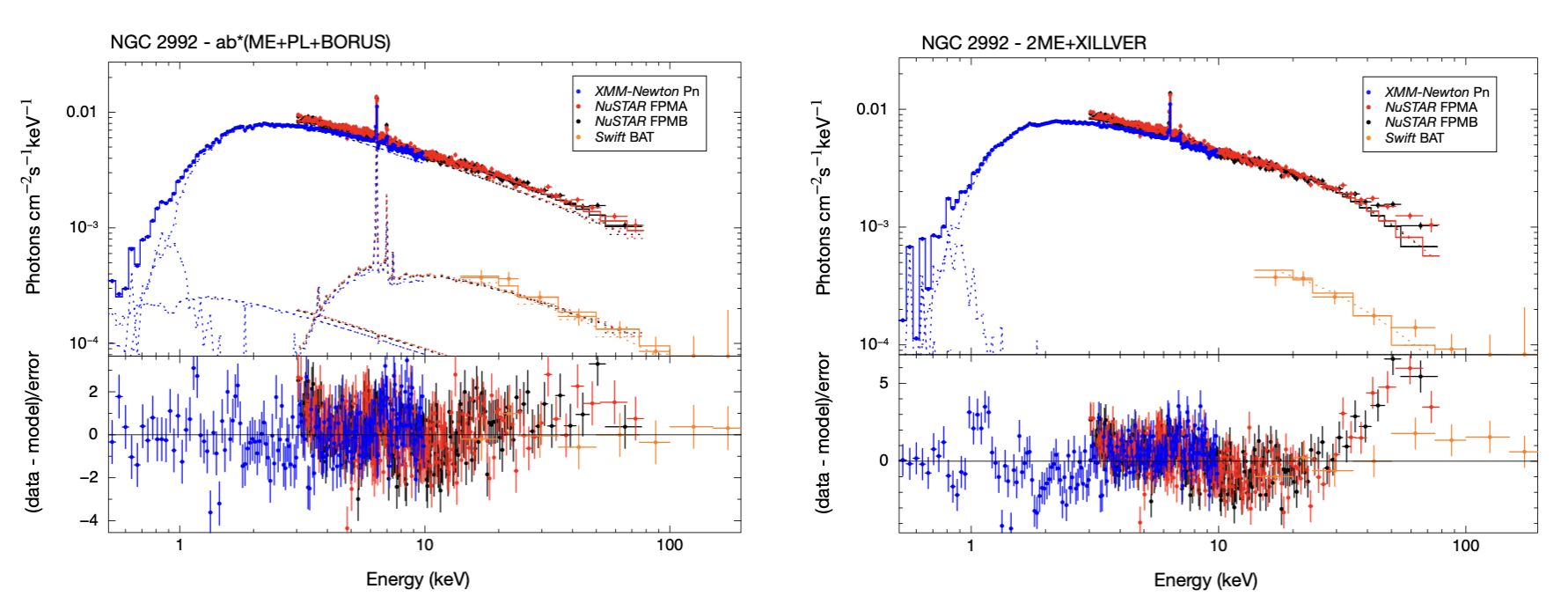}
    \includegraphics[width=0.850\textwidth]{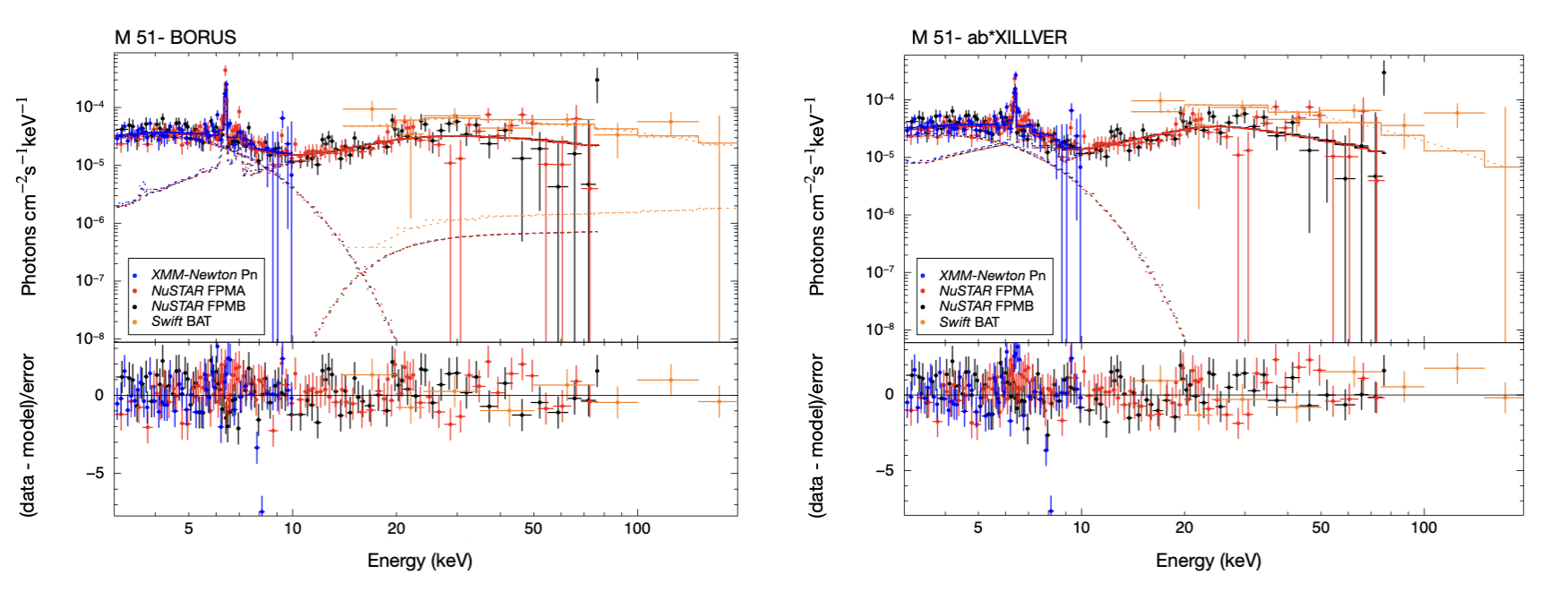}
    \includegraphics[width=0.850\textwidth]{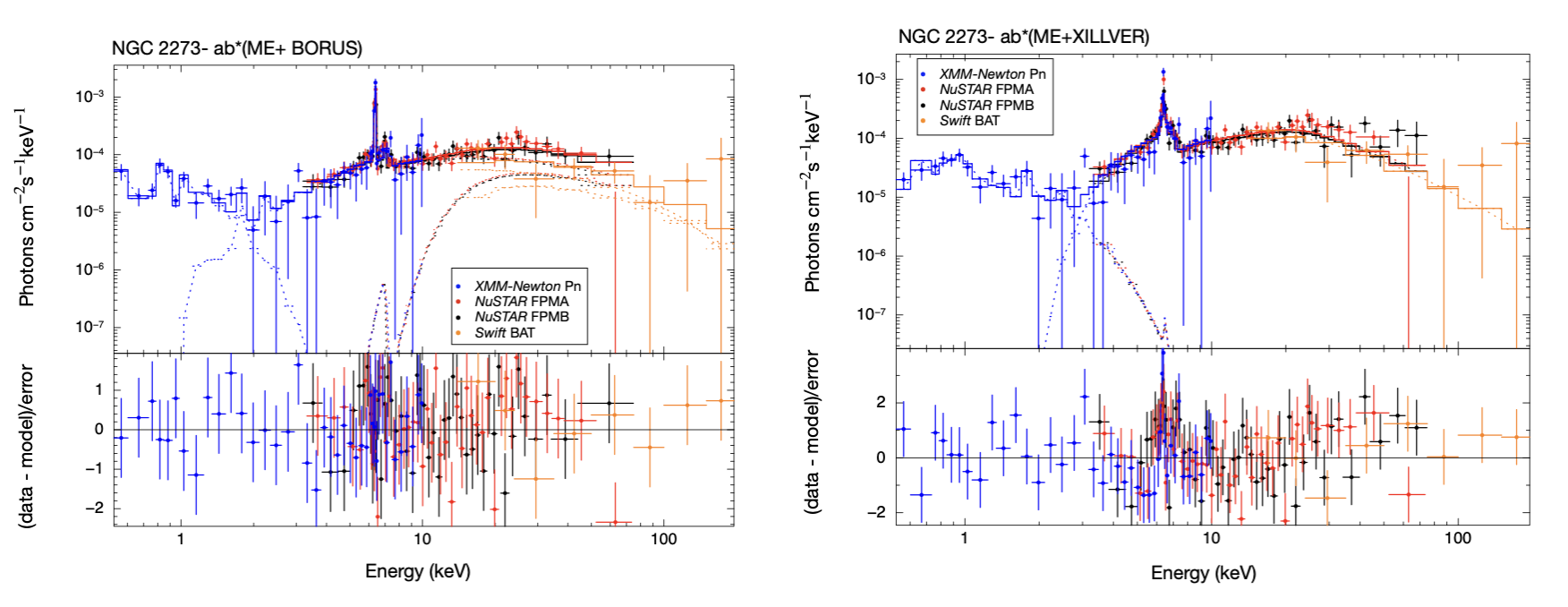}
    \caption{Spectral modelling of LEDA\,96373, NGC\,2992, M\,51 and NGC\,2273. The plots correspond to \texttt{borus02} (left) and \texttt{XILLVER} (right). }
     \label{fig:spectr_3}
\end{figure*}


\begin{figure*}
\centering
    \includegraphics[width=0.850\textwidth]{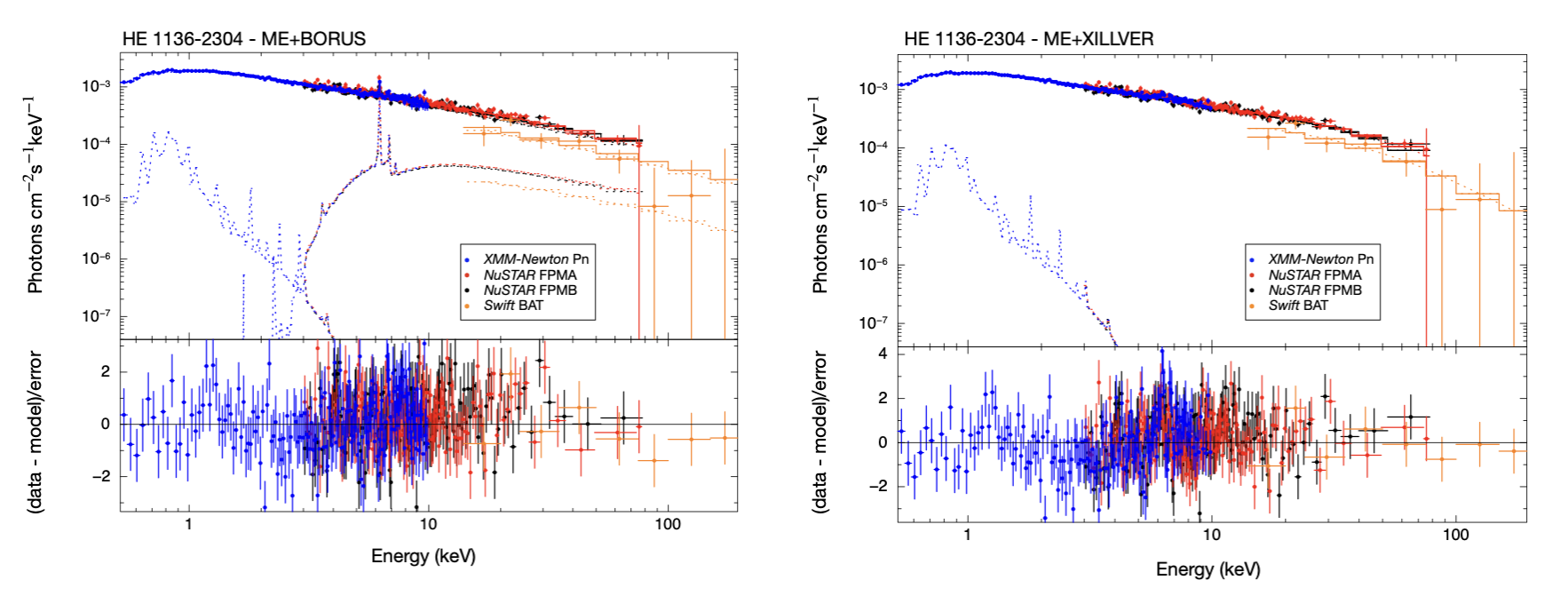}
    \includegraphics[width=0.850\textwidth]{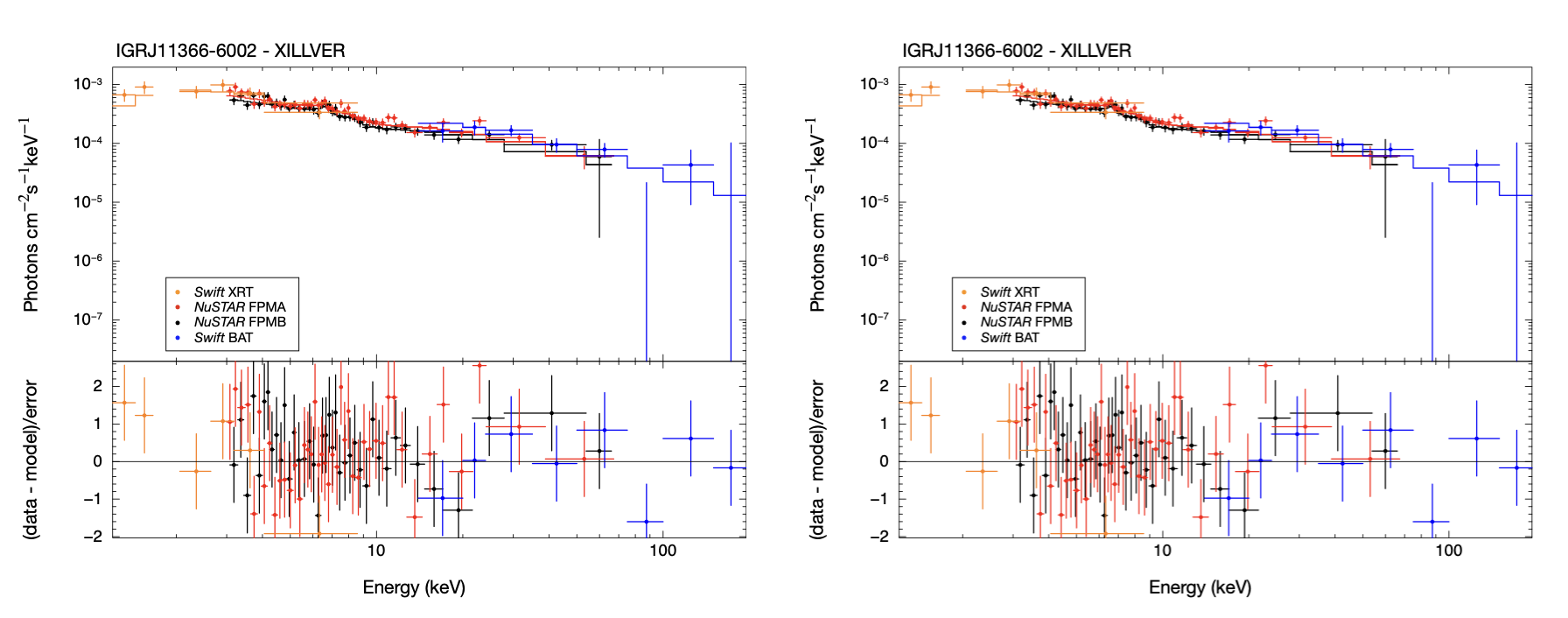}
    \includegraphics[width=0.850\textwidth]{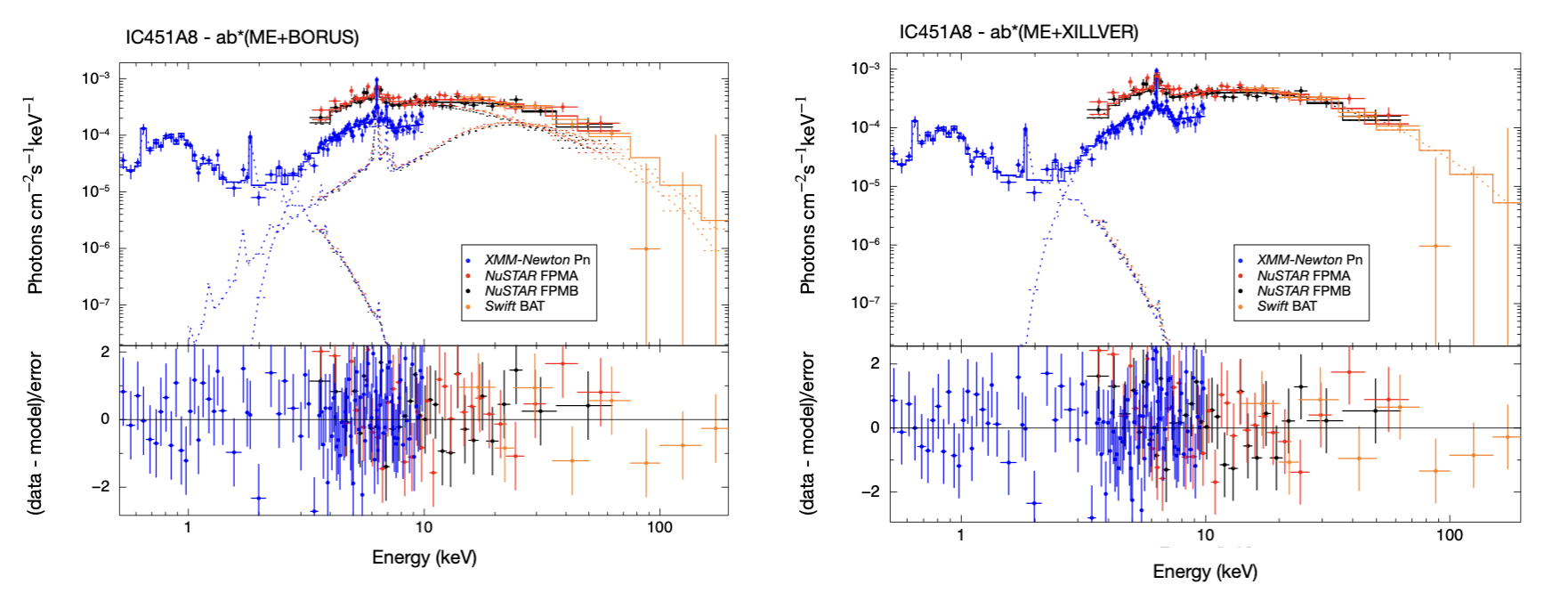}
    \includegraphics[width=0.850\textwidth]{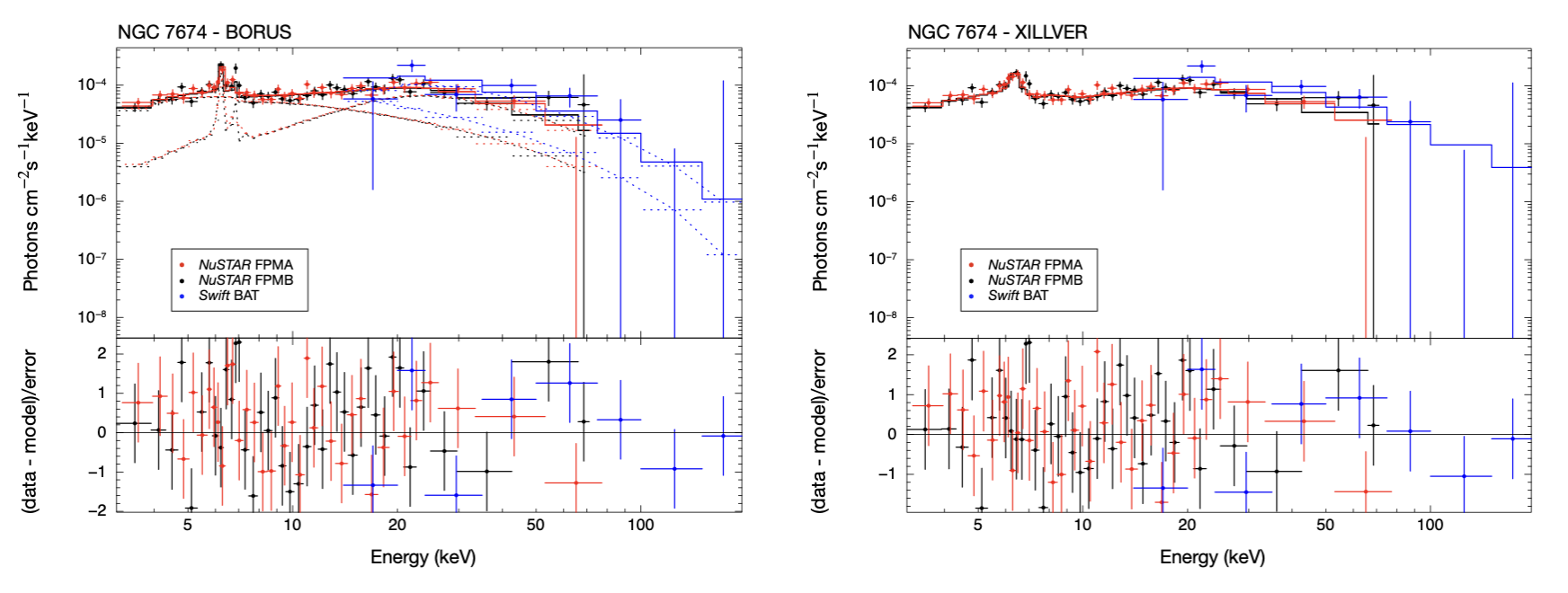}
    \caption{Spectral modelling of HE\,1136-2304, IGRJ\,11366-6002, IC4518 and NGC\,7674. The plots correspond to \texttt{borus02} (left) and \texttt{XILLVER} (right). }
     \label{fig:spectr_4}
\end{figure*}


\begin{figure*}
\centering
    \includegraphics[width=0.850\textwidth]{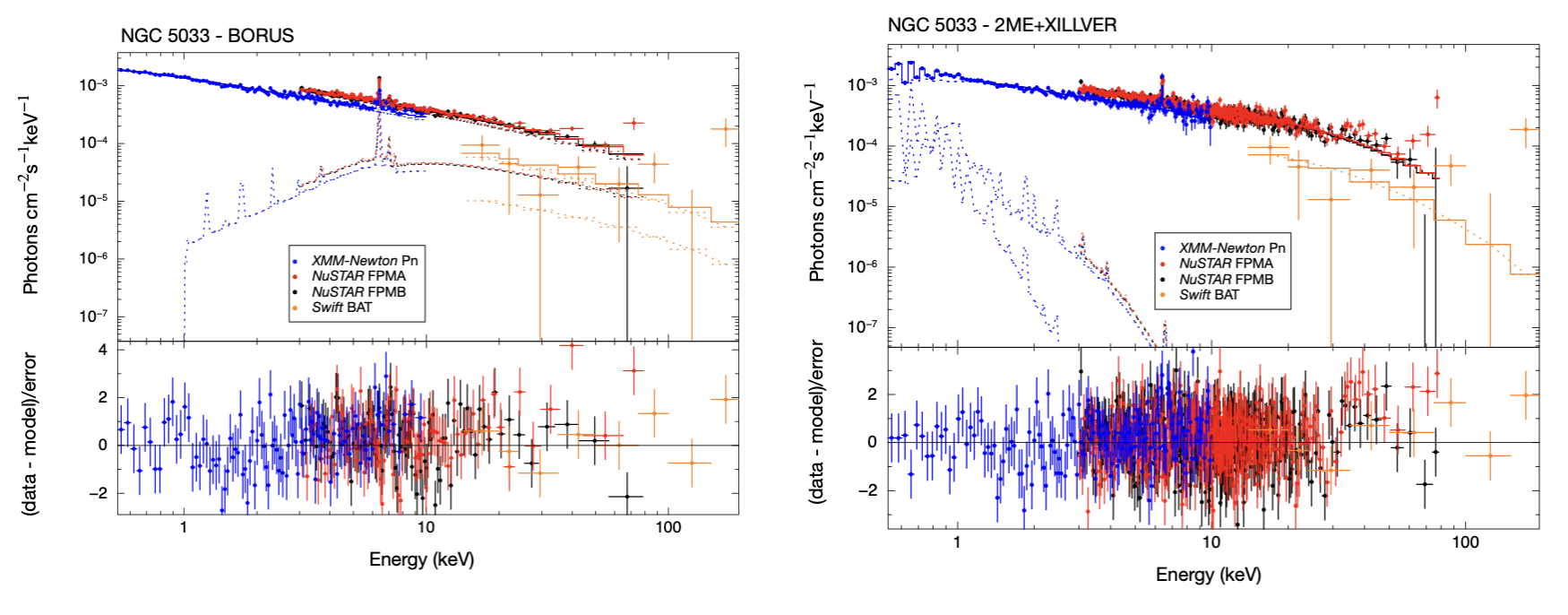}

    \caption{Spectral modelling of NGC\,5033. The plots correspond to \texttt{borus02} (left) and \texttt{XILLVER} (right). }
     \label{fig:spectr_5}
\end{figure*}

\end{appendix}

\end{document}